\documentclass[12pt,a4paper]{article}
\usepackage[utf8]{inputenc}
\usepackage[english]{babel}
\usepackage{amsmath}
\usepackage{subfigure}
\usepackage{amsfonts}
\usepackage{amssymb}
\usepackage{graphicx}
\newcommand{\be}{\begin{equation}}
\newcommand{\ee}{\end{equation}}
\newcommand{\bea}{\begin{eqnarray}}
\newcommand{\eea}{\end{eqnarray}}

\newcommand{\1}{\chi_{{}_{{}_1}}}
\newcommand{\2}{\chi_{{}_{{}_2}}}
\newcommand{\3}{\chi_{{}_{{}_3}}}

\catcode`@=12
\def\la{\mathrel{\mathchoice {\vcenter{\offinterlineskip\halign{\hfil
$\displaystyle##$\hfil\cr<\cr\sim\cr}}}
{\vcenter{\offinterlineskip\halign{\hfil$\textstyle##$\hfil\cr<\cr\sim\cr}}}
{\vcenter{\offinterlineskip\halign{\hfil$\scriptstyle##$\hfil\cr<\cr\sim\cr}}}
{\vcenter{\offinterlineskip\halign{\hfil$\scriptscriptstyle##$\hfil\cr<\cr\sim
\cr}}}}}

\begin{document}
%\begin{flushright}
%SINP-APC-15/--
%\end{flushright}
\thispagestyle{empty}
\begin{center}
{\Large \bf
{Nonthermal Two Component Dark Matter Model for Fermi-LAT $\gamma$-ray
excess and 3.55 keV X-ray Line}}\\
\vspace{0.25cm}
\begin{center}
{{\bf Anirban Biswas}$^{\dagger}$ \footnote{email: anirban.biswas@saha.ac.in},
{\bf Debasish Majumdar}$^{\dagger}$ \footnote{email: debasish.majumdar@saha.ac.in},
{\bf Probir Roy}$^{\ddagger}$ \footnote{email: probirrana@gmail.com}}\\
\vspace{0.5cm}
$^\dagger$ \it Astroparticle Physics and Cosmology Division, \\
\it Saha Institute of Nuclear Physics, Kolkata 700064, India \\
\vspace{0.25 cm}
$^\ddagger$\it Saha Institute of Nuclear Physics, Kolkata 700064, India \\
\&\\
\it Centre for Astroparticle Physics and Space science,\\
\it Bose Institute, Kolkata 700091, India
\end{center}
\vspace{1cm}
%%%%%%%%%%%%%%%%%%%%%%%%%%%%%%%%%%%%%%%%%%%%%%%%%%%%%%%%%%%%%%%%%%%%%%%%%%%%%%%%%%%%%%
{\bf ABSTRACT} \\
%%%%%%%%%%%%%%%%%%%%%%%%%%%%%%%%%%%%%%%%%%%%%%%%%%%%%%%%%%%%%%%%%%%%%%%%%%%%%%%%%%%%%%
\end{center}
A two component model of nonthermal dark matter is formulated to
simultaneously explain the Fermi-LAT results indicating a $\gamma$-ray excess
observed from our Galactic Centre in the 1-3 GeV energy range
and the detection of an X-ray line at 3.55 keV from extragalactic
sources. Two additional Standard Model singlet scalar fields $S_2$ and $S_3$
are introduced. These fields couple among themselves and with the
Standard Model Higgs doublet $H$. The interaction terms among the
scalar fields, namely $H$, $S_2$ and $S_3$, are constrained by the
application of a discrete $\mathbb{Z}_2\times \mathbb{Z}^\prime_2$ symmetry
which breaks softly to a remnant $\mathbb{Z}^{\prime \prime}_2$
symmetry. This residual discrete symmetry is then spontaneously
broken through an MeV order vacuum expectation value $u$ of the
singlet scalar field $S_3$. The resultant physical scalar spectrum
has the Standard Model like Higgs as $\chi_{{}_{{}_1}}$ with $M_{\chi_{{}_{{}_1}}}\sim 125$
GeV, a moderately heavy scalar $\chi_{{}_{{}_2}}$ with $50 \,\,{\rm GeV} \leq M_{\chi_{{}_{{}_2}}}\leq
80\,\,{\rm GeV}$ and a light $\chi_{{}_{{}_3}}$ with $M_{\chi_{{}_{{}_3}}} \sim 7$
keV. There is only tiny mixing between $\1$ and $\2$ as well as
between $\1$ and $\3$. The lack of importance of domain wall formation in the present
scenario from the spontaneous breaking of the discrete symmetry ${\mathbb{Z}_2^{\prime\prime}}$,
provided $u\leq 10$ MeV, is pointed out.
We find that our proposed two component dark matter model is able to
explain successfully both the above mentioned phenomena $-$ the
Fermi-LAT observed $\gamma$-ray excess (from the $\chi_{{}_{{}_2}} \rightarrow {\rm b}
\bar{\rm b}$ decay mode) and the observation of the X-ray line
(from the decay channel $\chi_{{}_{{}_3}}\rightarrow\gamma \gamma$)
by the XMM-Newton observatory. 
%%%%%%%%%%%%%%%%%%%%%%%%%%%%%%%%%%%%%%%%%%%%%%%%%%%%%%%%%%%%%%%%%%%%%%%%%%%%%%%%%%%%%%%%
%\begin{abstract}
%\end{abstract}
\vskip 1cm
%\qquad\quad Pacs: 95.35.+d, 98.80.Cq
%\vskip 1cm
%\quad\,\, Dark Matter, Beyond SM
\newpage
%%%%%%%%%%%%%%%%%%%%%%%%%%%%%%%%%%%%%%%%%%%%%%%%%%%%%%%%%%%%%%%%%%%%%%%%%%%%%%%%%%%%%%%%
\section{Introduction}
\label{intro}
%%%%%%%%%%%%%%%%%%%%%%%%%%%%%%%%%%%%%%%%%%%%%%%%%%%%%%%%%%%%%%%%%%%%%%%%%%%%%%%%%%%%%%%%
The presence of Dark Matter (DM) in the Universe is now an accepted reality. 
So far, its existence has been inferred only from its gravitational effects.
The latter include rotation curves of galaxies, gravitational lensing,
observations of the Bullet cluster etc. However, a very attractive proposition is that
DM consists of Weakly Interacting Massive Particles (WIMPs) \cite{Bertone:2004pz}
and this is the paradigm that we adopt.
The possibility of detecting DM WIMPs through direct and indirect 
processes is now being vigorously pursued by experimenters. Meanwhile, the amount
of DM present in the Universe has been precisely determined by
means of results from the PLANCK satellite \cite{Ade:2013zuv}
whose instruments have probed and analysed anisotropies
in the smooth cosmic microwave background.
One can think of two possible scenarios for the production of DM particles
in the early Universe. (1) This could have been due to thermal processes 
with Standard Model particles interacting with 
the thermal plasma in the expanding soup ball whereby all
possible particle-antiparticle pairs were produced in a
reverse chemical process. The decoupling of 
the DM particles occurred when their interaction rates fell short of 
the expansion rate of the Universe. Being out of equilibrium, 
the dark matter particles ``froze" to a particular relic density.
(2) The DM particles could have originated nonthermally. In this mode,
they got produced from out of equilibrium decays of heavier particles.
That could have occurred at the stage of the preheating of the
Universe from the inflaton energy.

Dark Matter, if accumulated in considerable measure at highly dense
regions of celestial bodies, such as at the Galactic Centre (GC) or
other sites in a galaxy, may undergo self annihilation. That would lead
to high energy photons or fermion-antifermion pairs. Such
hard photons may also arise from the radiative decay of a metastable
DM particle such as a sterile neutrino or a scalar. Therefore, the observation
and analysis of such high energy photons could not
only lead to an indirect detection of DM particles but
also provide insights into the physics of Dark Matter.
 
Recently, there have been two major observations of such high energy photons:
$\gamma$-rays from the centre of our Milky Way galaxy and X-rays from other
galaxies and galaxy clusters. The observation of a weak unidentified line
in the X-ray spectrum obtained from the XMM-Newton observatory, has been
reported by Bulbul et al \cite{Bulbul:2014sua}, from an analysis of
data from 73 galaxy clusters including Perseus and others. It is also
reported in same article \cite{Bulbul:2014sua} that the best fit value of the observed
X-ray flux obtained from the XMM-Newton MOS(PN) observation is $4.0^{+0.8}_{-0.8}
\times 10^{-6}$($3.9^{+0.6}_{-1.0}\times 10^{-6}$) photons cm$^{-2}$ s$^{-1}$ while
the energy of the X-ray line is $E= 3.57 \pm 0.02(3.51 \pm 0.03)$ keV.
This has been confirmed later by Boyarsky et al \cite{Boyarsky:2014jta}. 
In Ref. \cite{Boyarsky:2014jta} this 3.55 keV line has been claimed to
have been observed from the Andromeda as well as other galaxies in the
Local Group. Separately, analyses of the Fermi-LAT data by several
groups \cite{Goodenough:2009gk}-\cite{Calore:2014nla} during the last few years have been
indicating the presence of a significant excess at an energy range 1-3 GeV 
in the $\gamma$-ray spectrum observed from regions close to
the centre of our Milky way galaxy.
%Neither the 3.55 keV X-ray emission
%line nor the $\gamma$-ray excess around 1-3 GeV can be explained
%by known astrophysical processes occurring at these sites.
There have been attempts to explain the 3.55 keV line from
other astrophysical processes. For example, in Ref. \cite{Jeltema:2014qfa}
the authors claimed from their analysis of XMM-Newton data
that a 3.55 keV line from the Galactic Centre could be interpreted
from a known plasma line of astrophysical origin. In another
work by Carlson et. al. \cite{Carlson:2014lla} the 3.55 keV line from the Perseus
cluster is related to the cool core of the cluster. There have also been
attempts \cite{Malyshev:2014xqa} to obtain the XMM-Newton X-ray emission
line from dwarf galaxies. Authors have considered
sterile neutrino dark matter to explain the signature of this emission and
accordingly have given upper bounds on the mixing angle of the relevant
sterile neutrino DM. X-ray emission lines of the order of 3.5 keV
have also been probed from galaxy clusters, the Galactic Centre
and M31 where analyses \cite{Anderson:2014tza} have been made considering sterile
neutrino dark matter with a few keV mass. But the authors could not report any 
$\sim$ 3.5 keV excess. For the case of the $1$-$3$ GeV $\gamma$-ray excess from the Galactic Centre,
the authors of Ref. \cite{Petrovic:2014uda} attempted to demonstrate
the origin of the spectral and the angular features of this excess
from inverse compton scattering of high energy electrons available from
a burst event in the distant past. A possible millisecond pulsar origin
of this $\gamma$-ray excess from the Galactic Centre has been discussed
in Ref. \cite{Petrovic:2014xra}. But none of the above interpretations has 
addressed both the phenomena of the X-ray line and the $\gamma$-ray excess together. 
%Though there have been attempts to explain the 3.55 keV X-ray
%emission line \cite{Jeltema:2014qfa}-\cite{Carlson:2014lla} and the
%$\gamma$-ray excess around 1-3 GeV \cite{Petrovic:2014uda}-\cite{Petrovic:2014xra}
%by known astrophysical processes, such as two different millisecond pulsars
%or a plasma line, unlike the present work, none of these can account for both together.  
Hence it is natural to conjecture a common origin of both
phenomena from processes involving DM particles.
There have been many dark matter models in the literature explaining either the Fermi-LAT
observed $\gamma$-ray excess \cite{Boucenna:2011hy}-\cite{Cerdeno:2015ega}
or the X-ray line \cite{Krall:2014dba}-\cite{Babu:2014uoa} detected by the
XMM-Newton observatory. However, we are among the first two \cite {Cheung:2014tha}
to propose a new single dark matter model explaining both the phenomena simultaneously,
although our model of two SM-singlet scalars is very different
from that of Ref. \cite{Cheung:2014tha}. While our explanation of the $\gamma$-ray excess
is new, our explanation of the X-ray line uses the same mechanism as that of Ref. \cite{Babu:2014pxa}.
Once again, there have been previous works on two component dark matter models
\cite{Biswas:2014hoa}, \cite{Cao:2007fy}-\cite{Bian:2014cja},
however none of those has been aimed at explaining the Fermi-LAT $\gamma$-ray excess
and the 3.55 keV X-ray line simultaneously like we do.  

In this work we propose a single two-component model for nonthermal dark matter
%aiming to simultaneously explain the above two phenomena which occur in 
applying it to two completely different parts of the electromagnetic spectrum.
We have two distinct scalar dark matter particles, one being light and hence
``warm" ($\sim$ 7 keV in mass) and another which is
moderately heavy and ``cold", being in the mass range 50$-$80 GeV. Thus, we extend the
scalar sector of the Standard Model (SM) of particle physics by two real
scalar field $S_2$ ans $S_3$, both of which are singlets
under the Standard Model gauge group ${\rm SU}(2)_{\rm L}\times {\rm U}(1)_{\rm Y}$.
%Decays of the former is responsible for the X-ray line and those of the
%latter for the $\gamma$-ray excess.
We envisage nonthermal WIMPs and propose that both the DM particles were produced
from decays of the Standard Model Higgs boson as well as from the pair annihilation
of SM particles such as fermions, gauge bosons and Higgs bosons. The production
processes of these DM particles took place in the early stage of the Universe
when its temperature fell below the electroweak phase transition scale. 
We need not consider the chemical equilibrium in the thermal bath before
electroweak symmetry breaking but are concerned with the nonthermal
processes occurring afterwards that lead to the creation of the two lighter scalars.   
The interactions of the new scalar fields $S_2$, $S_3$ $-$ among themselves
as well as with the SM Higgs field $H$ $-$ are controlled by appropriate
discrete symmetries which are softly broken down to a residual
$\mathbb{Z}^{\prime\prime}_2$ symmetry under which both $S_2$ and $S_3$
are odd while all other fields are even. This residual discrete symmetry
gets spontaneously broken when $S_3$ develops a Vacuum Expectation Value (VEV)
$u$ of order MeV. This results in a $3 \times 3$ mass squared matrix in the three
dimensional space of the residual neutral SM Higgs field
and the two new singlet scalar fields. The diagonalisation
of that mass squared matrix leads to three physical scalar masses which are taken as
$M_{\chi_{{}_{{}_1}}} \sim 125$ GeV, $M_{\chi_{{}_{{}_2}}}\sim 50$-$80$ GeV
and $M_{\chi_{{}_{{}_3}}}\sim$ 7 keV corresponding respectively to the
SM-like Higgs $\chi_{{}_{{}_1}}$ and the two DM particles $\chi_{{}_{{}_2}}$
and $\chi_{{}_{{}_3}}$ \footnote{This mass range for the dark matter
component $\2$ is required to explain the Fermi-LAT excess $\gamma$-rays
from the decay channel $\2\rightarrow {\rm b}\bar{\rm b}$ ({\it cf.}
Section \ref{1-3gev-gamma} for a more detailed discussion) while the
mass of the other DM component $\3$ needs to be $7.1$ keV so that
its decay $\3\rightarrow \gamma \gamma$ produces two monoenergetic photons each with energy 3.55 keV.}.

The comoving number densities $Y_{\chi_{{}_{{}_2}}}$ and $Y_{\chi_{{}_{{}_3}}}$
of $\chi_{{}_{{}_2}}$, $\chi_{{}_{{}_3}}$ at the present epoch are calculated
by numerically solving the corresponding two coupled Boltzmann equations describing
their temperature evolution. These equations take into the account the roles played
by the decays $\chi_{{}_{{}_1}} \rightarrow \chi_{{}_{{}_j}} \chi_{{}_{{}_j}}$
($j =$ 2, 3) as well as the pair annihilation processes
$x\bar{x} \rightarrow \chi_{{}_{{}_j}} \chi_{{}_{{}_j}}$
where $x$ can be $W^\pm,\,\, Z,\,\, f(\bar{f}), \,\,\chi_{{}_{{}_1}},\,\,
\chi_{{}_{{}_2}}$, $f$ being any SM fermion. The individual relic densities
$\Omega_{\chi_{{}_{{}_2}}} h^2$ and $\Omega_{\chi_{{}_{{}_3}}} h^2$
at the present temperature then follow in a straightforward way once the above are
computed. The temperature variation of these, as well as of the total relic density
$\Omega_{\rm T} h^2$, are numerically studied for three cases: $\Omega_{\chi_{{}_{{}_2}}} >
\Omega_{\chi_{{}_{{}_3}}}$, $\Omega_{\chi_{{}_{{}_2}}} < \Omega_{\chi_{{}_{{}_3}}}$
and $\Omega_{\chi_{{}_{{}_2}}} \sim \Omega_{\chi_{{}_{{}_3}}}$.
The dependence on the mass $M_{\chi_{{}_{{}_2}}}$, which is not pinpointed, is also studied.
The total relic density is always found to lie in the range $0.1172 \leq {\Omega_{\rm T} h^2}
\leq 0.1226$ in conformity with the latest PLANCK results \cite{Ade:2013zuv}. 

We further consider the issue of domain wall formation from the restoration
of the discrete symmetry $\mathbb{Z}^{\prime\prime}_2$ and argue, {\it $\grave{a}$ la}
Babu and Mahapatra \cite{Babu:2014pxa}, to conclude that the corresponding energy
density would be too little to have any effect on the possible overclosing
of the Universe or the near-isotropy of the CMB so long as $u$ is bounded
from above by $\sim $10 MeV. Finally, we take up the two main issues towards
which this paper is addressed: the $\gamma$-ray excess observed by Fermi-LAT
at 1-3 GeV energies from our GC and the detection of an anomalous
3.55 keV X-ray line from extragalactic sources by the XMM Newton observatory.
For the former, we argue that $\chi_{{}_{{}_2}} \rightarrow {\rm b} \bar{\rm b}$
is the dominant decay mode acting as a source of GeV energy $\gamma$-rays
from decays of neutral pions created during the hadronisation of the
b and the $\bar{\rm b}$. The computation of the resultant $\gamma$-ray flux
from the decay of the dark matter candidate $\chi_{{}_{{}_2}}$ is carried through
using the NFW \cite{Navarro:1996gj} halo profile near
the GC. A comparison with the Fermi-LAT data provides an excellent fit
for the mass range of $\chi_{{}_{{}_2}}$ chosen by us. The observed 3.55 keV X-ray
line arises in our model from the decay $\chi_{{}_{{}_3}} \rightarrow \gamma \gamma$.
We first compute the partial decay width $\Gamma_{\chi_{{}_{{}_3}} \rightarrow \gamma \gamma}$
of the DM particle $\chi_{{}_{{}_3}}$ utilising three operative one loop diagrams. This partial
decay width of $\chi_{{}_{{}_3}}$ for the channel ${\chi_{{}_{{}_3}} \rightarrow \gamma \gamma}$,
however, needs to be modified by the factor $\frac{\Omega_{\chi_{{}_{{}_3}}}}{\Omega_{\rm T}}$
for the purpose of computing the differential X-ray flux since we are working
in a multicomponent dark matter scenario. We find that the modified decay width
for the channel $\chi_{{}_{{}_3}} \rightarrow \gamma \gamma$ lies within the range
predicted in Refs. \cite{Bulbul:2014sua, Boyarsky:2014jta, Higaki:2014qua} so long as
the VEV $u$ of the singlet scalar field $S_3$ is bounded from below by $2.4$ MeV.
The decay modes $\chi_{{}_{{}_2}} \rightarrow {\rm b} \bar{\rm b}$ and
$\chi_{{}_{{}_3}} \rightarrow \gamma \gamma$ arise respectively through the nonzero
mixing of both the dark matter candidates $\chi_{{}_{{}_2}}$ and $\chi_{{}_{{}_3}}$
with the SM-like Higgs boson $\chi_{{}_{{}_1}}$. Moreover,
the $\chi_{{}_{{}_1}} \chi_{{}_{{}_2}} \chi_{{}_{{}_2}}$
and $\chi_{{}_{{}_1}} \chi_{{}_{{}_3}} \chi_{{}_{{}_3}}$ couplings are sufficiently small (owing
to the nonthermal origin of $\chi_{{}_{{}_2}}$ and $\chi_{{}_{{}_3}}$) to evade all
the existing constraints from dark matter detection experiments
\cite{Aprile:2012nq, Akerib:2013tjd}.
\paragraph{}
The rest of the paper is organised as follows. In Section \ref{Model} we
describe the model and discuss its interscalar interactions as well as
the emergent $3 \times 3$ mass squared matrix in the scalar sector.
Section \ref{relic-density} presents the calculation of the relic densities
$\Omega_{\chi_{{}_{{}_2}}} h^2$, $\Omega_{\chi_{{}_{{}_3}}} h^2$ at the present temperature and
a discussion of their variation with temperature. In Section \ref{domenwall}
we discuss the issue of possible domain wall formation and point out why it
is unimportant in the present context. The spin independent elastic scattering
cross sections of both the dark matter particles are computed in Section \ref{derect}.
Section \ref{1-3gev-gamma} contains our calculation of the $\gamma$-ray flux arising from the decay of
the dark matter component $\chi_{{}_{{}_2}}$ and its comparison with the available
Fermi-LAT data. In Section \ref{xray} we compute the decay width
$\Gamma_{\chi_{{}_{{}_3}} \rightarrow \gamma \gamma}$ for the channel
${\chi_{{}_{{}_3}} \rightarrow \gamma \gamma}$ which is required to
produce the observed 3.55 keV X-ray line from the decay
of the dark matter particle $\chi_{{}_{{}_3}}$.
The final Section \ref{conclusion} summarises our conclusions. The two
appendices (\ref{a1} and \ref{a2}) contain the algebraic expressions for the couplings and masses
of the physical scalars $\chi_{{}_{{}_1}}$, $\chi_{{}_{{}_2}}$ and $\chi_{{}_{{}_3}}$.   
%%%%%%%%%%%%%%%%%%%%%%%%%%%%%%%%%%%%%%%%%%%%%%%%%%%%%%%%%%%%%%%%%%%%%%%%%%%%%%%%%%%%%%%%%%
\section{The Model}
\label{Model}
%%%%%%%%%%%%%%%%%%%%%%%%%%%%%%%%%%%%%%%%%%%%%%%%%%%%%%%%%%%%%%%%%%%%%%%%%%%%%%%%%%%%%%%%%%
We start with the Standard Model fields including, in particular, the SU(2)$_{\rm L}$
Higgs doublet
\begin{eqnarray}
H = \left(\begin{array}{c}
     h^+\\
      \frac{s_{{}_1} +ip_{{}_1} + v}{\sqrt{2}}\\
    \end{array}\right) \,\, .
\label{higgs}
\end{eqnarray}
Here $s_{{}_1}$ is the residual neutral SM Higgs field, $h^+$ and $p_{{}_1}$ are
unphysical charged and neutral pseudoscalar Higgs fields while $v$ is the
Vacuum Expectation Value $\simeq$ 246 GeV. Thus
\begin{eqnarray}
\langle H\rangle = \left(\begin{array}{c}
0\\
\frac{v}{\sqrt{2}}\\
\end{array}\right) \,\, .
\end{eqnarray}
Our DM sector consists of two real scalar fields $S_2$ and $S_3$ which are
singlets under the SM gauge group ${\rm SU}(2)_{\rm L}\times {\rm U}(1)_{\rm Y}$.
The Lagrangian of the spin zero sector
of the present model can then be written in terms of the
${\rm SU}(2)_{\rm L}\times {\rm U}(1)_{\rm Y}$ gauge covariant derivative
$D_{\mu}$ and a scalar potential function $V(H, S_2, S_3)$: 
\begin{eqnarray}
\mathcal{L} &=& (D_{\mu}H)^{\dagger}(D^{\mu}H) +
\frac{1}{2}\partial_{\mu} {S_2} \partial^{\mu} {S_2} +
\frac{1}{2}\partial_{\mu} {S_3} \partial^{\mu} {S_3} - V(H, S_2, S_3) \,\, .
\label{lagrangian}
\end{eqnarray}
The stability of the newly added scalars $S_{2,3}$ is ensured by the
postulated invariance of the model
under the discrete symmetry $\mathbb{Z}_2 \times \mathbb{Z}^{\prime}_2$ with
respect to which only these two fields transform nontrivially. The
$\mathbb{Z}_2 \times \mathbb{Z}^{\prime}_2$ charges of $S_2$ and $S_3$ are (-1, 1)
and (1, -1) respectively, whereas the corresponding charge of every other
field in the model is (1, 1). Thus, while $S_2$ ($S_3$) is odd (even) under $\mathbb{Z}_2$,
the reverse is the case for $S_2$ ($S_3$) with respect to $\mathbb{Z}^\prime_2$. 
The function $V(H, S_2, S_3)$ of Eq. (\ref{lagrangian})
contains all possible $\mathbb{Z}_2 \times \mathbb{Z}^{\prime}_2$
invariant interaction terms among
the Standard Model Higgs doublet field $H$ and the singlet fields
$S_2$, $S_3$ including all allowed self interactions and mass
terms. We write it  explicitly in Eq. (\ref{potential}):
\begin{eqnarray}
V(H, S_2, S_3) &=& \kappa_{{}_1} \left(H^{\dagger}H-
\frac{v^2}{2}\right)^2 + \frac{\kappa_{{}_2}}{4}S_2^4
+ \frac{\kappa_{{}_3}}{4}(S_3^2-u^2)^2 + \frac{\rho_{{}_2}^2}{2}S_2^2
+ \lambda_{12} (H^{\dagger}H) S_2^2 
+ \lambda_{23} S_2^2 S_3^2 \nonumber \\
&& +~\lambda_{13}\left(H^{\dagger}H-\frac{v^2}{2}\right)(S_3^2 - u^2)\,\,.
\label{potential}
\end{eqnarray}
The required invariance under the symmetry $\mathbb{Z}_2 \times \mathbb{Z}^{\prime}_2$
excludes terms such as $(H^\dagger H)S_2 S_3$, $S_2^3 S_3$, $S_2 S_3^3$.
We also assume that, due to high scale physics, a `soft' term
\begin{equation}
V^\prime = \alpha S_2 S_3 
\label{v-soft}
\end{equation}
gets added to $V$, explicitly breaking the $\mathbb{Z}_2 \times \mathbb{Z}^{\prime}_2$
invariance down to $\mathbb{Z}^{\prime\prime}_2$ under which both
$S_2$ and $S_3$ are odd. We have introduced six `hard' couplings in
$V$ $-$ $\kappa_{{}_{1,2,3}}$, and $\lambda_{12}$, $\lambda_{23}$, $\lambda_{13}$ $-$
all of which describe quartic field interactions in their leading terms.
Besides the six quartic couplings, we have also a dimensional coupling $\rho_{{}_2}$.
The nonleading terms in the potential $V$ have been chosen in a way that
$V$ is manifestly minimised at $\langle H \rangle=\frac{v}{\sqrt{2}}$,
$\langle S_2\rangle = 0$, $\langle S_3 \rangle = u$. The VEV $u$ needs to be in the
range $2\,\,{\rm MeV}\,\,<u \leq 10\,\,{\rm MeV}$ (see Sections \ref{domenwall}, \ref{xray}
for details). As a result, the residual discrete symmetry
$\mathbb{Z}^{\prime\prime}_2$ gets spontaneously broken.
The stability of the potential $V$ in Eq. (\ref{potential}) can be
investigated following the procedure of Ref. \cite{Kannike:2012pe}.
The required conditions are:
\begin{eqnarray}
&&\kappa_{{}_{1,2,3}} > 0~, \\
&&\rho_{{}_2} > 0~,\\
&&\lambda_{12} + \sqrt{\kappa_{{}_1} \kappa_{{}_2}} > 0~,\\
&&\lambda_{13} + \sqrt{\kappa_{{}_1} \kappa_{{}_3}} > 0~,\\
&&\lambda_{23} + \frac{1}{2}\sqrt{\kappa_{{}_2} \kappa_{{}_3}} > 0~,
\end{eqnarray}
\begin{eqnarray}
&&\sqrt{2\left(\lambda_{12} + \sqrt{\kappa_{{}_1} \kappa_{{}_2}}\right)
\left(\lambda_{13} + \sqrt{\kappa_{{}_1} \kappa_{{}_3}}\right)
\left(2\lambda_{23} + \sqrt{\kappa_{{}_2} \kappa_{{}_3}}\right)} \nonumber \\
&& + \sqrt{\kappa_{{}_1}\kappa_{{}_2}\kappa_{{}_3}} + \lambda_{12}\sqrt{\kappa_{{}_3}} +
\lambda_{13}\sqrt{\kappa_{{}_2}} + 2\lambda_{23}\sqrt{\kappa_{{}_1}} > 0~.
\end{eqnarray}

In analogy with the residual neutral SM Higgs field
$s_{{}_1}$, we can define the corresponding
$s_{{}_2} \equiv S_2$ and $s_{{}_3} = S_3 - u$. Now, in the
$s_{{}_1}$-$s_{{}_2}$-$s_{{}_3}$ system, the squared mass matrix
is given by
\begin{eqnarray}
\mathcal{M}^2=\left( \begin{array}{ccc}
2 \kappa_{{}_1} v^2 & 0 & 2\lambda_{13}u v\\
0 & \rho_{{}_2}^2 + \lambda_{12} v^2 + 2\lambda_{23} u^2 & \alpha \\
2\lambda_{13}u v  & \alpha & 2 \kappa_{{}_3} u^2 \\
\end{array}\right) \,\, .
\label{mass-matrix}
\end{eqnarray}
The eigenvalues of $\mathcal{M}^2$ are designated $M^2_{\chi_{{}_{{}_1}}}$, $M^2_{\chi_{{}_{{}_2}}}$
and $M^2_{\chi_{{}_{{}_3}}}$. The eigenstates, namely the physical scalars $\chi_{{}_{{}_1}}$, 
$\chi_{{}_{{}_2}}$, $\chi_{{}_{{}_3}}$, are linearly related to $S_1$, $S_2$, $S_3$ by an orthogonal
transformation which may be characterised in the usual way by the mixing
angles $\theta_{12}$, $\theta_{23}$ and $\theta_{13}$, The latter are the
rotation angles needed to rotate the $S_1$-$S_2$, $S_2$-$S_3$ and $S_3$-$S_1$
mass squared submatrices ($2\times 2$ matrices) sequentially such that the whole
mass squared matrix $\mathcal{M}^2$ (Eq. (\ref{mass-matrix})) becomes diagonal.
The approximate expressions for $M^2_{\chi_{{}_{{}_1}}}$, $M^2_{\chi_{{}_{{}_2}}}$ and
$M^2_{\chi_{{}_{{}_3}}}$ are given in the Appendix \ref{a2}.
We only note that $M_{\chi_{{}_{{}_1}}}>M_{\chi_{{}_{{}_2}}}>M_{\chi_{{}_{{}_3}}}$
with $M_{\chi_{{}_{{}_1}}}\sim 125\,\,{\rm GeV}$,
$M_{\chi_{{}_{{}_2}}}\simeq 50-80$ GeV and $M_{\chi_{{}_{{}_3}}}\sim 7\,\,{\rm keV}$.
\footnote{The light mass $\mathcal{O}$(keV) for the dark matter component
$\chi_{{}_{{}_3}}$ arises from an order MeV value of $u$ ({\it cf.} Sections
\ref{domenwall}, \ref{xray}) so long as the parameter $\kappa_{{}_3}$ lies in the
range $\sim 2\times 10^{-7}$ to $4\times 10^{-6}$.}
%%%%%%%%%%%%%%%%%%%%%%%%%%%%%%%%%%%%%%%%%%%%%%%%%%%%%%%%%%%%%%%%%%%%%%%%%%%%%%%%%%%%%%%
\section{Relic Density Calculation of Two Component Nonthermal Dark Matter}
\label{relic-density}
%%%%%%%%%%%%%%%%%%%%%%%%%%%%%%%%%%%%%%%%%%%%%%%%%%%%%%%%%%%%%%%%%%%%%%%%%%%%%%%%%%%%%%%
In the present scenario both the heavier and lighter dark matter components
would be produced nonthermally in the early Universe. We assume 
that, when the temperature of the Universe was above that of electroweak symmetry
breaking, there was no source of production of the DM
particles $\chi_{{}_{{}_2}}$ and $\chi_{{}_{{}_3}}$.
\begin{figure}[h!]
\includegraphics[height=3.0cm,width=4.0cm]{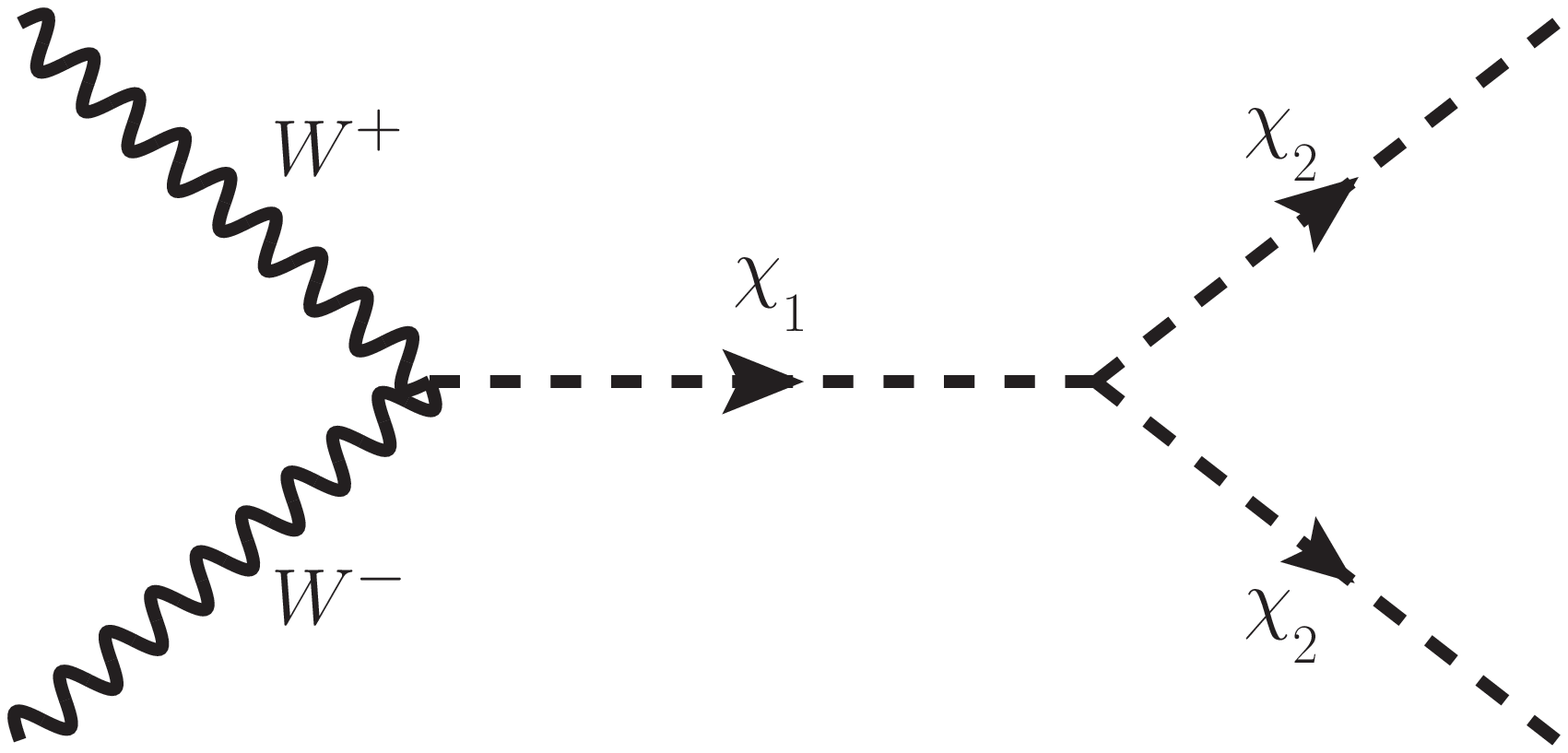}
\hspace{0.1 cm}
\includegraphics[height=3.0cm,width=4.0cm]{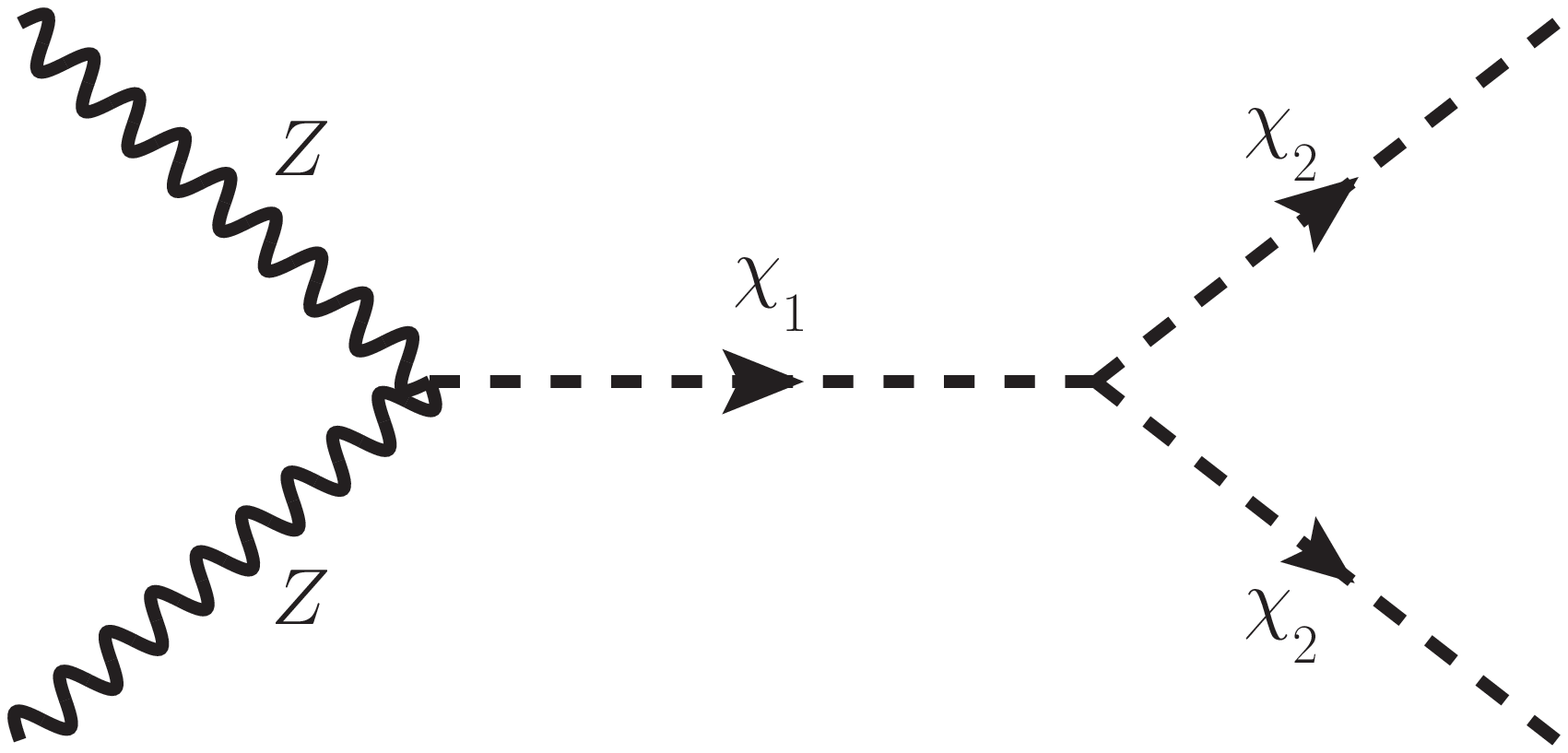}
\hspace{0.1 cm}
\includegraphics[height=3.0cm,width=4.0cm]{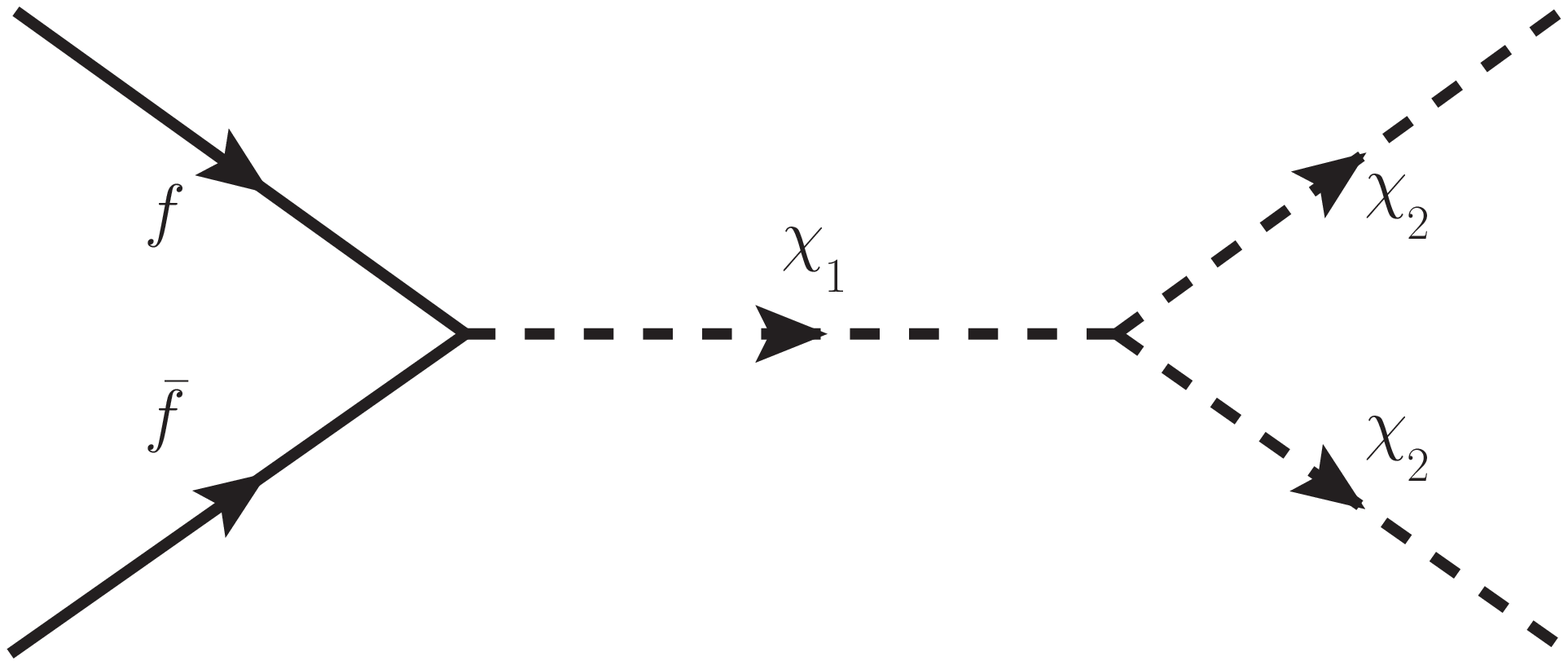}
\hspace{0.1 cm}
\includegraphics[height=3.0cm,width=4.0cm]{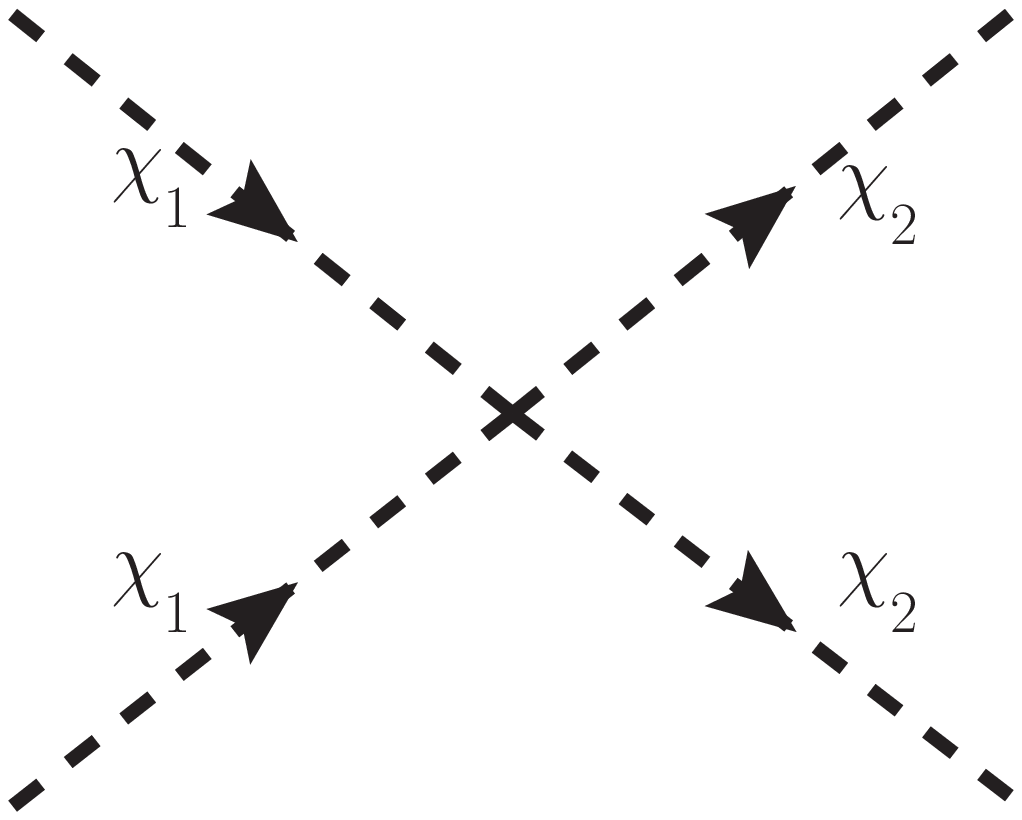}\\
\includegraphics[height=3.0cm,width=4.0cm]{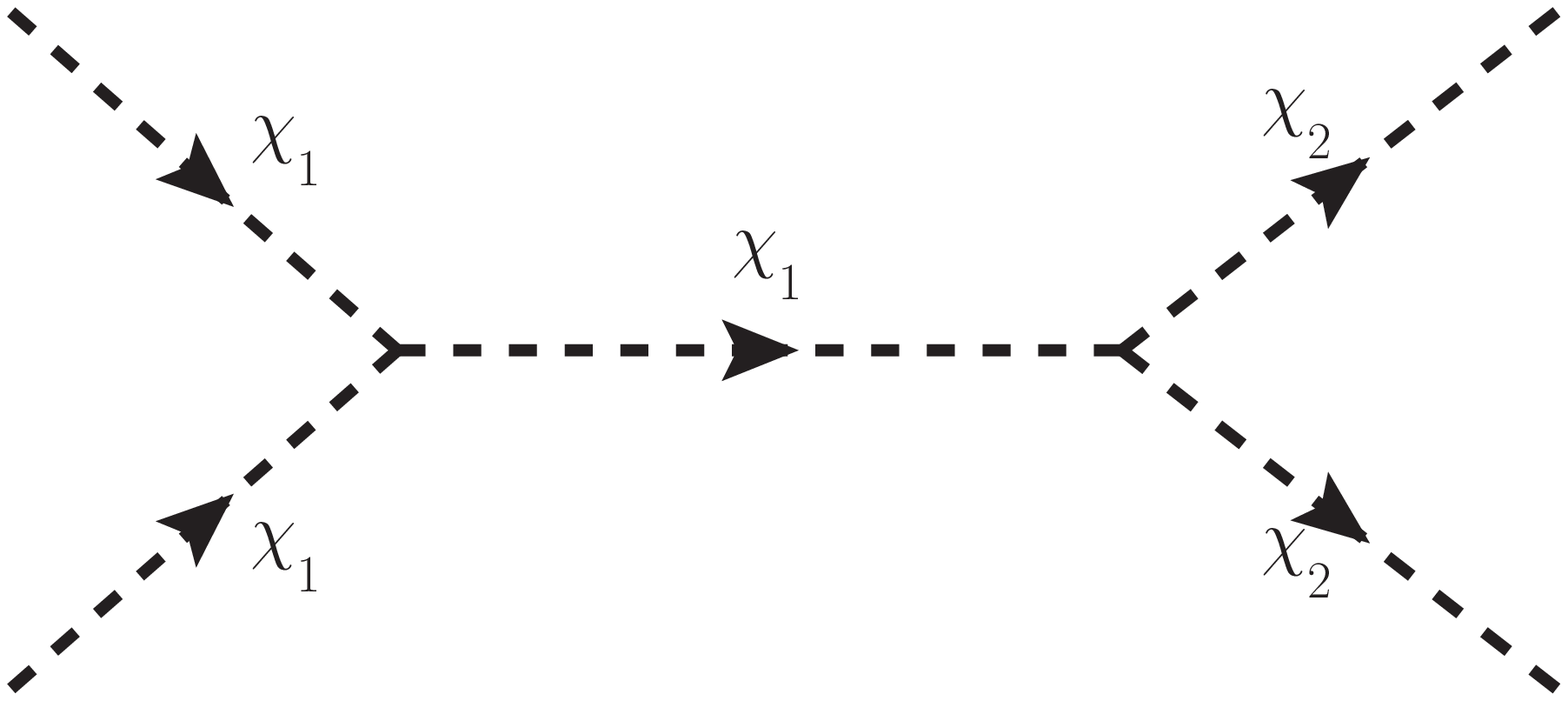}
\hspace{0.1cm}
\includegraphics[height=3.0cm,width=4.0cm]{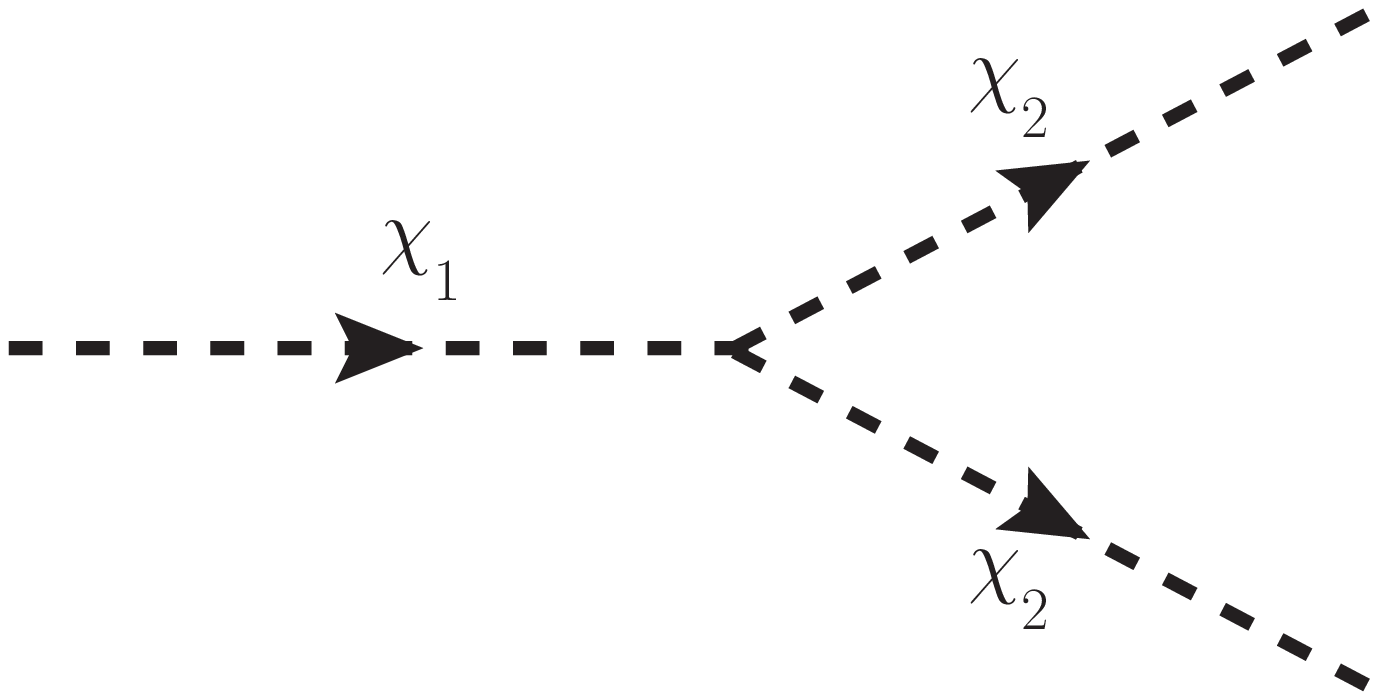}
\hspace{0.1cm}
\includegraphics[height=3.0cm,width=4.0cm]{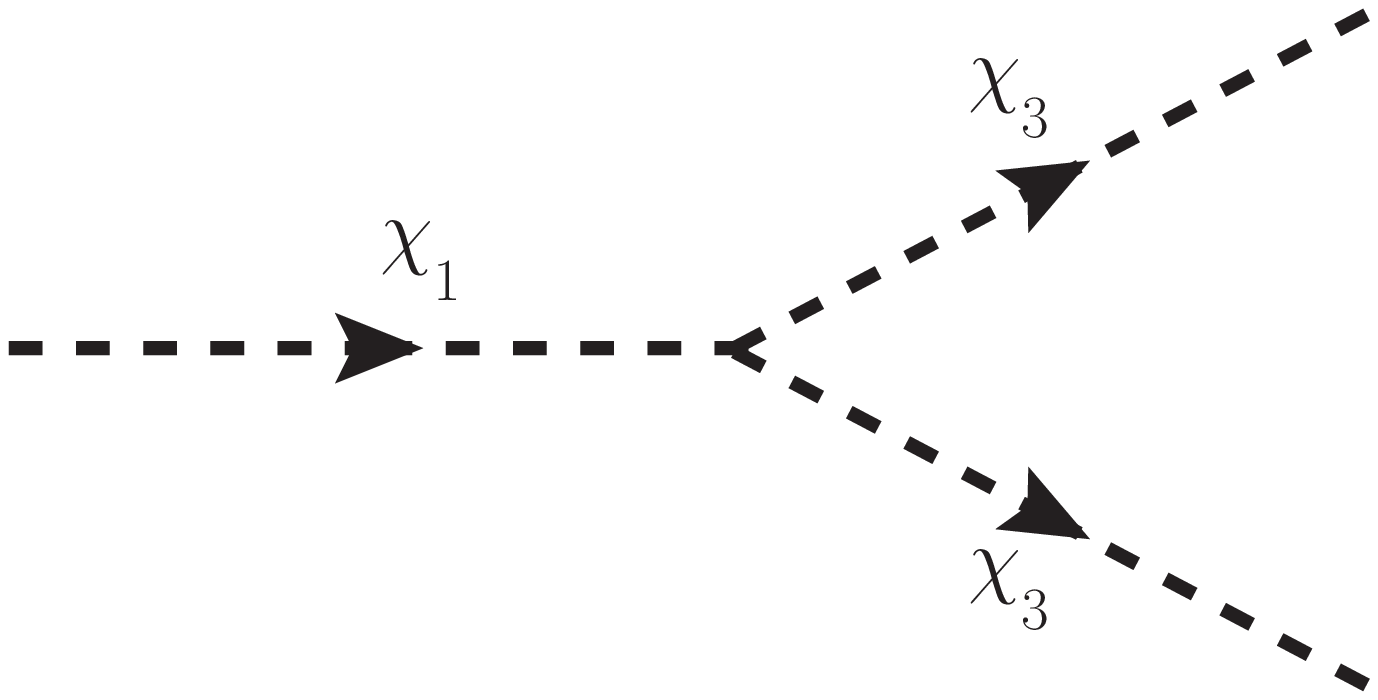}
\hspace{0.1cm}
\includegraphics[height=3.0cm,width=4.0cm]{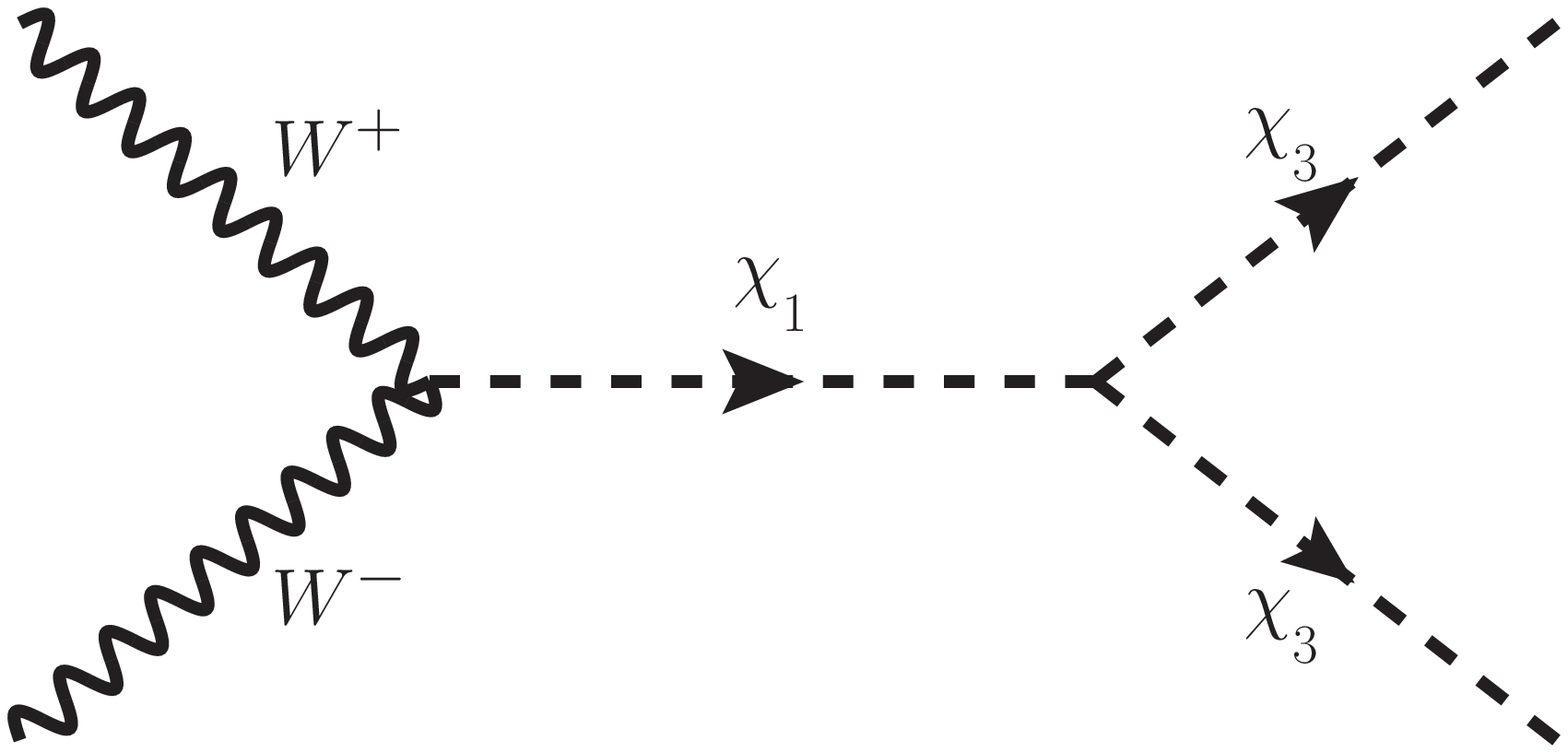}\\
\includegraphics[height=3.0cm,width=4.0cm]{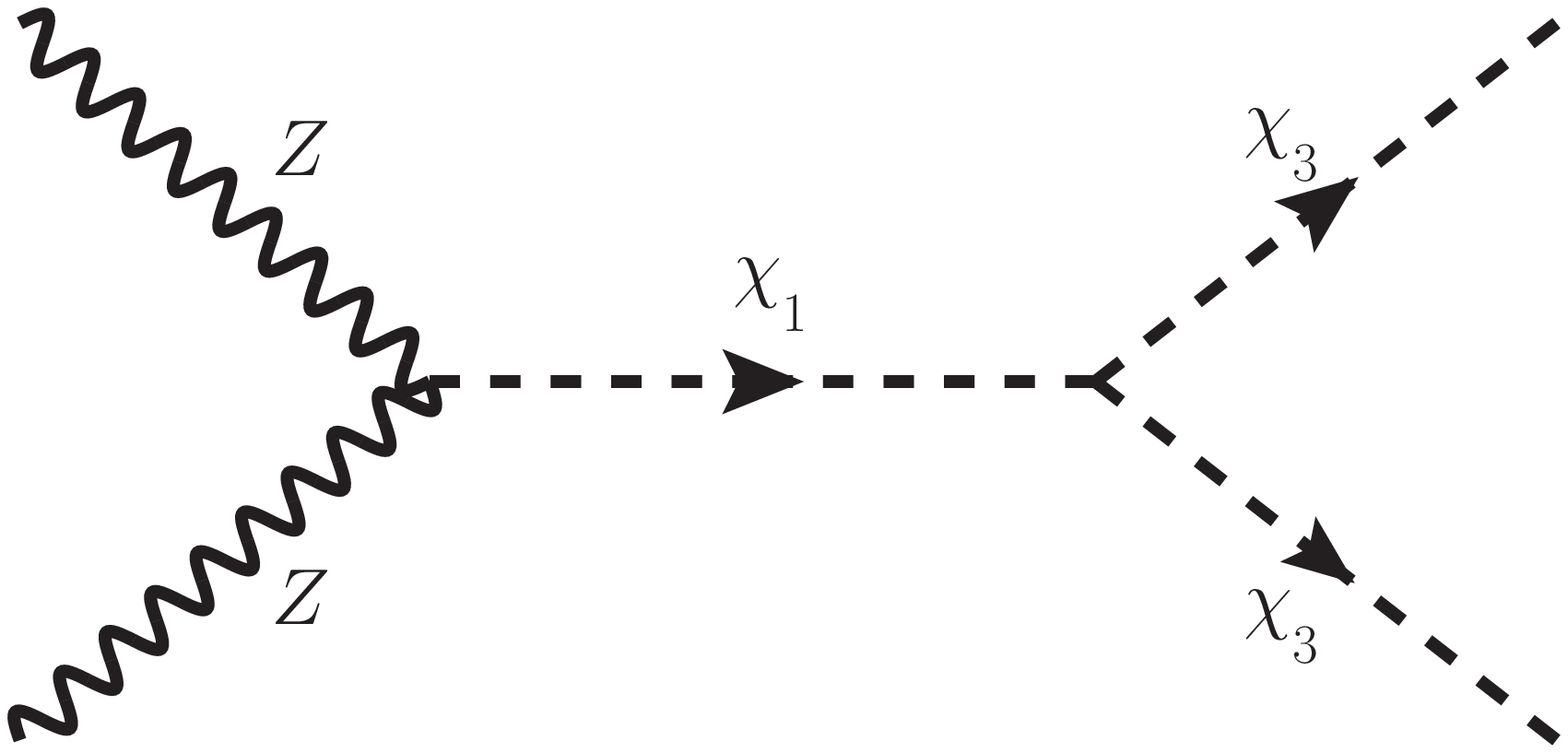}
\hspace{0.1cm}
\includegraphics[height=3.0cm,width=4.0cm]{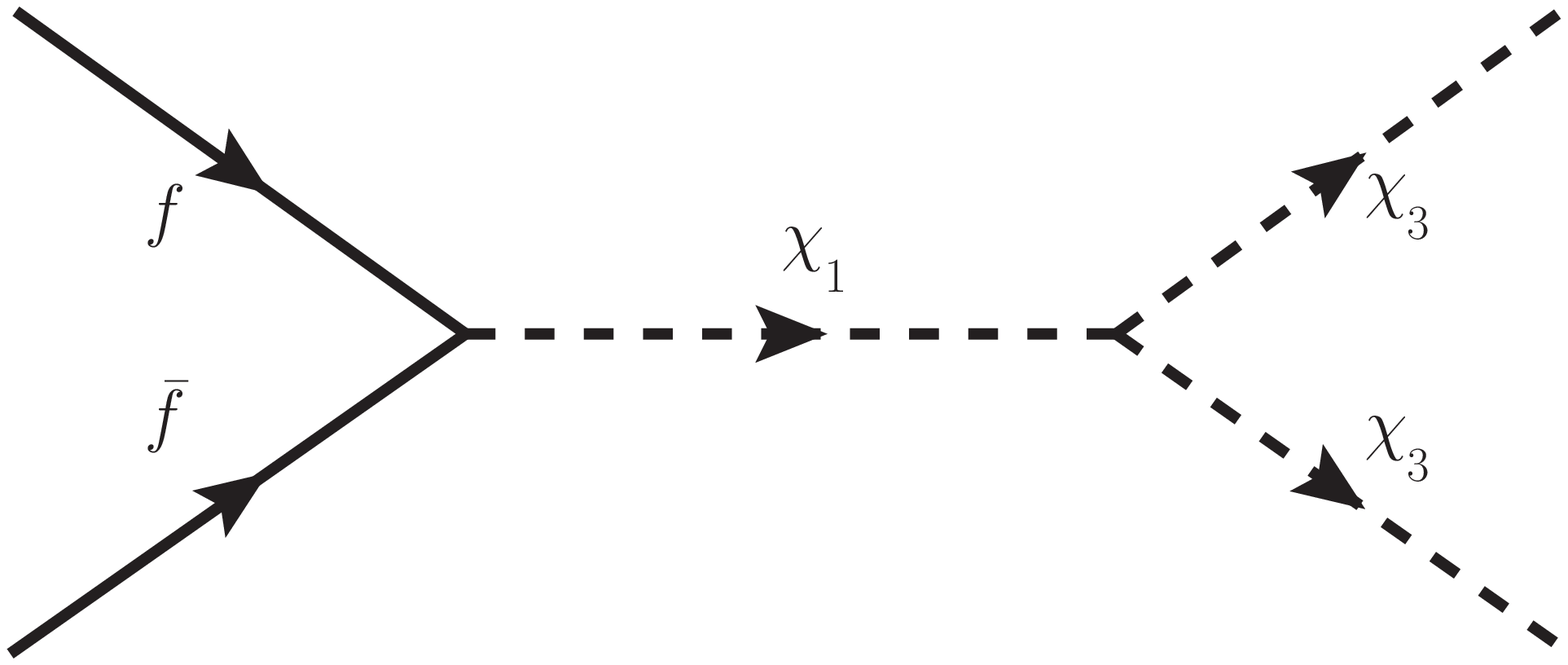}
\hspace{0.1cm}
\includegraphics[height=3.0cm,width=4.0cm]{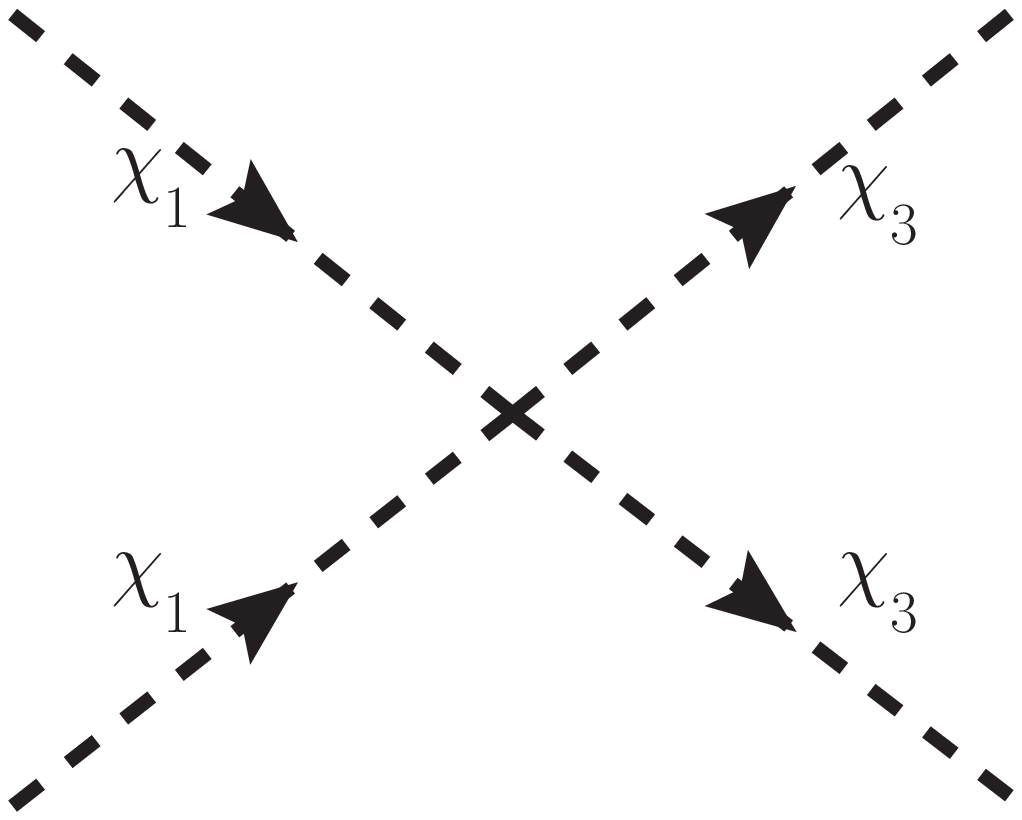}
\hspace{0.1 cm}
\includegraphics[height=3.0cm,width=4.0cm]{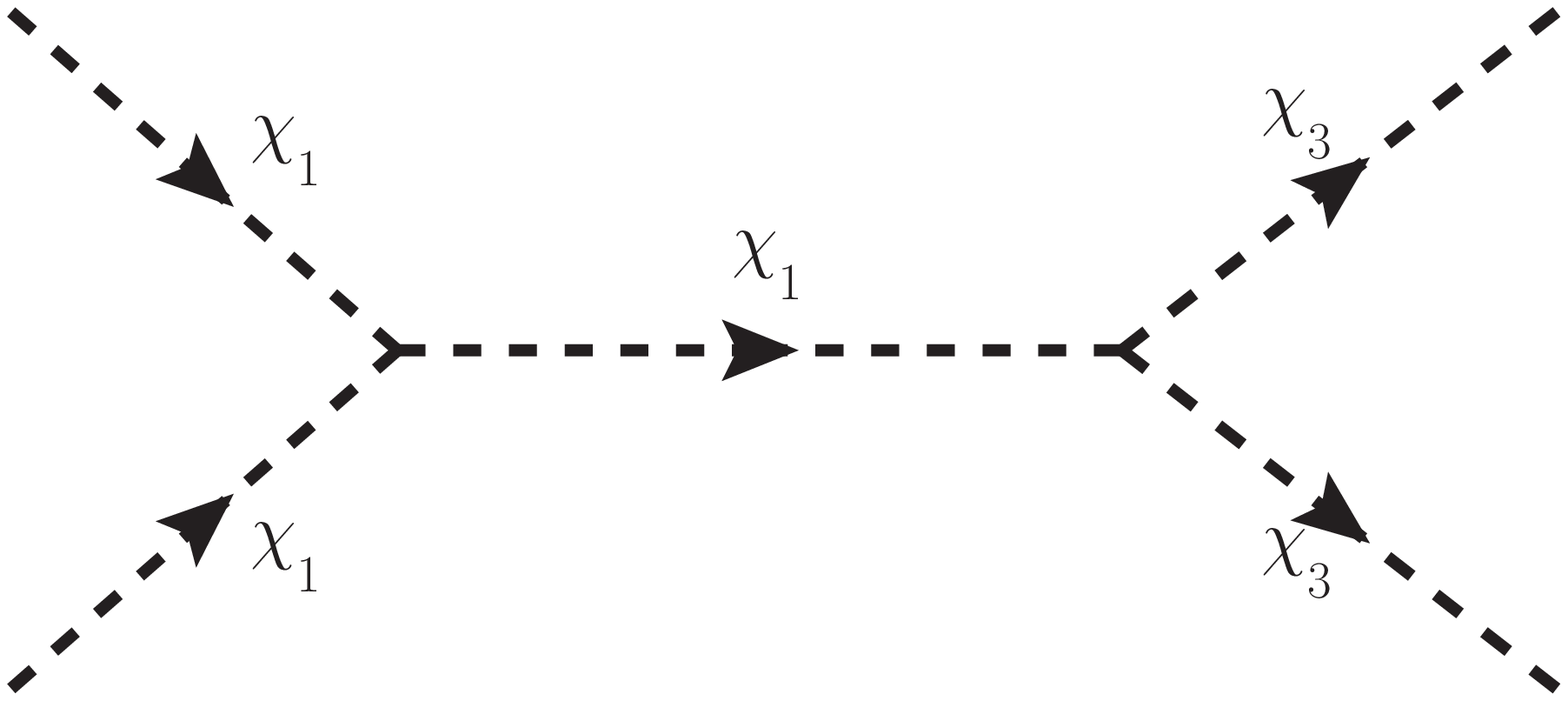}\\
\includegraphics[height=3.0cm,width=5.0cm]{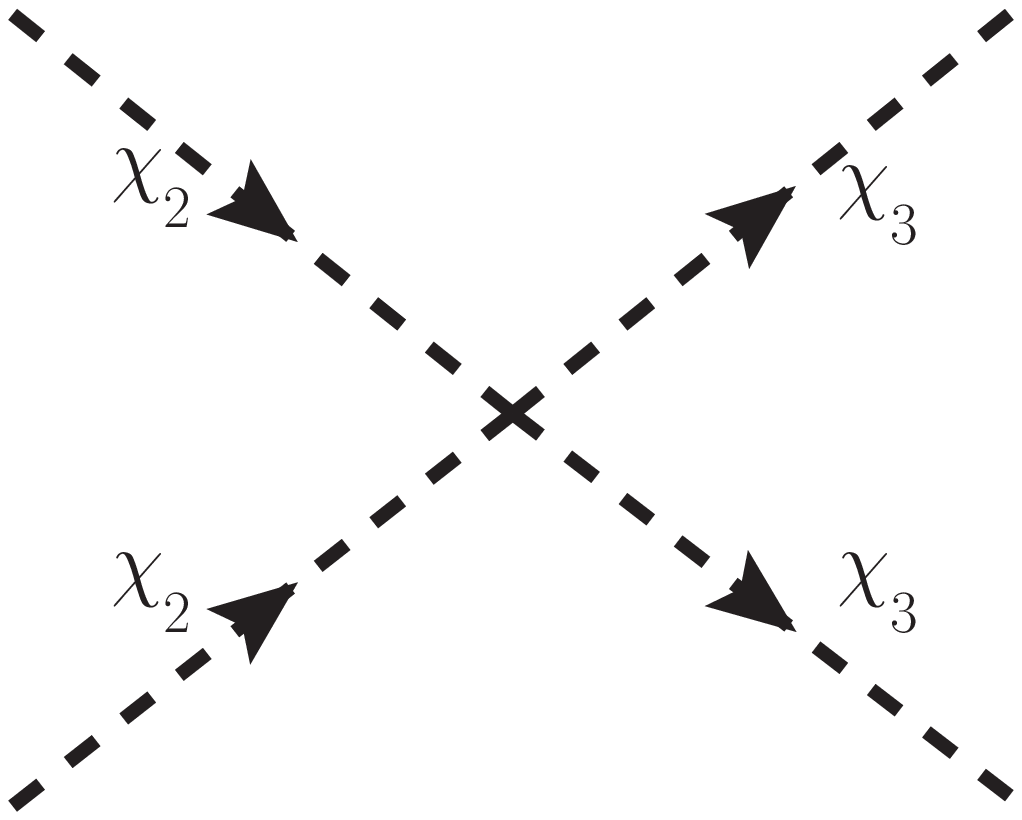}
\hspace{0.5cm}
\includegraphics[height=3.0cm,width=5.0cm]{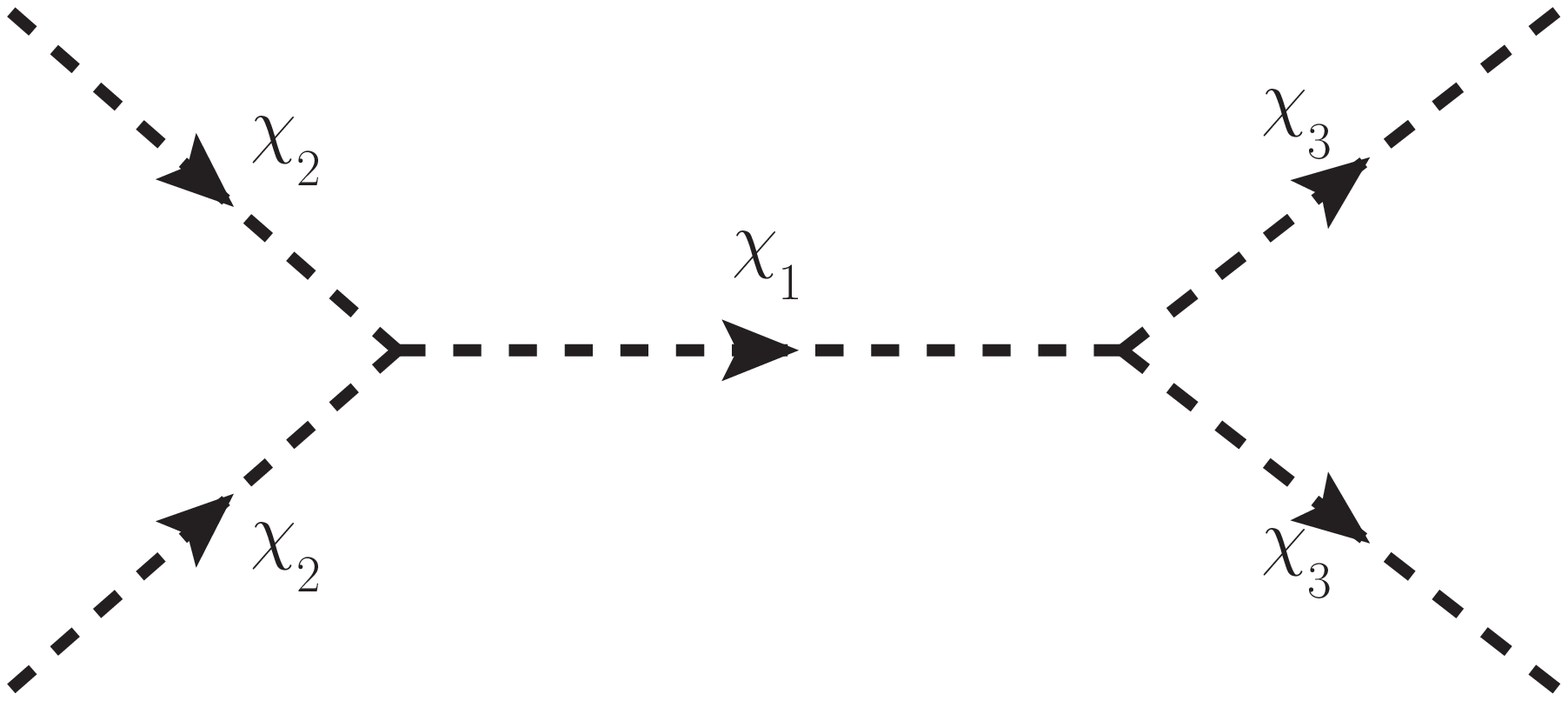}
\caption{Feynman diagrams for dominant production channels of
both the dark matter components $\chi_{{}_{{}_2}}$ and $\chi_{{}_{{}_3}}$}
\label{fynman-dia}
\end{figure} 
After the breaking of SU(2)$_{\rm L} \times$ U(1)$_{\rm Y}$ symmetry,
the self-annihilation of SM particles such as W, Z, Higgs, t-quark
and/or decays of the Higgs boson act as primary sources of
the dark matter particles $\chi_{{}_{{}_2}}$ and $\chi_{{}_{{}_3}}$.
The Feynman diagrams of the above mentioned processes, which are relevant for the evolution of
the number densities of $\chi_{{}_{{}_2}}$ and $\chi_{{}_{{}_3}}$, are shown in Fig. \ref{fynman-dia}.
We compute the number densities of both the dark matter components $\chi_{{}_{{}_2}}$
and $\chi_{{}_{{}_3}}$ at the present temperature ($T_0 \sim 10^{-13}$ GeV) by numerically
solving two coupled Boltzmann equations which are given below.  
\begin{eqnarray}
\frac{dY_{\chi_{{}_{{}_2}}}}{dz} &=& 
-\frac{2M_{pl}}{1.66 M^2_{\chi_{{}_{{}_1}}}}
\frac{z\,\sqrt{g_{\star}(T)}}{g_{\rm s}(T)} 
\Bigg(\langle \Gamma_{\chi_{{}_{{}_1}} \rightarrow \chi_{{}_{{}_2}} \chi_{{}_{{}_2}}}
\rangle (Y_{\chi_{{}_{{}_2}}} - Y^{eq}_{\chi_{{}_{{}_1}}})
%+\frac{1}{2}\langle \Gamma_{\chi_{{}_{{}_2}} \rightarrow \chi_{{}_{{}_3}} \chi_{{}_{{}_3}}} \rangle 
%Y_{\chi_{{}_{{}_2}}}
\Bigg) -\frac{4 \pi^2}{45}
\frac{M_{pl} M_{\chi_{{}_{{}_1}}}}{1.66} \frac{\sqrt{g_{\star}(T)}}{z^2} \times
\nonumber \\ &&
\Bigg(\sum_{x = W, Z, f, H}
\langle {\sigma {\rm v}}_{x\bar{x}\rightarrow \chi_{{}_{{}_2}} \chi_{{}_{{}_2}}} \rangle
\,\,{({{Y}_{\chi_{{}_{{}_2}}}}^2 -{Y^{eq}_{\chi_{{}_{{}_1}}}}\,^2)} \, +
\langle {\sigma {\rm v}}_{\chi_{{}_{{}_2}}\chi_{{}_{{}_2}}\rightarrow
\chi_{{}_{{}_3}} \chi_{{}_{{}_3}}} \rangle Y^2_{\chi_{{}_{{}_2}}} \Bigg)
\,\,, \nonumber\\
\label{boltz-eq1}
\end{eqnarray}
\begin{eqnarray}
\frac{dY_{\chi_{{}_{{}_3}}}}{dz} &=& 
-\frac{2M_{pl}}{1.66 M^2_{\chi_{{}_{{}_1}}}} 
\frac{z\,\sqrt{g_{\star}(T)}}{g_{\rm s}(T)}
\Bigg(\langle \Gamma_{\chi_{{}_{{}_1}} \rightarrow \chi_{{}_{{}_3}} \chi_{{}_{{}_3}}} \rangle  
\left(Y_{\chi_{{}_{{}_3}}} - {Y^{eq}_{\chi_{{}_{{}_1}}}}\right)
%-\langle \Gamma_{\chi_{{}_{{}_2}} \rightarrow \chi_{{}_{{}_3}} \chi_{{}_{{}_3}}} \rangle 
%Y_{\chi_{{}_{{}_2}}}
\Bigg)\,
-~\frac{4 \pi^2}{45} 
\frac{M_{pl} M_{\chi_{{}_{{}_1}}}}{1.66}
\frac{\sqrt{g_{\star}(T)}}{z^2} \times
\nonumber \\ &&
\Bigg(\sum_{x = W, Z, f, H}
\langle {\sigma {\rm v}}_{x\bar{x}\rightarrow \chi_{{}_{{}_3}} \chi_{{}_{{}_3}}} \rangle
\,\,{({{Y}_{\chi_{{}_{{}_3}}}}^2 -{Y^{eq}_{\chi_{{}_{{}_1}}}}\,^2)} \,
%\nonumber\\&&
-~\langle {\sigma {\rm v}}_{\chi_{{}_{{}_2}}\chi_{{}_{{}_2}}\rightarrow
\chi_{{}_{{}_3}} \chi_{{}_{{}_3}}} \rangle Y^2_{\chi_{{}_{{}_2}}} \Bigg)\,\,.\nonumber \\
\label{boltz-eq2} 
\end{eqnarray}
In Eqs. (\ref{boltz-eq1}) and (\ref{boltz-eq2}), $Y_{\chi_{{}_{{}_2}}} =
\frac{n_{\chi_{{}_{{}_2}}}}{\rm s}$ ($Y_{\chi_{{}_{{}_3}}} = \frac{n_{\chi_{{}_{{}_3}}}}{\rm s}$) is
the comoving number density of $\chi_{{}_{{}_2}}$ ($\chi_{{}_{{}_3}}$),
$z=\frac{M_{\chi_{{}_{{}_1}}}}{T}$ and $T$ is the photon
temperature while ${\rm s}$ is the entropy density of the Universe
\footnote{It is to be noted that in the 
above two Boltzmann equations (Eqs. (\ref{boltz-eq1}), (\ref{boltz-eq2}))
we have neglected a term involving
$\langle {\sigma {\rm v}}_{\3\3\rightarrow\2\2} \rangle Y^2_{\3}$.
At an earlier epoch during the initial stage of production of $\3$
(mainly from the decay of $\1$), the number density of $\3$ was 
very low and hence this term could be neglected. On the other hand, at a later 
epoch when the Universe cools down to a temperature lower than the mass 
of $\2$, the process $\3\3 \rightarrow \2\2$
will not have any significant contribution to the term
$\langle {\sigma {\rm v}}_{\3\3\rightarrow\2 \2} \rangle Y^2_{\3}$
even though the number density for $\3$ is higher.}.
Here the number densities of $\chi_{{}_{{}_2}}$, $\chi_{{}_{{}_3}}$ are denoted
by $n_{\chi_{{}_{{}_2}}}$, $n_{\chi_{{}_{{}_3}}}$ respectively. Further, $M_{pl}$
is the Planck mass and $g_{\star}$ is given by
\begin{eqnarray}
\sqrt{g_\star(T)} = \frac{g_{\rm s}(T)}{\sqrt{g_{\rho}(T)}}
\left(1 + \frac{1}{3}\frac{d\,{\rm ln}\,g_{\rm s}(T)}{d\,{\rm ln}T}\right) \,\, .
\end{eqnarray}
In the above, $g_{\rho}(T)$ and $g_{\rm s}(T)$ are the effective 
degrees of freedom related to
the energy density $\rho$ and the entropy density $\rm s$ respectively
of the Universe through the relations
$\rho = g_{\rho}(T)\frac{\pi^2}{30}T^4, \, {\rm s} = g_{\rm s}(T)\frac{2\pi^2}{45}T^3$. 
Thus $g_{\star}$ is a function of the stated effective degrees of freedom. 
The thermal averages of decay widths ($\Gamma$) and annihilation 
cross sections times relative velocities ($\sigma {\rm v}$)
for various processes, that occur in Eqs. \ref{boltz-eq1} and \ref{boltz-eq2}, 
can be expressed as
\begin{eqnarray}
&&\langle \Gamma_{\chi_{{}_{{}_1}}\rightarrow\chi_{{}_{{}_j}}\chi_{{}_{{}_j}}} \rangle =
\Gamma_{\chi_{{}_{{}_1}} \rightarrow \chi_{{}_{{}_j}} \chi_{{}_{{}_j}}}
\frac{K_1(z)}{K_2(z)}, ~~~~j = 2,~3,
%~j, ~k = 2,~3~{\rm and}~i\neq j,~k,
\label{gamma-avr}\\
&&\langle {\sigma {\rm v}_{x\bar{x}\rightarrow \chi_{{}_{{}_j}}\chi_{{}_{{}_j}}}} \rangle =
\frac{1}{8 M_x^4 T K_2^2\left(\frac{M_x}{T}\right)}
\int_{4M_x^2}^\infty \,\sigma_{x x\rightarrow \chi_{{}_{{}_j}} \chi_{{}_{{}_j}}}\,
(s-4M_x^2)\,\sqrt{s}\,K_1\left(\frac{\sqrt{s}}{T}\right)\,ds \,\,, \nonumber\\
&&j = 2,3, \,\,\,\,\, x = W^\pm,~Z,~f,~\chi_{{}_{{}_1}},~\chi_{{}_{{}_2}}\,\, .
\label{sigmav-avr}
\end{eqnarray}
In Eqs. (\ref{gamma-avr}), (\ref{sigmav-avr}) $K_i$ is the modified Bessel
function of order $i$ and $s$ is the Mandelstam variable. The decay widths 
$\Gamma_{\chi_{{}_{{}_1}} \rightarrow \chi_{{}_{{}_j}} \chi_{{}_{{}_j}}}$ and 
annihilation cross sections $\sigma_{x \bar{x}\rightarrow \chi_{{}_{{}_j}} \chi_{{}_{{}_j}}}$
(for $j=2,\,\,3$, $x = W^\pm,~Z,~f,~\chi_{{}_{{}_1}},~\chi_{{}_{{}_2}}$)
of the processes mentioned in the subscripts of $\Gamma$ and $\sigma$ are given below:
\begin{eqnarray}
\Gamma_{\chi_{{}_{{}_1}}\rightarrow \chi_{{}_{{}_j}}\chi_{{}_{{}_j}}} &=&
\frac{g^2_{{\!}_{\chi_{{}_{{}_1}}\chi_{{}_{{}_j}}\chi_{{}_{{}_j}}}}}{8\pi M_{\chi_{{}_{{}_1}}}}
\sqrt{1-\frac{4M^2_{\chi_{{}_{{}_j}}}}{M^2_{\chi_{{}_{{}_1}}}}}\,\, ,\\
~~~~~~~~~\nonumber \\
~~~~~~~~~\nonumber \\
%\end{eqnarray}
%%%%%%%
%\Gamma_{\chi_{{}_{{}_2}}\rightarrow \chi_{{}_{{}_3}}\chi_{{}_{{}_3}}} &=&
%\frac{g^2_{{\!}_{\chi_{{}_{{}_2}}\chi_{{}_{{}_3}}\chi_{{}_{{}_3}}}}}{8\pi M_{\chi_{{}_{{}_2}}}}
%\sqrt{1-\frac{4M^2_{\chi_{{}_{{}_3}}}}{M^2_{\chi_{{}_{{}_2}}}}}\,\, ,\\
%%%%%%%
%\begin{eqnarray}
{\sigma}_{{\!}_{\chi_{{}_{{}_1}} \chi_{{}_{{}_1}}
\rightarrow \chi_{{}_{{}_j}} \chi_{{}_{{}_j}}}} &=&
\frac{1}{2\pi s}\sqrt{\frac{s-4M^2_{\chi_{{}_{{}_j}}}}{s-4M^2_{\chi_{{}_{{}_1}}}}}
\Bigg\{g^2_{{\!}_{\chi_{{}_{{}_1}}\chi_{{}_{{}_1}}\chi_{{}_{{}_j}}\chi_{{}_{{}_j}}}}
+ \frac{9\,g^2_{{\!}_{\chi_{{}_{{}_1}}\chi_{{}_{{}_1}}\chi_{{}_{{}_1}}}}
g^2_{{\!}_{\chi_{{}_{{}_1}}\chi_{{}_{{}_j}}\chi_{{}_{{}_j}}}}}
{\left[(s-M^2_{\chi_{{}_{{}_1}}})^2 + (\Gamma_{\chi_{{}_{{}_1}}}
M_{\chi_{{}_{{}_1}}})^2\right]} \nonumber \\
&&-\frac{6\,g_{{\!}_{\chi_{{}_{{}_1}}\chi_{{}_{{}_1}}\chi_{{}_{{}_j}}
\chi_{{}_{{}_j}}}}~g_{{\!}_{\chi_{{}_{{}_1}}\chi_{{}_{{}_1}}\chi_{{}_{{}_1}}}}
~g_{{\!}_{\chi_{{}_{{}_1}}\chi_{{}_{{}_j}}\chi_{{}_{{}_j}}}}
(s-M^2_{\chi_{{}_{{}_1}}})}{\left[(s-M^2_{\chi_{{}_{{}_1}}})^2 +
(\Gamma_{\chi_{{}_{{}_1}}}M_{\chi_{{}_{{}_1}}})^2\right]}\Bigg\}\,\,, 
\end{eqnarray}
%%%%%%%%
\begin{eqnarray}
{\sigma}_{{\!}_{\chi_{{}_{{}_2}} \chi_{{}_{{}_2}}
\rightarrow \chi_{{}_{{}_3}} \chi_{{}_{{}_3}}}} &=&
\frac{1}{2\pi s}\sqrt{\frac{s-4M^2_{\chi_{{}_{{}_3}}}}{s-4M^2_{\chi_{{}_{{}_2}}}}}
\Bigg\{g^2_{{\!}_{\chi_{{}_{{}_2}}\chi_{{}_{{}_2}}\chi_{{}_{{}_3}}\chi_{{}_{{}_3}}}}
+ \frac{\,g^2_{{\!}_{\chi_{{}_{{}_2}}\chi_{{}_{{}_2}}\chi_{{}_{{}_1}}}}
g^2_{{\!}_{\chi_{{}_{{}_1}}\chi_{{}_{{}_3}}\chi_{{}_{{}_3}}}}}
{\left[(s-M^2_{\chi_{{}_{{}_1}}})^2 + (\Gamma_{\chi_{{}_{{}_1}}}
M_{\chi_{{}_{{}_1}}})^2\right]} \nonumber \\
&&-\frac{2\,g_{{\!}_{\chi_{{}_{{}_2}}\chi_{{}_{{}_2}}\chi_{{}_{{}_3}}
\chi_{{}_{{}_3}}}}~g_{{\!}_{\chi_{{}_{{}_2}}\chi_{{}_{{}_2}}\chi_{{}_{{}_1}}}}
~g_{{\!}_{\chi_{{}_{{}_1}}\chi_{{}_{{}_3}}\chi_{{}_{{}_3}}}}
(s-M^2_{\chi_{{}_{{}_1}}})}{\left[(s-M^2_{\chi_{{}_{{}_1}}})^2 +
(\Gamma_{\chi_{{}_{{}_1}}}M_{\chi_{{}_{{}_1}}})^2\right]}\Bigg\}\,\,, \\
~~~~~~~~~\nonumber \\
~~~~~~~~~\nonumber \\
%\end{eqnarray}
%%%%%%%%
%\begin{eqnarray}
{\sigma}_{{\!}_{WW\rightarrow \chi_{{}_{{}_j}} \chi_{{}_{{}_j}}}} &=&
\frac{g^2_{{\!}_{WW\chi_{{}_{{}_1}}}}g^2_{{\!}_{\chi_{{}_{{}_1}}
\chi_{{}_{{}_j}}\chi_{{}_{{}_j}}}}}{72\pi s} 
\sqrt{\frac{s-4M^2_{\chi_{{}_{{}_j}}}}{s-4M^2_{W}}}
\frac{\left(3 - \frac{s}{M^2_{W}} + \frac{s^2}{4 M^4_{W}}\right)}
{(s-M^2_{\chi_{{}_{{}_1}}})^2}\,\, ,\\
~~~~~~~~~\nonumber \\
~~~~~~~~~\nonumber \\
%\end{eqnarray}
%%%%%%%%
%\begin{eqnarray}
{\sigma}_{{\!}_{ZZ\rightarrow \chi_{{}_{{}_j}} \chi_{{}_{{}_j}}}} &=&
\frac{g^2_{{\!}_{ZZ\chi_{{}_{{}_1}}}}g^2_{{\!}_{\chi_{{}_{{}_1}}
\chi_{{}_{{}_j}}\chi_{{}_{{}_j}}}}}{18\pi s}
\sqrt{\frac{s-4M^2_{\chi_{{}_{{}_j}}}}{s-4M^2_{Z}}}
\frac{\left(3 - \frac{s}{M^2_{Z}} + \frac{s^2}{4 M^4_{Z}}\right)}
{(s-M^2_{\chi_{{}_{{}_1}}})^2}\,\, ,\\
~~~~~~~~~\nonumber \\
~~~~~~~~~\nonumber \\
%\end{eqnarray}
%%%%%%%%%
%\begin{eqnarray}
{\sigma}_{{\!}_{f\bar{f}\rightarrow \chi_{{}_{{}_j}} \chi_{{}_{{}_j}}}} &=&
\frac{n_c~g^2_{{\!}_{ff\chi_{{}_{{}_1}}}}g^2_{{\!}_{\chi_{{}_{{}_1}}
\chi_{{}_{{}_j}}\chi_{{}_{{}_j}}}}}{16\pi s}
\frac{\sqrt{(s-4M^2_{\chi_{{}_{{}_j}}})(s-4M^2_{f})}}
{(s-M^2_{\chi_{{}_{{}_1}}})^2} \,\, . 
\end{eqnarray} 
In the above equations $g_{ijk}$ and $g_{ijkl}$ are couplings of the vertices involving
the fields $i$, $j$, $k$ as well as $i$, $j$, $k$, $l$ respectively while $M_W$, $M_Z$
and $M_f$ are the masses of W boson, Z boson and fermion $f$ ($f$ is any SM fermion).
All the couplings which are necessary to calculate the decay widths and annihilation
cross sections are given in the Appendix \ref{a1}.   

Finally, the total relic density ($\Omega_{\rm T} h^2$) of dark matter
in the Universe is given in terms of the normalised Hubble constant
$h = \frac{H_o}{100 {\rm km}/{\rm Mpc}/{\rm s}}$ as
\begin{eqnarray}
\Omega_{\rm T} h^2 = \Omega_{\chi_{{}_{{}_2}}} h^2 + \Omega_{\chi_{{}_{{}_3}}} h^2\,\,.
\end{eqnarray}
Once we obtain the comoving number densities $Y_{\chi_{{}_{{}_2}}}(T_0)$,
$Y_{\chi_{{}_{{}_3}}}(T_0)$ of both the dark matter components $\chi_{{}_{{}_2}}$
and $\chi_{{}_{{}_3}}$ at the present temperature $T_0$
by numerically solving the two coupled Boltzmann equations
(Eqs. \ref{boltz-eq1}, \ref{boltz-eq2}), the individual
relic densities ($\Omega_{\chi_{{}_{{}_2}}} h^2$ and $\Omega_{\chi_{{}_{{}_3}}} h^2$)
of each of the components can be obtained from \cite{Edsjo:1997bg, Biswas:2011td}
\begin{eqnarray}
\Omega_{\chi_{{}_{{}_i}}} h^2 = 2.755\times 10^8
\left(\frac{M_{\chi_{{}_{{}_i}}}}{\rm GeV}\right) Y_{\chi_{{}_{{}_i}}}(T_0)~~~(i=2,\,3)\,\,.
\end{eqnarray}
While solving these two coupled equations (Eqs. (\ref{boltz-eq1}), (\ref{boltz-eq2})),
we have adopted the following boundary condition: at the electroweak phase
transition temperature, which corresponds to $z\,(= \frac{M_{\chi_{{}_{{}_1}}}}{T})
\simeq 0.83$, the number densities of both the dark matter candidates are zero.
\begin{figure}[h!]
\centering
\subfigure[$\Omega_{\chi_{{}_{{}_2}}} > \Omega_{\chi_{{}_{{}_3}}}$]
{\includegraphics[height=8cm,width=6cm,angle=-90]{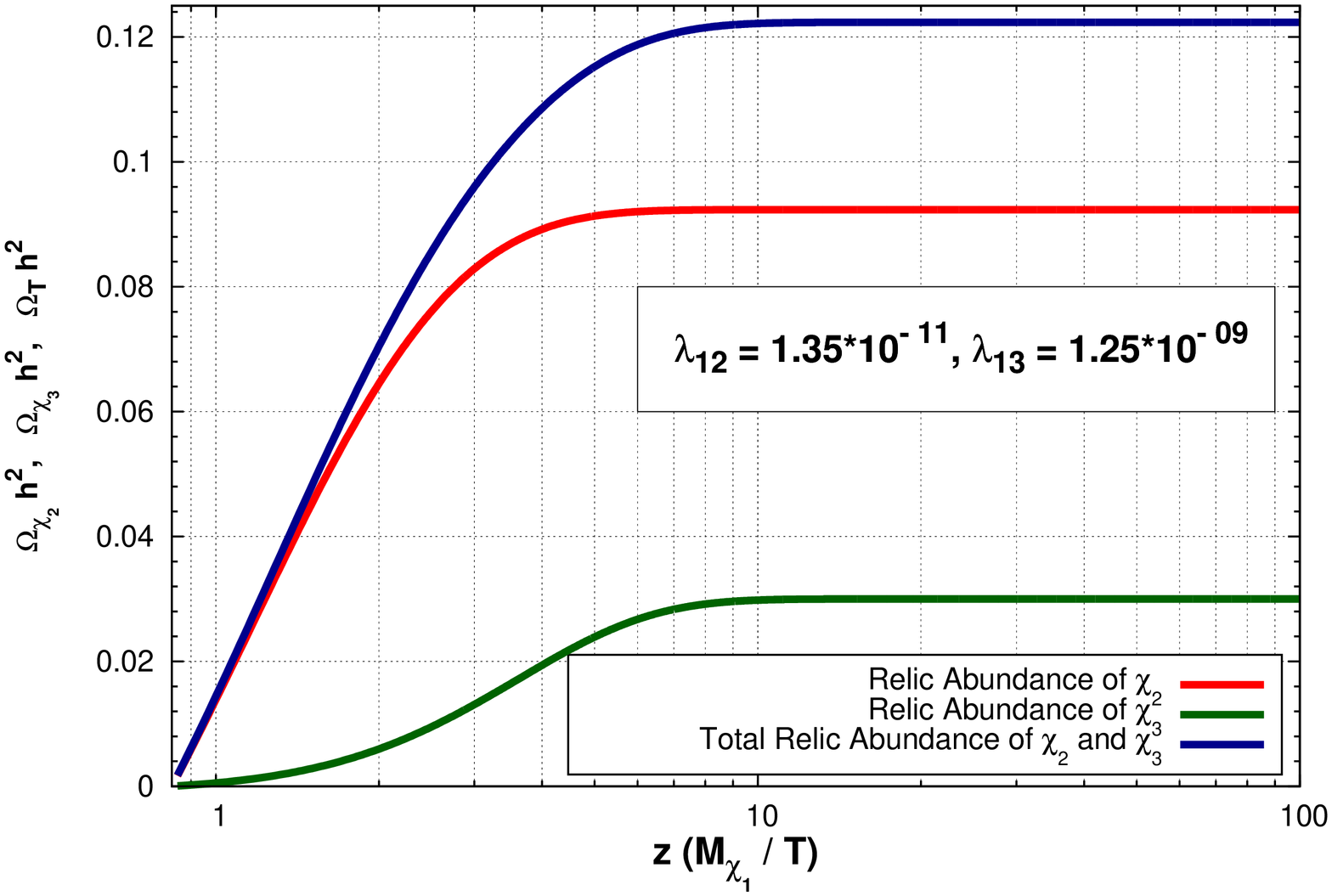}}
\subfigure[$\Omega_{\chi_{{}_{{}_2}}} < \Omega_{\chi_{{}_{{}_3}}}$]
{\includegraphics[height=8cm,width=6cm,angle=-90]{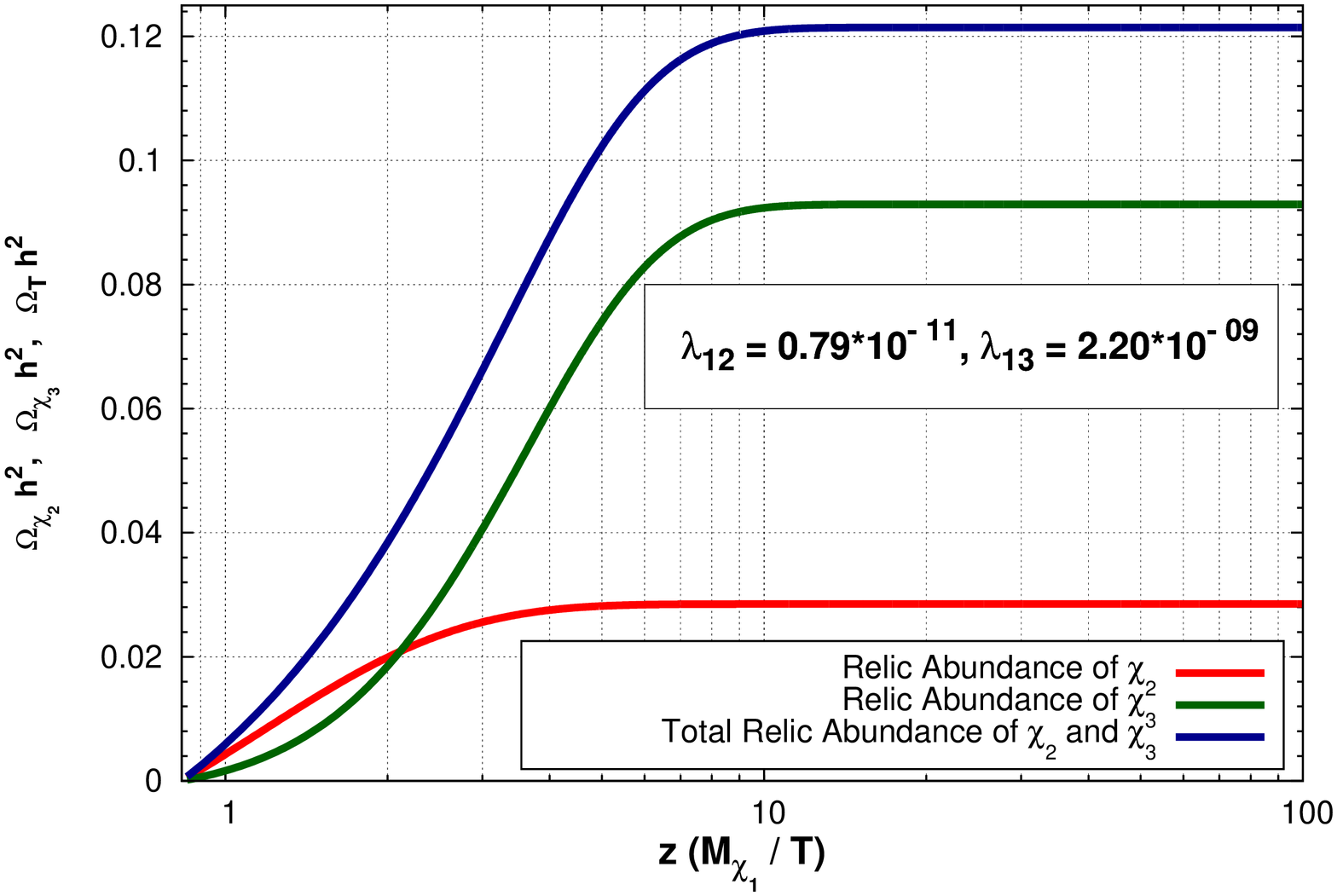}}\\
\subfigure[$\Omega_{\chi_{{}_{{}_2}}} \simeq \Omega_{\chi_{{}_{{}_3}}}$]
{\includegraphics[height=9cm,width=6cm,angle=-90]{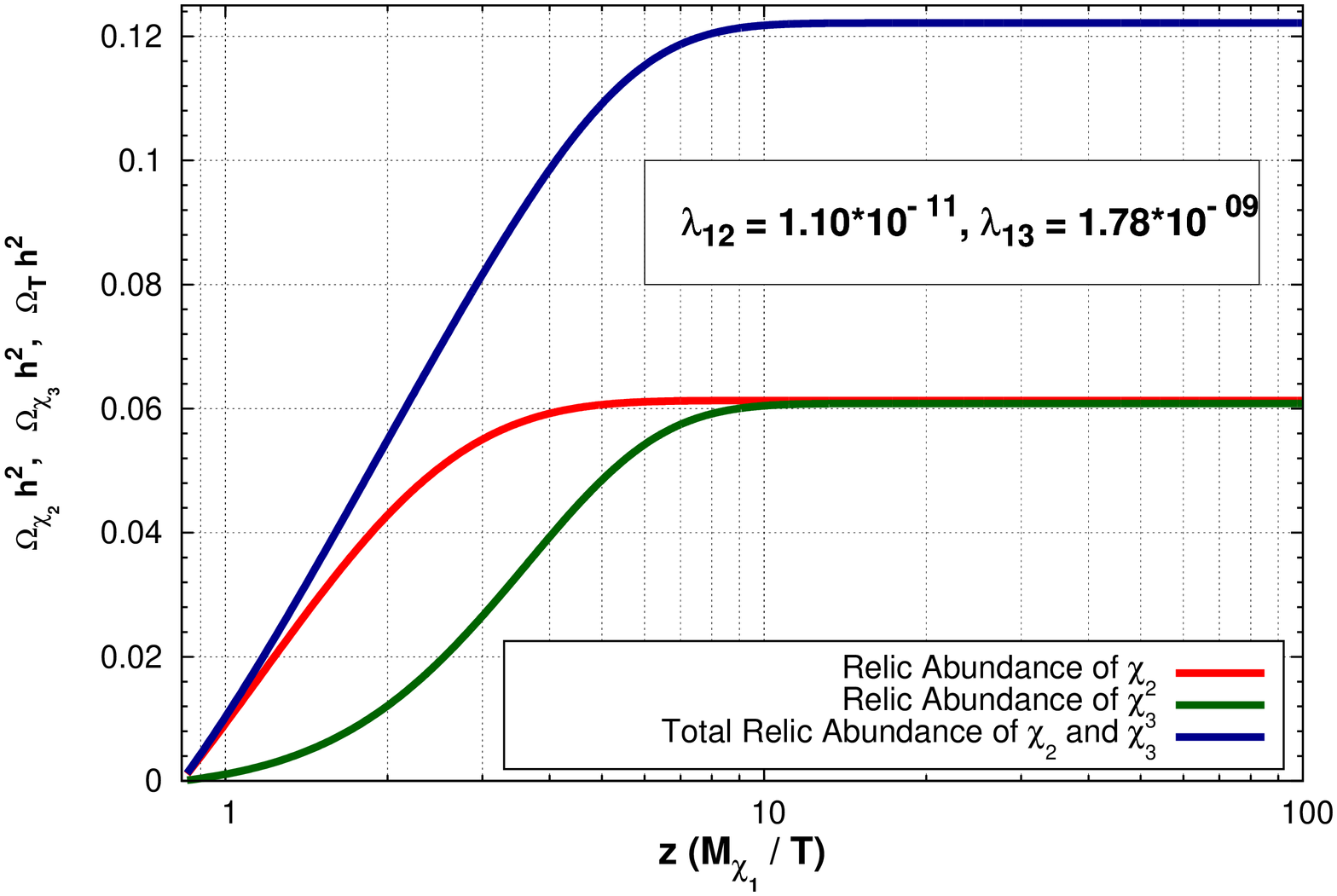}}
\caption{Variation of relic densities of both the dark matter candidates
with $z$.}
\label{relic-density-plot}
\end{figure}
The plots (a$-$c) in Fig. \ref{relic-density-plot} show the
variation of the relic densities of both the dark matter candidates
$\chi_{{}_{{}_2}}$ and $\chi_{{}_{{}_3}}$ with $z$ (inverse of temperature)
for different values of model parameters, namely $\lambda_{12}$, $\lambda_{13}$.
It is to be noted that these parameters turn out to be of order $ \sim 10^{-9}
$-$10^{-11}$ on account of the smallness of the comoving number density
$Y_{\chi_{{}_{{}_i}}}$ required to satisfy the observed dark matter relic density in the Universe.
All these plots (a$-$c) of Fig. \ref{relic-density-plot} are drawn for the case 
with $M_{\chi_{{}_{{}_2}}} = 70$ GeV and $M_{\chi_{{}_{{}_3}}} = 7.1$ keV. The red solid line
in each plot of Fig \ref{relic-density-plot} represents the relic density
of the heavier dark matter component ($\chi_{{}_{{}_2}}$) while the green and blue solid
lines denote the relic densities of the lighter DM candidate $\chi_{{}_{{}_3}}$ and
the total density of both the dark matter components respectively. In plot (a)
of Fig. \ref{relic-density-plot} we have chosen the values of $\lambda_{12}$,
$\lambda_{13}$ in such a way that the DM particle $\chi_{{}_{{}_2}}$ becomes the dominant
component within the dark sector in terms of its contributions
towards the total dark matter relic density ($\Omega_{\rm T} h^2$). However, for
the other two plots, different sets of chosen values of $\lambda_{12}$ and $\lambda_{13}$
result in the situations where $\Omega_{\chi_{{}_{{}_2}}} h^2 < \Omega_{\chi_{{}_{{}_3}}} h^2$ (plot b)
and $\Omega_{\chi_{{}_{{}_2}}} h^2 \sim \Omega_{\chi_{{}_{{}_3}}} h^2$ (plot c). It appears
from each plot of Fig. \ref{relic-density-plot} that 
%the individual relic density of
the relic density of each dark matter candidate starts growing 
from an initial value zero (due to the adopted boundary condition discussed earlier),
thereafter, as the temperature of
the Universe decreases ($z$ increases), the relic densities of both 
the DM components increase since more and more dark matter particles are produced 
by the decay and/or self annihilation of the SM particles. Finally, the relic 
densities of both the dark matter particles saturate to the respective particular
values at $z\sim 10$ (corresponding to a temperature $T \sim 12$ GeV of the Universe)
which depend upon the values of the parameters $\lambda_{12}$ and $\lambda_{13}$.
It is also seen from Fig. \ref{relic-density-plot} that in all three cases
the saturation values of the total relic density of the two dark matter candidates
always lie within the range $0.1172\leq \Omega_{\rm DM} h^2 \leq 0.1226$, as
predicted by the PLANCK experiment \cite{Ade:2013zuv} at a 68\% C.L.  
\begin{figure}[h!]
\centering
\includegraphics[height=8cm,width=6cm,angle=-90]{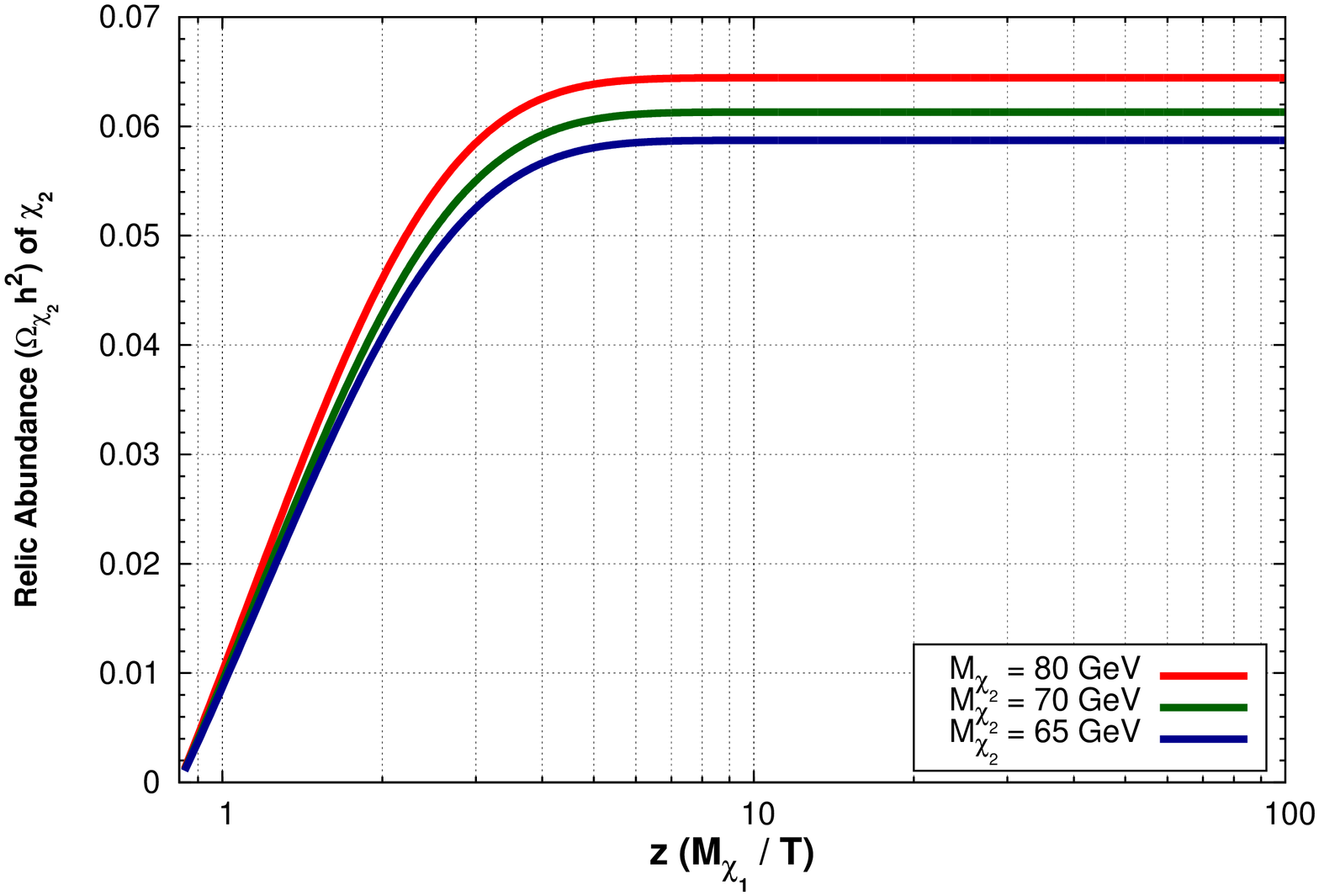}
\caption{Variation of relic density of the dark matter component $\chi_{{}_{{}_2}}$
with $z$ for three different values of $M_{\chi_{{}_{{}_2}}} = $ 65, 70, 80 GeV.}
\label{relic-density-chi2}
\end{figure}
In Fig. \ref{relic-density-chi2} we show the variation of the relic density of
the heavier dark matter component $\chi_{{}_{{}_2}}$ with $z$ for three different
values of $M_{\chi_{{}_{{}_2}}} = $ 65, 70, 80 GeV with $\lambda_{12} = 1.10\times 10^{-11}$.  
\begin{figure}[h!]
\centering
\subfigure[]
{\includegraphics[height=8cm,width=6cm,angle=-90]{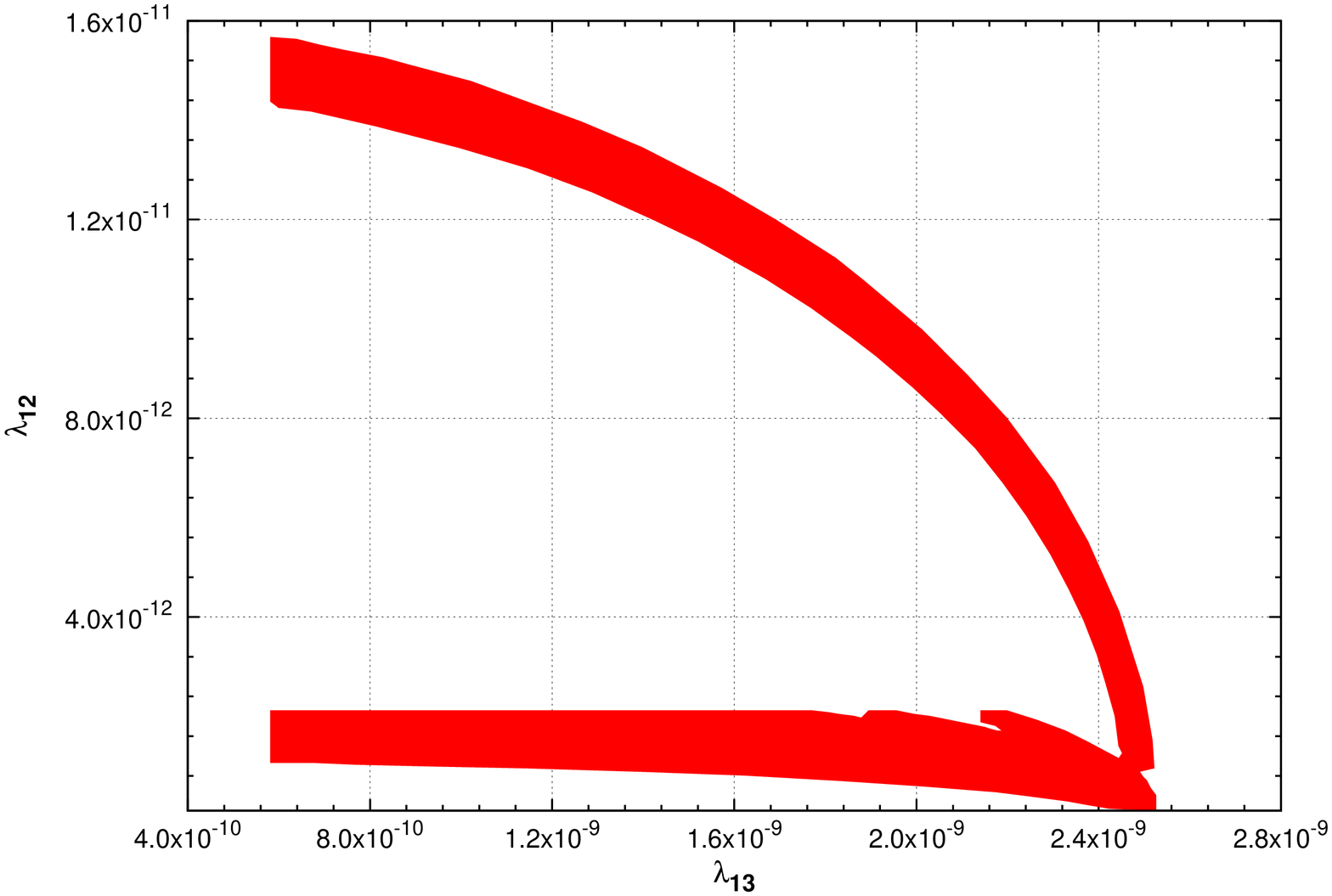}}
\subfigure[]
{\includegraphics[height=8cm,width=6cm,angle=-90]{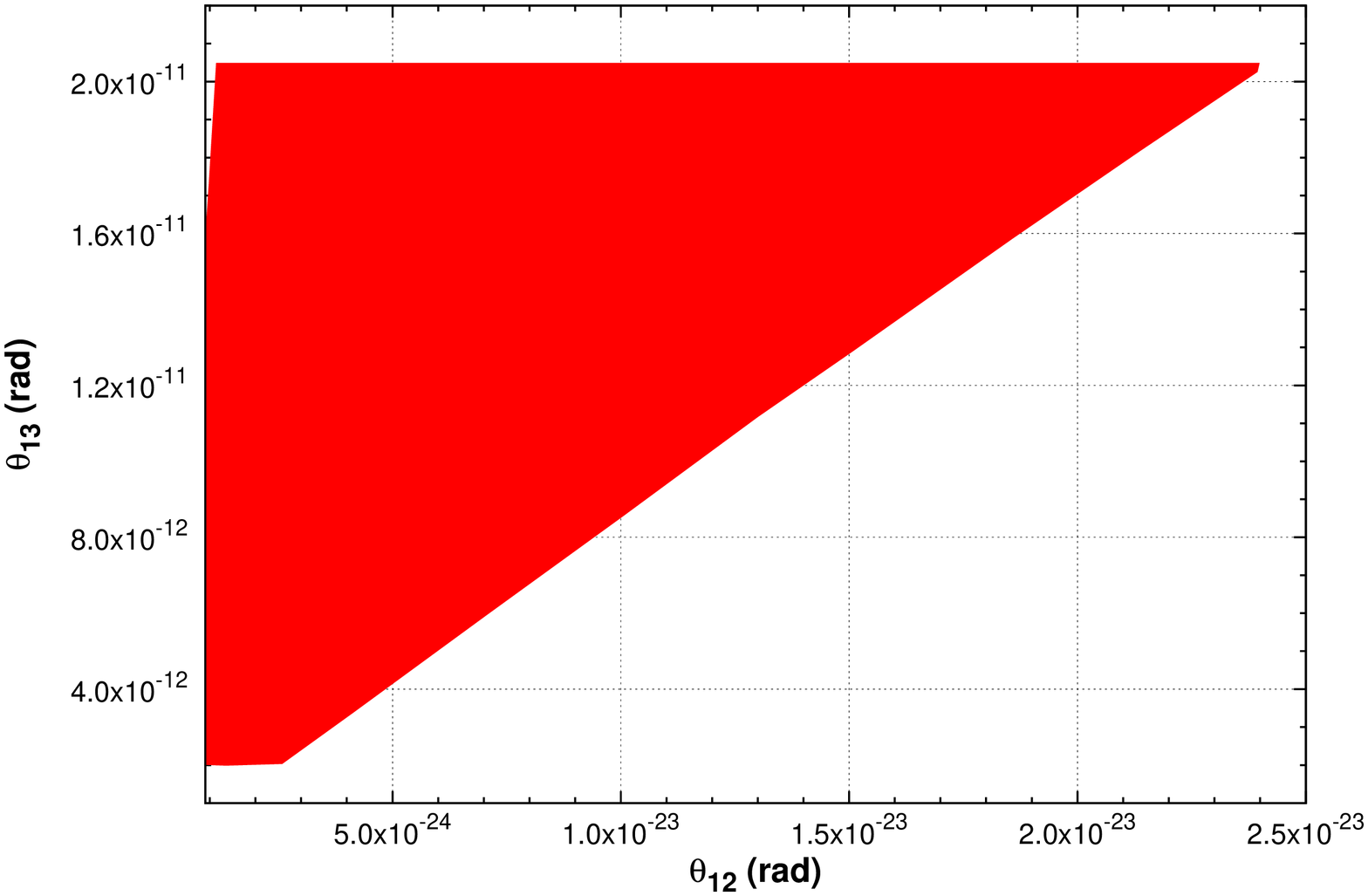}}\\
\subfigure[]
{\includegraphics[height=8cm,width=6cm,angle=-90]{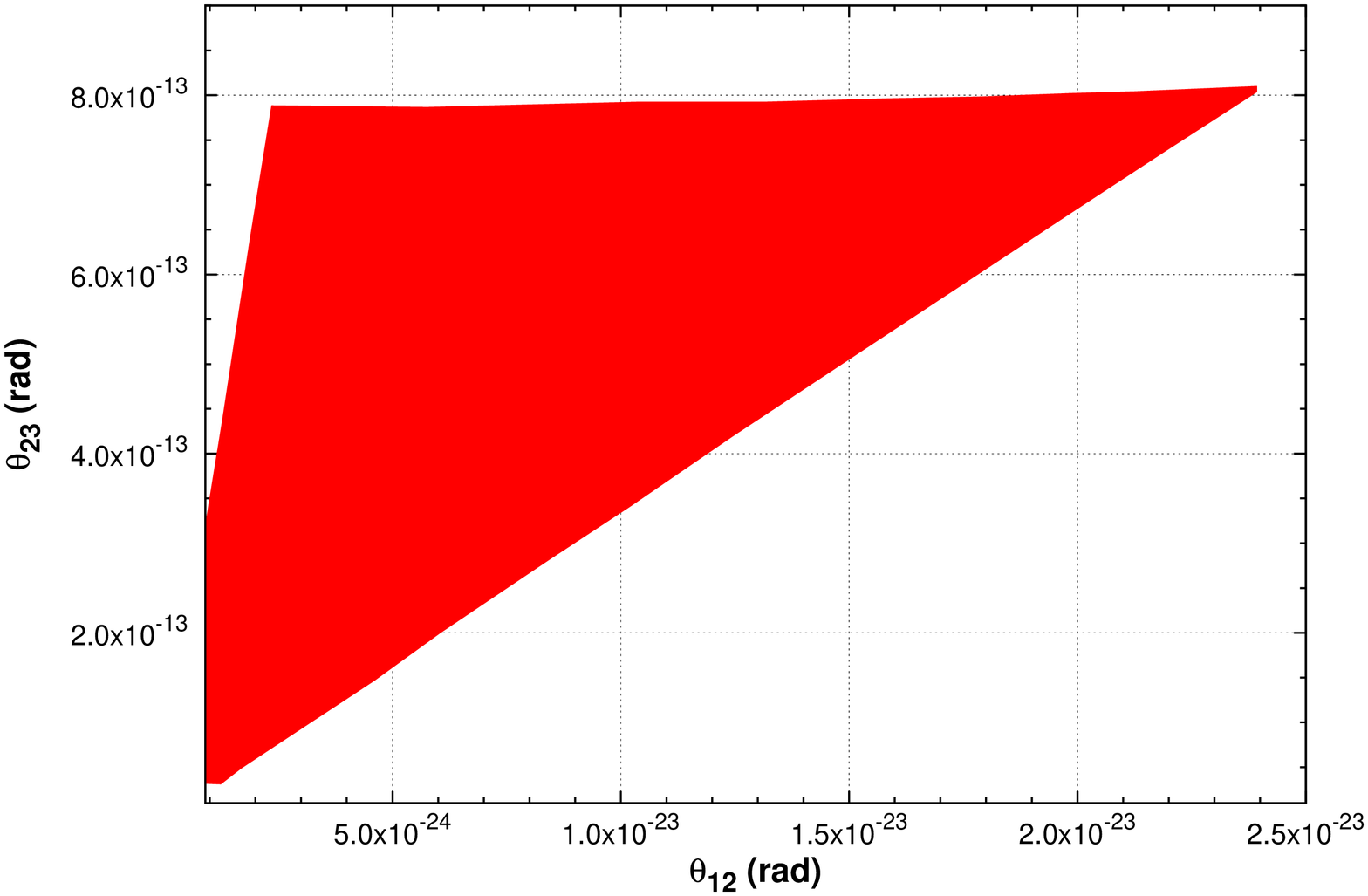}}
\caption{Allowed ranges of model parameters namely, $\lambda_{12}$, $\lambda_{13}$,
$\theta_{12}$, $\theta_{23}$ and $\theta_{13}$.}
\label{parameter-plot}
\end{figure}

In the three plots (a$-$c) of Fig. \ref{parameter-plot}, we show the allowed regions in 
the parameter space $\lambda_{12}$-$\lambda_{13}$, $\theta_{12}$-$\theta_{13}$ and
$\theta_{12}$-$\theta_{23}$ respectively. These plots (a$-$c of Fig. \ref{parameter-plot})
are drawn for $50\,\,{\rm GeV} \leq M_{\chi_{{}_{{}_2}}}\leq 80\,\,{\rm GeV}$,
$M_{\chi_{{}_{{}_3}}} = 7.1$ keV, $2\,\,{\rm MeV} < u\leq 10\,\,{\rm MeV}$, and
$10^{-9}\,\,{\rm GeV^2} \la \alpha \la 10^{-7}\,\,{\rm GeV^2}$.
The choice of the numerical ranges of these parameters will be justified
in Sections \ref{domenwall}, \ref{1-3gev-gamma} and \ref{xray}.
In order to avoid the late time decay of the heavier dark matter
particle $\chi_{{}_{{}_2}}$ into two lighter DM component $\chi_{{}_{{}_3}}$,
we adopt the value of the parameter $\lambda_{23} \leq 10^{-6}$. 
The constraint for all the plots of Fig. \ref{parameter-plot}
is that the calculations of relic densities with
the allowed values of the parameters must satisfy the 
PLANCK result for the total relic density of dark matter in the Universe.
It is also to be noted that, for plot (a) in Fig. \ref{parameter-plot},
we obtain two allowed regions in the parameter space 
$\lambda_{12}$-$\lambda_{13}$. The nonthermal production of the present 
dark matter candidates proceeds through two processes, namely the decay 
of the SM Higgs and pair annihilation of SM particles. But, 
if the mass of the heavier dark matter is higher than $M_{\chi_{{}_{{}_1}}}/2$, then 
it will not be produced through the decay of the SM Higgs boson. In this case 
the heavier dark matter can only be produced by the pair annihilation 
of SM particles. The lower allowed region in plot (a) 
between $\lambda_{12}$ and $\lambda_{13}$ is for the case when the mass of the 
heavier dark matter is less than $M_{\chi_{{}_{{}_1}}}/2$ while the other region 
is for the case when the nonthermal production of the same is through the
pair annihilation of SM particles. Needless to mention here that the 
lighter component can always be produced from the decay of the SM Higgs boson.
\begin{table}[h!]
\begin{center}
%\begin{flushleft}
\begin{tabular} {|c|c|c|c|c|}
\hline
{$\lambda_{12}$} & {$\lambda_{13}$}&{$\theta_{12}$}&$\theta_{13}$&$\theta_{23}$\\
&&(rad)&(rad)&(rad)\\
\hline
$\sim (0.18 - 1.6)$ &$\sim (0.56 - 2.6)$&$\sim(0.1-2.4)$&$\sim (0.2-2.0)$&$\sim (0.5-8.0)$\\
$\times10^{-11}$&$\times10^{-9}$&$\times 10^{-23}$&$\times10^{-11}$&$\times10^{-13}$\\
\hline
\end{tabular}
\end{center}
%\end{flushleft}
\caption{Allowed ranges for the model parameters $\lambda_{12}$, $\lambda_{13}$,
$\theta_{12}$, $\theta_{13}$, $\theta_{23}$.}
\label{table}
\end{table}
The allowed ranges for the model parameters $\lambda_{12}$, $\lambda_{13}$,
$\theta_{12}$, $\theta_{13}$, $\theta_{23}$, obtained from the plots a$-$c of
Fig. \ref{parameter-plot}, are given in a tabular form (Table \ref{table}).
Thus $\theta_{13}$, $\theta_{23}$ and $\theta_{12}$ turn out to be very small angles
in the range $\sim 10^{-11}$ rad, $10^{-13}$ rad and $10^{-23}$ rad respectively,
as the off-diagonal elements of the mass squared matrix of Eq. (\ref{mass-matrix})
are constrained to be very small. 
%%%%%%%%%%%%%%%%%%%%%%%%%%%%%%%%%%%%%%%%%%%%%%%%%%%%%%%%%%%%%%%%%%%%%%%%%%%%%%%%%%%%%%%%%
\section {Domain wall formation from restoration of $\mathbb{Z}_2^{\prime\prime}$}
\label{domenwall}
%%%%%%%%%%%%%%%%%%%%%%%%%%%%%%%%%%%%%%%%%%%%%%%%%%%%%%%%%%%%%%%%%%%%%%%%%%%%%%%%%%%%%%%%%
In this present model, though the residual discrete ${\mathbb{Z}^{\prime\prime}_2}$
symmetry is spontaneously broken by the VEV $u$ of the scalar field $S_3$,
there still remains the possibility that the symmetry could be restored
at a high temperature. This would result in the formation of domain walls.
Since the scalar field $S_1$ is in thermal equilibrium with the Universe,
the interaction between the scalars $S_1$ and $S_3$ results in a temperature
dependent mass term for the field $S_3$. The expression for the temperature
dependent mass term $\mu^2_{S_3}(T)$ is given by \cite{Dolan:1973qd}, 
\begin{equation}
\mu^2_{S_3}(T) = \frac{\lambda_{13}\,T^2}{2 \pi^2} \int_0^\infty dx
\frac{x^2}{\sqrt{x^2+M_{S_1}^2/T^2}} \frac{1}{e^{\sqrt{x^2+M_{S_1}^2/T^2}}-1}~.
\label{temp}
\end{equation}
We are considering the epoch of the Universe when $T \ll M_{S_1}$. In this limit,
we can write the above equation (Eq. (\ref{temp})) in the following 
approximate form
\begin{eqnarray}
\mu^2_{S_3}(T) & \simeq& \left(\frac{\lambda_{13}M_{S_1} T} {2 \pi^2}\right)\,
K_1\left(\frac{M_{S_1}}{T}\right), ~~~T \ll M_h \nonumber \\
&\simeq& \frac{\lambda_{13}}{2 \sqrt{2} \pi^2} T^2
\sqrt{\frac{M_{S_1}}{T}\,}e^{-M_{S_1}/T} \,
\left(1 + \frac{3}{8} \frac{T}{M_{S_1}} - 
\frac{15}{128} \frac{T^2}{M_{S_1}^2} + ...  \right) \,.
\end{eqnarray}
From Eq. (\ref{potential}) we see that there is a wrong sign mass term
$-\frac{\kappa_{{}_3}}{2} u^2$ of the field $S_3$. Combining this bare mass
term with the temperature dependent mass term given in Eq. (\ref{temp}),
we can define a quantity which is
\begin{equation}
M^2_{S_3}(T) = \mu^2_{S_3}(T) -\frac{\kappa_{{}_3}}{2} u^2 \,\, .
\label{Lambda}
\end{equation} 
\begin{figure}[h!]
\centering
\includegraphics[height=10cm,width=7cm,angle=-90]{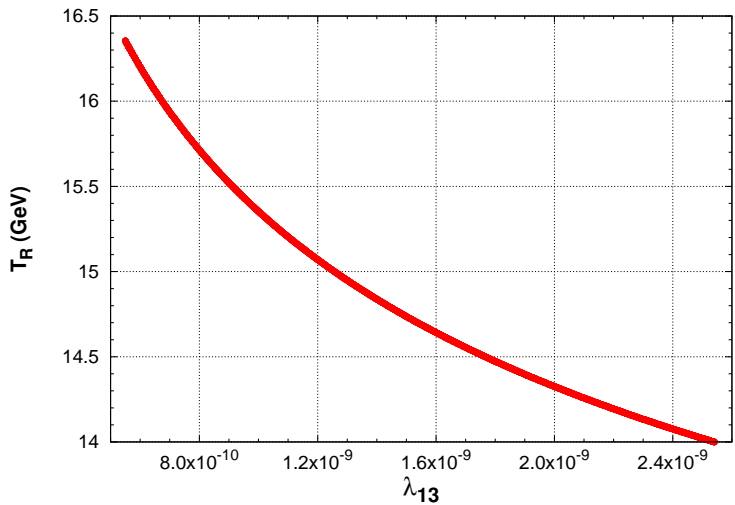}
\caption{Variation of discrete symmetry ($\mathbb{Z}^{\prime\prime}_2$)
restoration temperature with the parameter $\lambda_{13}$.}
\label{restemp-fig}
\end{figure}
The discrete symmetry $\mathbb{Z}^{\prime\prime}_2$ will be restored again if
the quantity $M^2_{S_3}(T)$ changes its sign and becomes positive. This will
happen only if the quartic coupling $\lambda_{13}$ between the fields $S_1$ and
$S_3$ (see Eq. (\ref{potential})) is positive. For negative values of $\lambda_{13}$,
which are equally allowed, there is no possibility that the spontaneously broken
discrete symmetry $\mathbb{Z}^{\prime\prime}_2$ could be restored again.  
The temperature at which this phase transition occurs
is defined as the symmetry restoration temperature $T_R$.
In Fig. \ref{restemp-fig}, we show the variation of $T_R$ with the 
allowed values of the coupling parameter $\lambda_{13}$. It is seen from 
Fig. \ref{restemp-fig} that for the allowed values of $\lambda_{13}$,
$T_R$ varies from $\sim 14$ GeV to $\sim 16.5$ GeV.  

We note that our $T_R$ is quite close to the discrete symmetry
restoration temperature $\sim 15.2$ GeV obtained in the model
of Babu and Mahapatra \cite{Babu:2014pxa}. Therefore, the properties of any domain
wall that is formed here will be similar to theirs. In consequence,
their discussion carries over to our case {\it mutatis mutandis} and the energy
density of such a domain wall would be too little either to overclose
the Universe or affect the observed near-isotropy of the CMB so long as the VEV
$u$ of the scalar field $S_3$ is bounded from above by $\sim 10$ MeV.
%%%%%%%%%%%%%%%%%%%%%%%%%%%%%%%%%%%%%%%%%%%%%%%%%%%%%%%%%%%%%%%%%%%%%%%%%%%%%%%%%%%%%%%%%%
\section{Direct detection of the dark matter candidates $\chi_{{}_{{}_2}}$ and $\chi_{{}_{{}_3}}$.}
\label{derect}
\begin{figure}[h!]
\centering
\includegraphics[height=3.5cm,width=4.5cm]{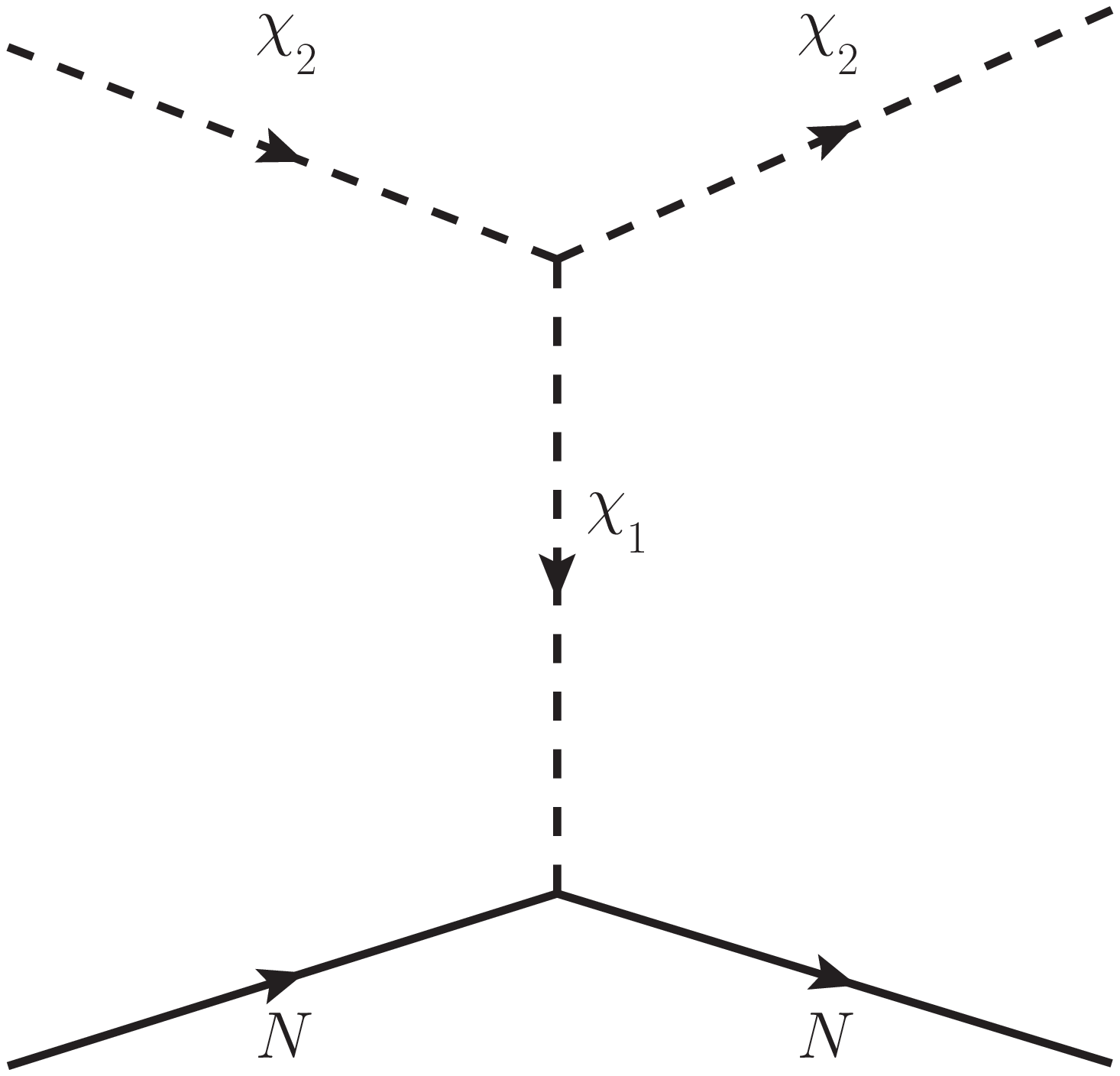}
\hspace{3cm}
\includegraphics[height=3.5cm,width=4.5cm]{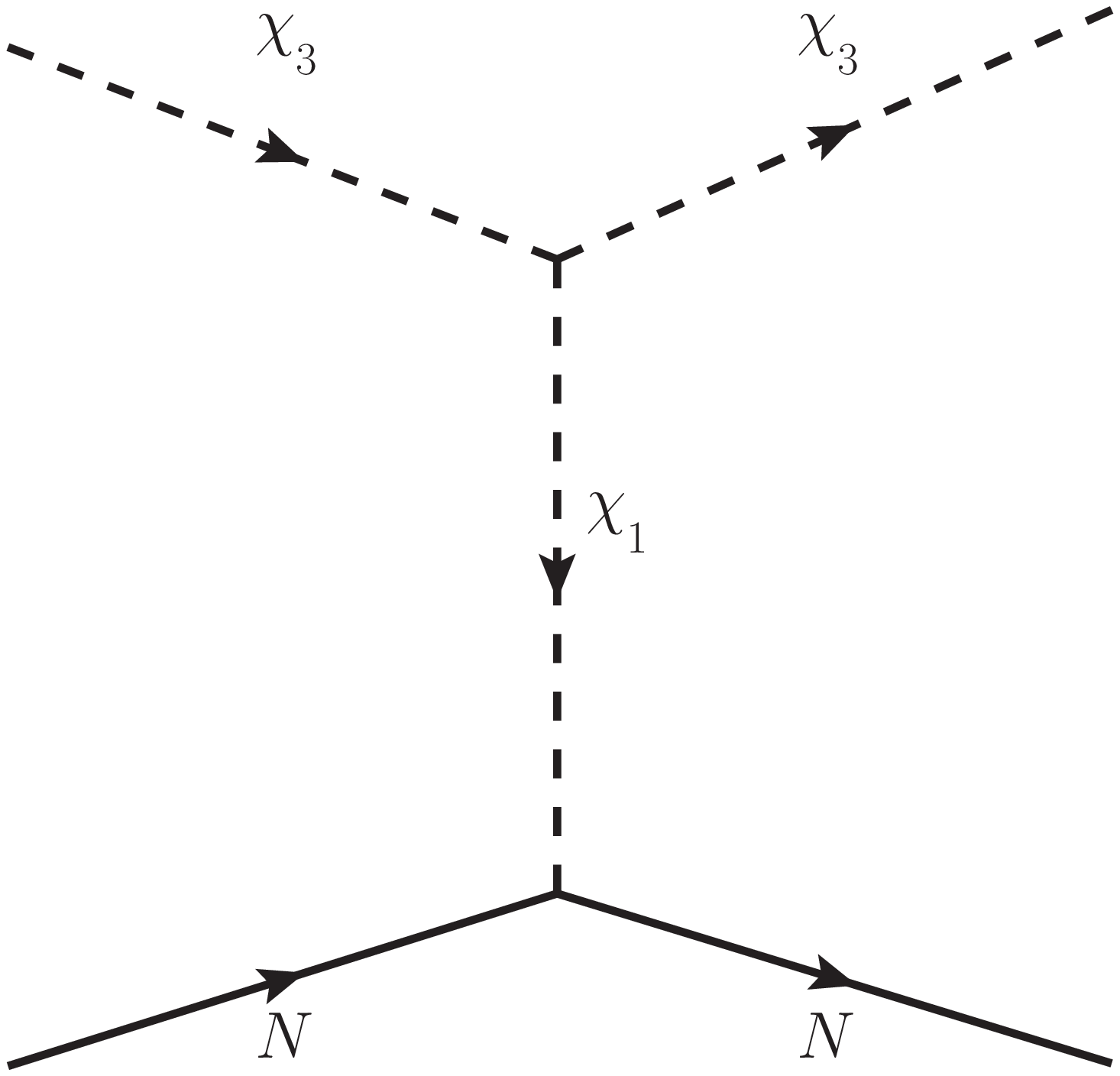}
\caption{Feynman diagrams for the spin independent elastic scattering between
each of the dark matter particles and the nucleon ($N$).}
\label{scatter}
\end{figure}

%%%%%%%%%%%%%%%%%%%%%%%%%%%%%%%%%%%%%%%%%%%%%%%%%%%%%%%%%%%%%%%%%%%%%%%%%%%%%%%%%%%%%%%%%% 
In the present two component dark matter model both the dark matter particles, namely
$\chi_{{}_{{}_2}}$ and $\chi_{{}_{{}_3}}$, scatter off the detector nuclei placed at
various underground laboratories. These scattering processes occur mainly through
the exchange of the SM-like Higgs boson $\chi_{{}_{{}_1}}$. The Feynman diagrams for
the spin independent scattering of both the dark matter particles with the
detector nucleon ($N$) are shown in Fig. \ref{scatter}.
The expression for spin independent elastic scattering cross section between the dark
matter component $\chi_{{}_{{}_i}}$ ($i = 2,\,3$)and the nucleon $N$ through the
exchange of $\chi_{{}_{{}_1}}$ is given by
\begin{eqnarray}
\sigma^{\chi_{{}_{{}_i}} N \rightarrow \chi_{{}_{{}_i}} N}_{{}_{SI}} =
\frac{\mu^2_{{}_{N\,i}}}{\pi}
\left(\frac{g_{\chi_{{}_{{}_1}}\chi_{{}_{{}_i}}\chi_{{}_{{}_i}}}}{v}\right)^2
\left(\frac{M_{{}_N}}
{M_{\chi_{{}_{{}_i}}}\,M^2_{\chi_{{}_{{}_1}}}}\right)^2\,f^2 \,\, ,
\label{scatter-cross}
\end{eqnarray}
where $\mu_{{}_{N\,i}}$ is the reduced mass between the dark matter component
$\chi_{{}_{{}_i}}$ and the nucleon $N$ of mass $M_{{}_N}$.
In Eq. (\ref{scatter-cross}) the coupling term among the fields
$\chi_{{}_{{}_1}}\,\chi_{{}_{{}_i}}\,\chi_{{}_{{}_i}}$
is represented by the quantity $g_{\chi_{{}_{{}_1}}\chi_{{}_{{}_i}}\chi_{{}_{{}_i}}}$
while $f \sim 0.3$ \cite{Barbieri:2006dq} is the usual nucleonic matrix element.
The expressions of the coupling term $g_{\chi_{{}_{{}_1}}\chi_{{}_{{}_i}}
\chi_{{}_{{}_i}}}$ for $i = 2,\,3$ are given in Appendix \ref{a1}.

We have already shown that in the present scenario the dark sector
is composed of two different scalar fields $\chi_{{}_{{}_2}}$ and $\chi_{{}_{{}_3}}$
whose masses as well as interaction strengths with the SM particles are different.
Therefore in order to compare the spin independent scattering cross section
between each of the dark matter particles and the nucleon, computed using
the present formalism with the results obtained from the present ongoing direct detection
experiments such as XENON 100 \cite{Aprile:2012nq}, LUX \cite{Akerib:2013tjd},
one needs to multiply the scattering cross section
$\sigma^{\chi_{{}_{{}_i}} N \rightarrow \chi_{{}_{{}_i}} N}_{{}_{SI}}$
by the factor $R_{\chi_{{}_{{}_i}}} = \frac{n_{\chi_{{}_i}}}{n_{\chi_{{}_2}} + n_{\chi_{{}_3}}}$
\cite{Biswas:2014hoa}, where $n_{\chi_{{}_i}}$ is the number density of the dark matter
candidate $\chi_{{}_{{}_i}}$. The rescaling of
$\sigma^{\chi_{{}_{{}_i}} N \rightarrow \chi_{{}_{{}_i}} N}_{{}_{SI}}$ is due
to the fact that the exclusion plots obtained from various dark matter direct
detection experiments are computed by assuming that the Universe contains only
one type of dark matter particle, which is not a valid assumption for the present scenario.
The spin independent scattering cross sections of both the dark matter
particles $\chi_{{}_{{}_2}}$, $\chi_{{}_{{}_3}}$ are given in Table \ref{tab2}.  
\begin{table}[h!]
\begin{center}
\begin{tabular} {|c|c|c|c|c|}
\hline
{Dark matter} & {Mass}&{${g_{\chi_{{}_{{}_1}}\chi_{{}_{{}_i}}\chi_{{}_{{}_i}}}}$}&
${\sigma^{\chi_{{}_{{}_i}} N \rightarrow \chi_{{}_{{}_i}} N}_{{}_{SI}}}$
&{$R_{\chi_{{}_{{}_i}}}$}\\
{candidate} ($\chi_{{}_{{}_i}}$)&{(GeV)}&{(GeV)}&{(pb)}&\\
\hline
$\chi_{{}_{{}_2}}$ &$50-80$&$(0.44-3.9)\times 10^{-9}$&
$(0.02-4.42)\times 10^{-27}$&$10^{-9}-10^{-6}$\\
\hline
$\chi_{{}_{{}_3}}$&$7.1\times 10^{-6}$&$(1.4-6.4)\times 10^{-7}$
&$(0.14-3.03)\times 10^{-19}$&$\sim 1$\\
\hline
\end{tabular}
\end{center}
\caption{Spin independent scattering cross sections of the dark
matter particles $\chi_{{}_{{}_2}}$ and $\chi_{{}_{{}_3}}$.}
\label{tab2}
\end{table}
It is seen from Table \ref{tab2} that the couplings for the vertices
$\chi_{{}_{{}_1}}\,\chi_{{}_{{}_2}}\,\chi_{{}_{{}_2}}$ and
$\chi_{{}_{{}_1}}\,\chi_{{}_{{}_3}}\,\chi_{{}_{{}_3}}$ are extremely small
(due to the nonthermal origin of $\chi_{{}_{{}_2}}$ and $\chi_{{}_{{}_3}}$).
Hence the spin independent scattering cross sections
of both the dark matter particles $\chi_{{}_{{}_2}}$ and $\chi_{{}_{{}_3}}$ lie well
below the present limit which is $\sim 10^{-9}$ pb to $10^{-10}$ pb for a dark matter
particle having mass in the range $10\,\,{\rm GeV}-100$ GeV \cite{Akerib:2013tjd}.    
%%%%%%%%%%%%%%%%%%%%%%%%%%%%%%%%%%%%%%%%%%%%%%%%%%%%%%%%%%%%%%%%%%%%%%%%%%%%%%%%%%%%%%%%%%
\section{$\gamma$-ray excess at 1-3 GeV energies from Galactic Centre.}
\label{1-3gev-gamma}
%%%%%%%%%%%%%%%%%%%%%%%%%%%%%%%%%%%%%%%%%%%%%%%%%%%%%%%%%%%%%%%%%%%%%%%%%%%%%%%%%%%%%%%%%%
Our endeavour in this section is to explain, within the framework 
of our proposed two component dark matter model, the $\gamma$-ray excess
from the Galactic Centre observed at an energy range 1 to 3 GeV  
by the Fermi-Large Area Telescope. One of our dark matter components,
namely $\chi_{{}_{{}_2}}$ having a mass in the range $\sim$ 50$-$80 GeV,
decays predominantly into the ${\rm b}\bar{\rm b}$ final state since it
has nonzero mixing with the SM-like Higgs boson $\chi_{{}_{{}_1}}$.
The $\chi_{{}_{{}_2}} {\rm b} \bar{\rm b}$ coupling ($g_{{\!}_{ff\2}}$, $f={\rm b}$)
can be read off from the expression given in Appendix \ref{a1}. It is noteworthy
that the small strength of the $\chi_{{}_{{}_1}} \chi_{{}_{{}_2}} \chi_{{}_{{}_2}}$ coupling
makes the pair annihilation of two $\chi_{{}_{{}_2}}$'s into ${\rm b} \bar{\rm b}$
(through an s channel $\chi_{{}_{{}_1}}$ exchange) a negligible competitor. We
believe that ours is the first model explaining the Fermi-LAT $\gamma$-ray excess
purely from the decay $\chi_{{}_{{}_2}} \rightarrow {\rm b} \bar{\rm b}$. 
The ${\rm b}$-quarks resulting from the decay of $\chi_{{}_{{}_2}}$ hadronise to produce
$\gamma$-rays, the spectrum of which should explain the 1-3 GeV excess as seen 
by Fermi-LAT. The differential $\gamma$-ray flux due to the decay of the
component $\chi_{{}_{{}_2}}$ in the channel
$\chi_{{}_{{}_2}}\rightarrow {\rm b} \bar{\rm b}$ in our
two component DM model is given by \cite{Cirelli:2010xx}            
\begin{eqnarray}
\frac{d\Phi_{\gamma}}{d \Omega d E} &=& \frac{r_{\odot}}{4\pi}
\frac{\rho_{\odot}}{M_{\2}}~\bar{J}~ \Gamma^\prime_{\chi_{{}_{{}_2}}
\rightarrow {\rm b} \bar{\rm b}} \frac{d N^{\rm b}_{\gamma}}{dE}\,\,.
\label{gamma-flux}
\end{eqnarray}
In Eq. (\ref{gamma-flux}), $\frac{dN^{\rm b}_{\gamma}}{dE}$ is the energy spectrum
of the photons produced with energy $E$ from the hadronisation of ${\rm b}$
quarks\footnote{These originate from the decay of the dark matter component
$\chi_{{}_{{}_2}}$ through the process $\chi_{{}_{{}_2}}
\rightarrow {\rm b} \bar{\rm b}$.}. We have used the numerical values
of the photon spectrum ($\frac{dN^{\rm b}_{\gamma}}{dE}$) for different photon
energies given in Ref. \cite{Cirelli:2010xx}. The value of the dark matter density
at the solar location, namely $\rho_{\odot}$, is taken to be 0.3 GeV$/$cm$^3$
while $r_{\odot}\simeq 8.5$ Kpc is the distance between the GC and the solar
location. We have averaged the $J$ factor over the opening solid angles:
\begin{eqnarray}
\bar {J} &=& \frac{4}{\Delta \Omega} \int \int db~dl \cos b~J( l, b) \,\,
\label{jbar}
\end{eqnarray}
with
\begin{eqnarray}
J(l, b) &=& \int_{\rm l.o.s} \frac{d\mathfrak{s}}{r_\odot} 
\left(\frac{\rho(r)}{\rho_{\odot}}\right) \,\, 
\label{j}
\end{eqnarray}
and
\begin{eqnarray}
\Omega = 4 \int dl \int db \cos b \,\, ,
\label{solidangle} 
\end{eqnarray}
\begin{eqnarray}
r &=& \left(r_{\odot}^2 + \mathfrak{s}^2 - 2\,r_{\odot}\,\mathfrak{s}\,
\cos b \cos l\right)^{{1}/{2}}\,\,\, .
\label{radius}
\end{eqnarray}
In Eqs. (\ref{jbar}), (\ref{solidangle}), (\ref{radius}), $l$ and $b$
denote the galactic longitude and latitude respectively and $\mathfrak{s}$ is the
line of sight distance. While computing
the values of $\bar{J}$, we have performed the integral over a region
which is situated within an angular distance of $5^o$ \cite{Daylan:2014rsa} around the GC.  
The integral over $\mathfrak{s}$ in Eq. (\ref{j}) is along the line of sight (l.o.s).
In the expression for differential $\gamma$-ray flux (Eq. (\ref{gamma-flux}))
the quantity ${\Gamma^\prime}_{\chi_{{}_{{}_2}}\rightarrow {\rm b} \bar{\rm b}}$
is the product of the decay width of the channel
$\chi_{{}_{{}_2}}\rightarrow {\rm b} \bar{\rm b}$ and the contribution
of the component $\chi_{{}_{{}_2}}$ to the total dark matter relic density
($\Omega_{\rm T} h^2$):
%{\it i.e.}
\begin{eqnarray}
{\Gamma^\prime}_{\chi_{{}_{{}_2}}\rightarrow {\rm b} \bar{\rm b}} =
\xi_{\chi_{{}_{{}_2}}} \Gamma_{\chi_{{}_{{}_2}}\rightarrow {\rm b} \bar{\rm b}} \,\, ,
\label{gamma'}
\end{eqnarray}
where 
\begin{eqnarray}
\xi_{\chi_{{}_{{}_2}}} = \frac{\Omega_{\chi_{{}_{{}_2}}}}{\Omega_{\rm T}} \,\,
\end{eqnarray}
is the fractional relic density for the component $\chi_{{}_{{}_2}}$.
The use of the modified decay width 
${\Gamma^\prime}_{\chi_{{}_{{}_2}}\rightarrow {\rm b} \bar{\rm b}}$ (Eq. \ref{gamma'}),
instead of the actual decay width of the channel
$\chi_{{}_{{}_2}}\rightarrow {\rm b} \bar{\rm b}$
(${\Gamma}_{\chi_{{}_{{}_2}}\rightarrow {\rm b} \bar{\rm b}}$)
in Eq. (\ref{gamma-flux}), is required here as we are dealing with a
two component dark matter model and the fractional amount of the
relevant component should be considered for the computation.  
Needless to mention here that, if the
entire dark sector consists of only one type of particle (say $\chi_{{}_{{}_2}}$),
then $\xi_{\chi_{{}_{{}_2}}} =1$, and consequently 
${\Gamma^\prime}_{\chi_{{}_{{}_2}}\rightarrow {\rm b} \bar{\rm b}}$
and ${\Gamma}_{\chi_{{}_{{}_2}}\rightarrow {\rm b} \bar{\rm b}}$ are identical. 
We calculate the decay width of the DM component $\chi_{{}_{{}_2}}$ for the
${\rm b} \bar{\rm b}$ final state (see Fig. \ref{x2bb} for the
Feynman diagram of this decay process) and the expression 
for this decay process is given below:
\begin{figure}[h!]
\centering
\includegraphics[height=4cm,width=6cm]{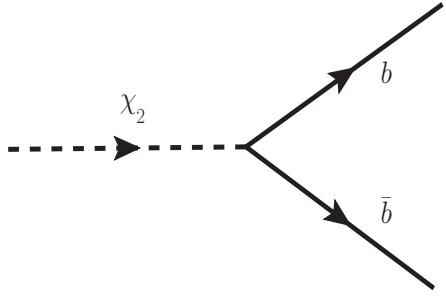}
\caption{Feynman diagram for the decay channel $\chi_{{}_{{}_2}}
\rightarrow {\rm b} \bar{\rm b}$.}
\label{x2bb}
\end{figure}
\begin{eqnarray}
\Gamma_{\chi_{{}_{{}_2}} \rightarrow {\rm b} \bar{\rm b}} = \frac{n_c~G_F}{4 \sqrt{2} \pi}
(\sin{\theta_{12}} \cos\theta_{23} + \cos\theta_{12} \sin\theta_{23} \sin\theta_{13})^2 
M_{\chi_{{}_{{}_2}}} M^2_{\rm b} \beta^3_{\rm b} \,\, , 
\end{eqnarray}
where
\begin{equation*}
\beta_{\rm b} = \left(1 - \frac{4 M^2_{\rm b}}{M^2_{\chi_{{}_{{}_2}}}}\right)^{{1}/{2}} \,\, .
\end{equation*}
In the above, $G_F=\frac{1}{\sqrt{2}v^2}$, with $v$ as defined in Eq. (\ref{higgs}),
is the Fermi constant and $M_b$ is the mass of the b quark.

For the computation of the $\gamma$-ray flux using Eq. (\ref{gamma-flux}),
the astrophysical input is the variation of dark matter density (as a function 
of the radial distance $r$) in the neighbourhood of the GC.
%In other words the functional dependence of $\rho(r)$ with $r$. 
This functional relation between $\rho(r)$ and $r$ is known as the halo profile
of DM. In this work, the computation of the gamma flux is done considering 
the NFW profile \cite{Navarro:1996gj}. The general expression of the NFW profile is given by, 
\begin{eqnarray}
\rho_{\rm NFW} = \rho_s \frac{\left(\frac{r}{r_s}\right)^{-\gamma}}
{\left(1+\frac{r}{r_s}\right)^{3-\gamma}} \,\,,
\end{eqnarray}
where $\gamma$ is a parameter (index) of order one. In the present calculation
$\gamma = 1$ is adopted. Now the halo profile reduces to the form
\begin{equation}
\rho_{\rm NFW} = \rho_s \frac{\left(\frac{r_s}{r}\right)}
{\left(1+\frac{r}{r_s}\right)^2}\ ,
\label{nfw_can}
\end{equation}
where $r_s =$ 20 Kpc. The normalisation constant $\rho_s$ is obtained
by demanding that the dark matter density at the solar 
location ($r = r_{\odot}$) is 0.3 GeV$/$cm$^3$.
\begin{figure}[h!]
\centering
\includegraphics[height=8.5cm,width=7cm,angle=-90]{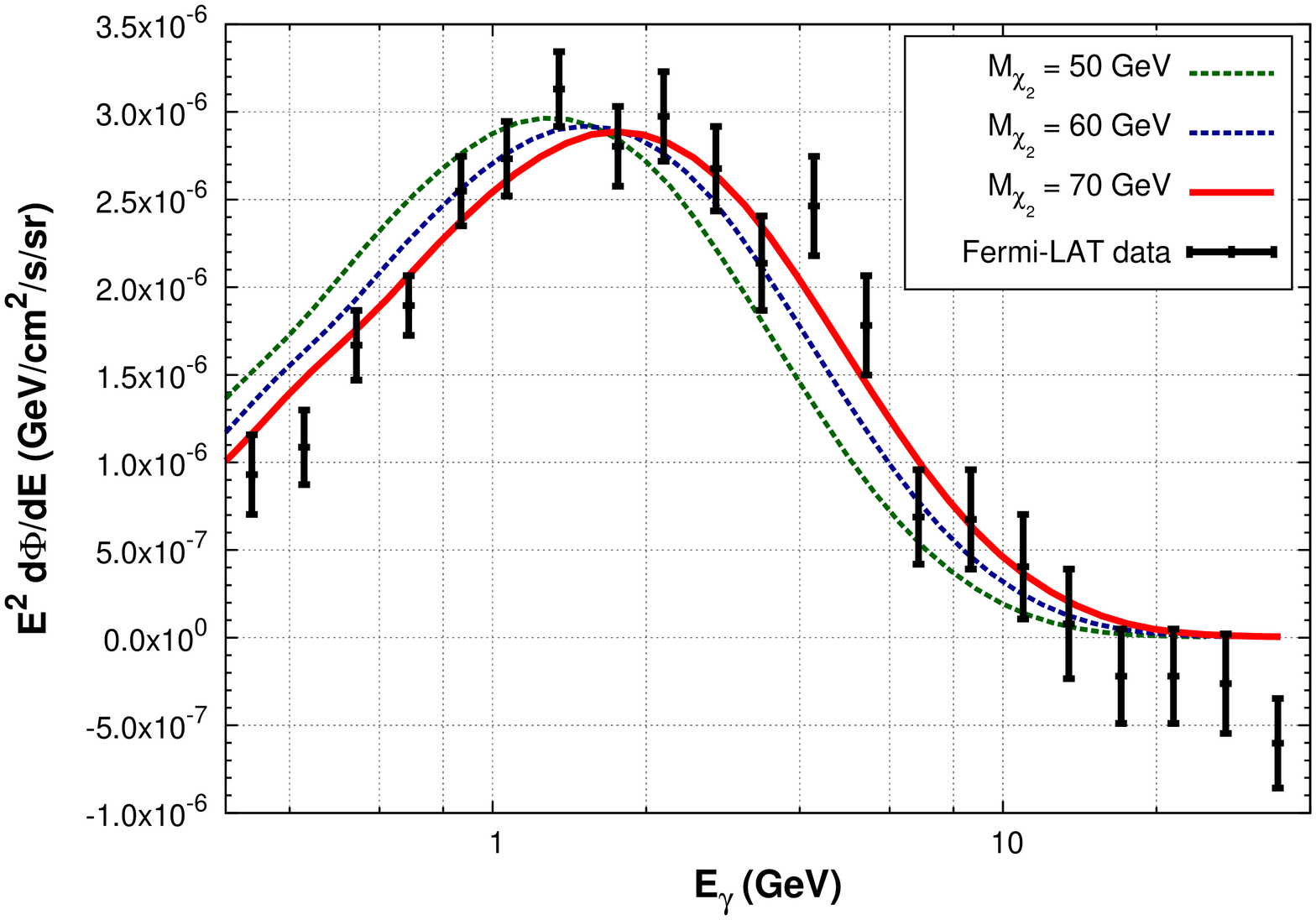}
\includegraphics[height=8.5cm,width=7cm,angle=-90]{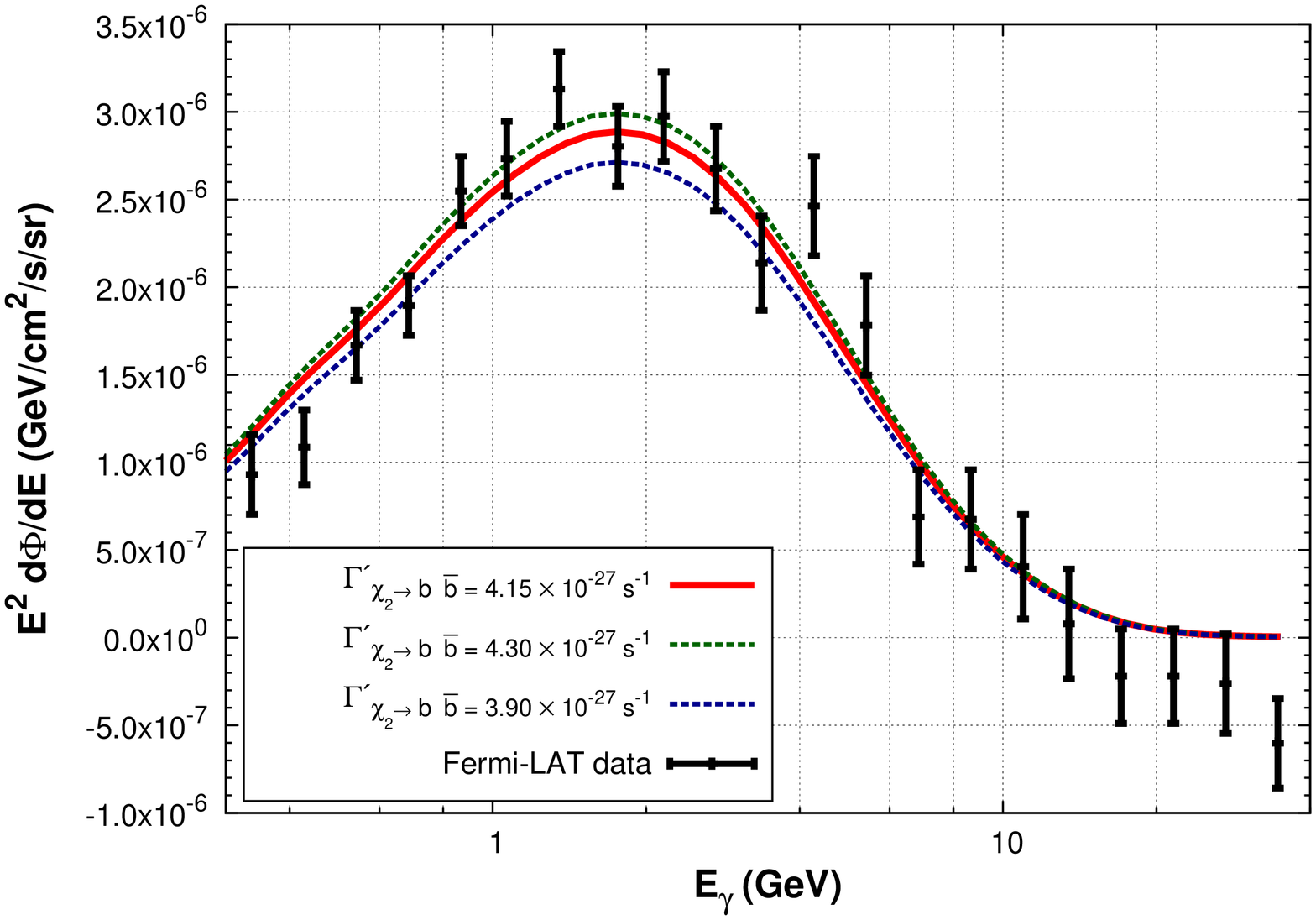}
\caption{Left Panel: $\gamma$-ray flux for three different values
of $M_{\chi_{{}_{{}_2}}} =$ 50, 60, 70 GeV. Right Panel: $\gamma$-ray flux
obtained from the decay of a 70 GeV dark matter particle ($\chi_{{}_{{}_2}}$) for
three different values of modified decay width of the channel
$\chi_{{}_{{}_2}} \rightarrow {\rm b} \bar{\rm b}$.}
\label{fermi-result-plot}
\end{figure}

The gamma-ray flux is calculated following the above discussion and the 
results are shown in Fig. \ref{fermi-result-plot}. The Fermi-LAT observational 
results (black vertical lines) are shown in Fig. \ref{fermi-result-plot} for comparison.
In the left panel of Fig. \ref{fermi-result-plot}, we plot the $\gamma$ flux from the Galactic 
Centre region for three values of $\chi_{{}_{{}_2}}$ namely $M_{\chi_{{}_{{}_2}}} = 50,\,\,60,\,\,70$
GeV. From the left panel of Fig. \ref{fermi-result-plot}, it is evident that 
the Fermi-LAT results are best described for the choice  
$M_{\chi_{{}_{{}_2}}} = 70$ GeV. In the right panel of Fig. \ref{fermi-result-plot}, 
we adopt the value $M_\chi = 70$ GeV and compare our calculated 
results (for the $\gamma$ flux) with the Femi-LAT 
observational data with three different values of modified decay widths 
${\Gamma^\prime}_{\chi_{{}_{{}_2}}\rightarrow {\rm b} \bar{\rm b}} =$
3.90$\times 10^{-27}$ s$^{-1}$ (blue dashed line), 4.15$\times 10^{-27}$ s$^{-1}$
(red solid line) and 4.30$\times 10^{-27}$ s$^{-1}$ (green dashed line) respectively
for the process ${\chi_{{}_{{}_2}}\rightarrow {\rm b} \bar{\rm b}}$.
The right panel of Fig. \ref{fermi-result-plot} shows that the decay width of
${\Gamma^\prime}_{\chi_{{}_{{}_2}}\rightarrow {\rm b} \bar{\rm b}} 
= 4.15\times10^{-27}$ s$^{-1}$ best represents the observational results from Fermi-LAT.
Here we note that obtaining ${\Gamma^\prime}_{\chi_{{}_{{}_2}}\rightarrow {\rm b} \bar{\rm b}}$
in the right ballpark of 3.90$\times 10^{-27}$ s$^{-1}$ to 4.30$\times 10^{-27}$ s$^{-1}$,
one requires the soft breaking parameter $\alpha$ to be in the range
$10^{-9}\,\,{\rm GeV}^{2}\la \alpha \la 10^{-7}\,\,{\rm GeV}^{2}$.
\begin{figure}[h!]
\centering
\includegraphics[height=8.5cm,width=7cm,angle=-90]{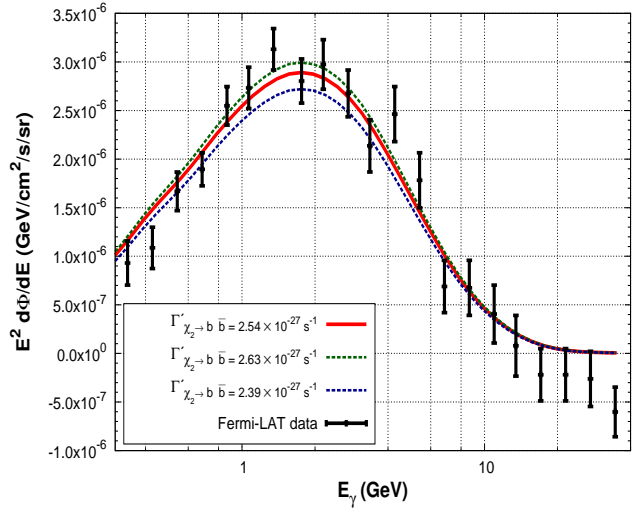}
\caption{$\gamma$-ray flux obtained from the decay of a 70 GeV
dark matter particle ($\chi_{{}_{{}_2}}$) for a
steeper NFW profile with $\gamma=1.26$. Three lines
represent the $\gamma$-ray flux for
three different values of modified decay width of the channel
$\chi_{{}_{{}_2}} \rightarrow {\rm b} \bar{\rm b}$.}
\label{fermi-result-plot-1.26}
\end{figure}

We had mentioned earlier that we have taken
the canonical NFW profile where the index $\gamma$ equals 1 in Eq. (\ref{nfw_can}).
But in Ref. \cite{Daylan:2014rsa}, where a model independent 
analysis of Fermi-LAT data with dark matter annihilation has been made 
to explain the gamma excess in the energy region 1-3 GeV,
a profile steeper than the canonical NFW profile
with $\gamma = 1.26$ was taken. The index $\gamma \sim 1.2$
was also required for explaining, within the framework of dark matter 
annihilation, the gamma signal from the Fermi bubble region in 
Ref. \cite{Hooper:2013rwa}. We therefore also compute the gamma flux 
from the Galactic Centre in the present framework using a steeper 
NFW profile with index $\gamma = 1.26$. The results are shown in Fig. \ref{fermi-result-plot-1.26}
where we plot the calculated gamma ray
flux (taking an NFW type profile where $\gamma = 1.26$)
with $M_{\2} = 70$ GeV and three values of the modified decay width
$\Gamma^\prime_{\2\rightarrow b \bar{b}}$. In this case the Fermi-LAT data 
are best represented when $\Gamma^\prime_{\2\rightarrow b \bar{b}} =
2.54 \times 10^{-27}$ s$^{-1}$.    
%%%%%%%%%%%%%%%%%%%%%%%%%%%%%%%%%%%%%%%%%%%%%%%%%%%%%%%%%%%%%%%%%%%%%%%%%%%%%%%%%%%%%%%%%%%%%
\section{3.55 keV X-ray line}
\label{xray}
\begin{figure}[h!]
\centering
\includegraphics[height=3cm,width=4.5cm]{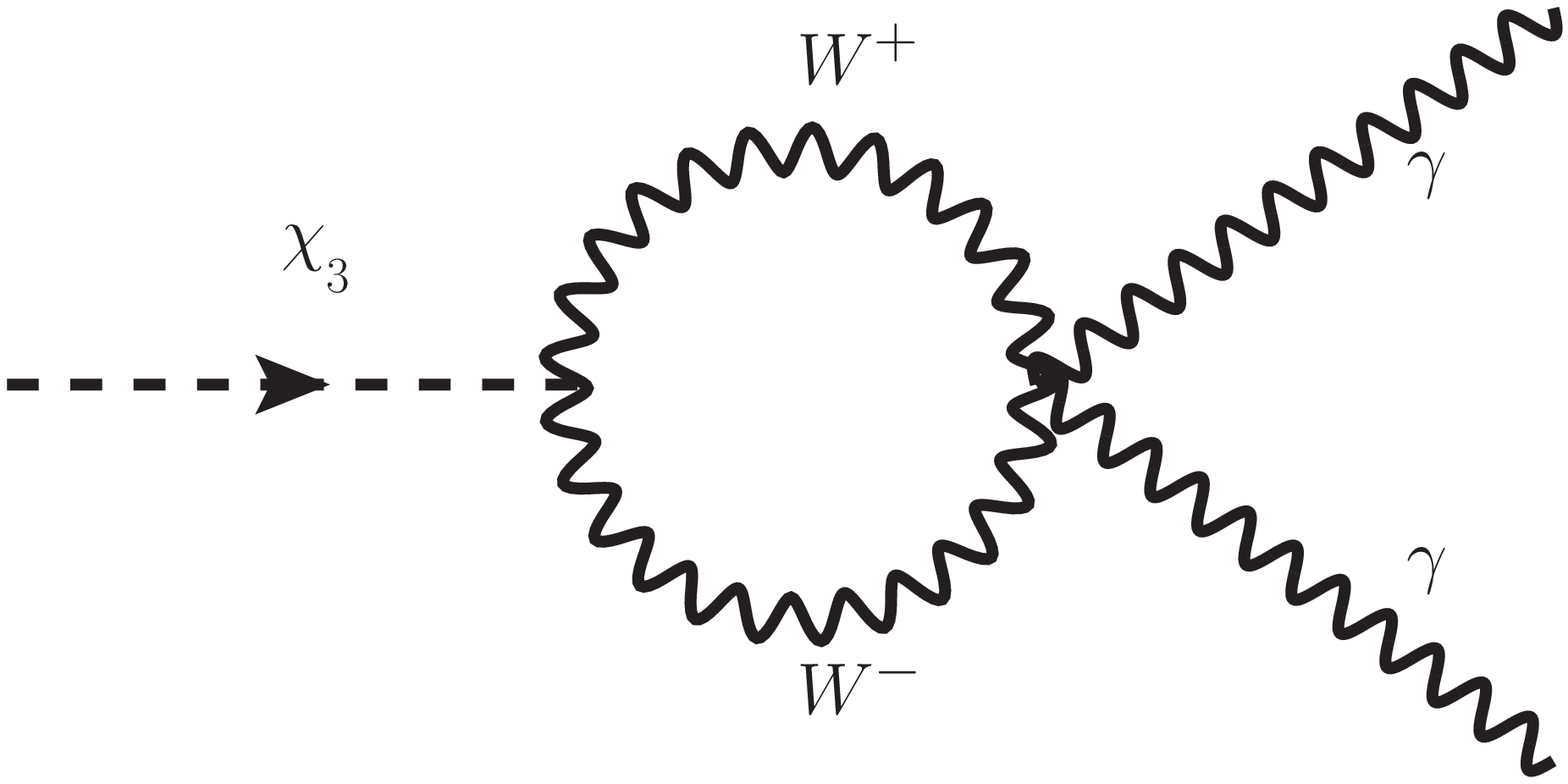}
\includegraphics[height=3cm,width=4.5cm]{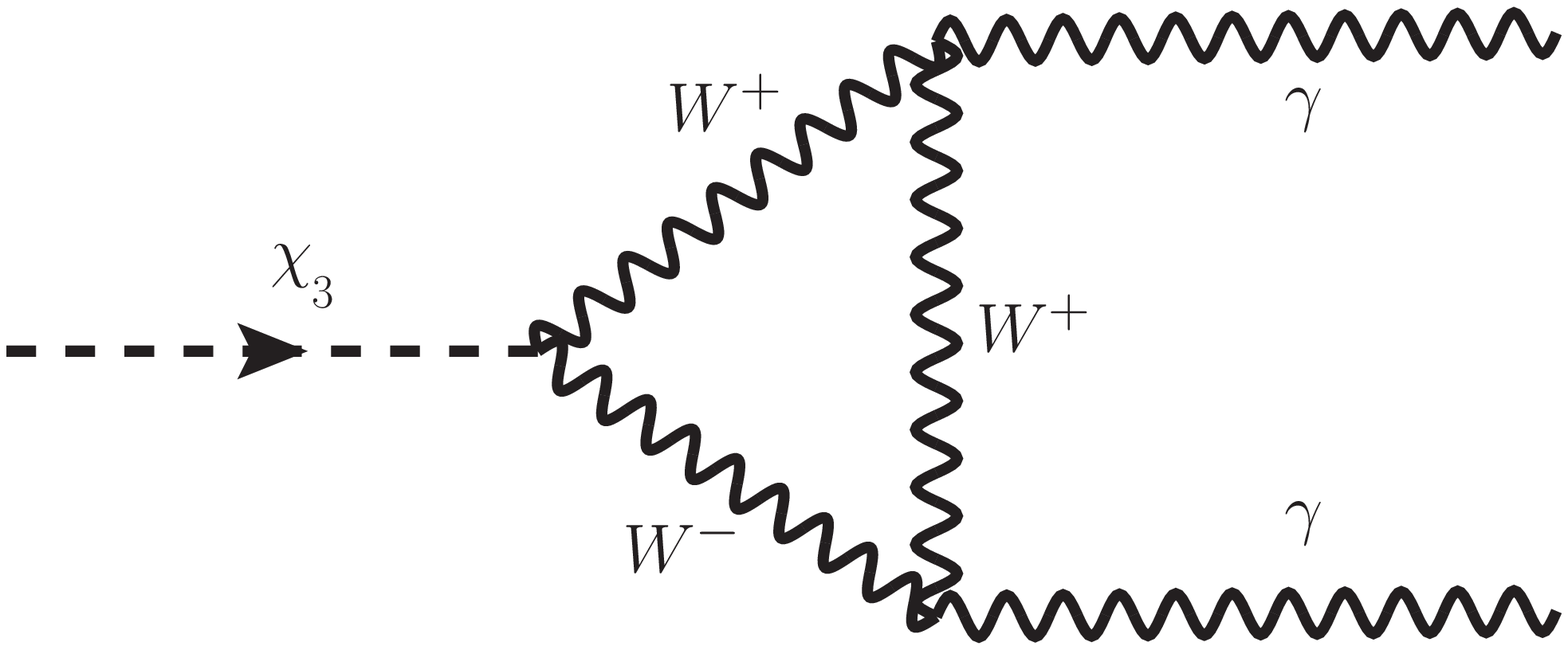}
\includegraphics[height=3cm,width=4.5cm]{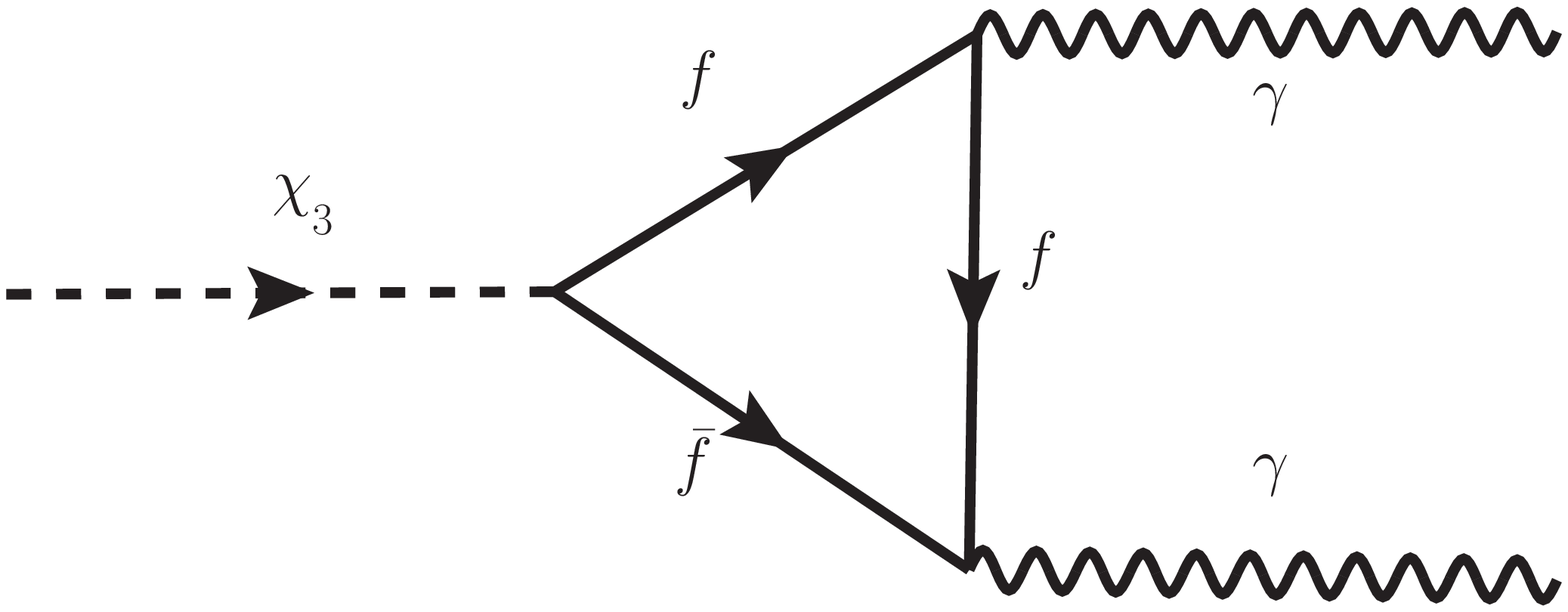}
\caption{One loop Feynman diagrams for the decay channel
$\chi_{{}_{{}_3}} \rightarrow \gamma \gamma$.}
\label{x3gg}
\end{figure}

%%%%%%%%%%%%%%%%%%%%%%%%%%%%%%%%%%%%%%%%%%%%%%%%%%%%%%%%%%%%%%%%%%%%%%%%%%%%%%%%%%%%%%%%%%%%%
We mentioned earlier that our present two component dark matter model
contains a scalar particle $\chi_{{}_{{}_3}}$, with a mass of order keV, which possesses 
a tiny mixing with the SM-like Higgs boson $\chi_{{}_{{}_1}}$. The two photon
decay mode of the dark matter candidate $\chi_{{}_{{}_3}}$ produces monoenergetic
photons (X-rays) which have been detected by the X-ray telescopes of the XMM-Newton
observatory. It has been reported in Refs. \cite{Bulbul:2014sua, Boyarsky:2014jta,
Higaki:2014qua} that, in order to produce the observed X-ray flux from the decay
of a dark matter particle, the corresponding decay width for the channel DM
$\rightarrow \gamma \gamma$ must be in the range
$\sim$ 2.5$\times 10^{-29}\,\,{{\rm s}^{-1}}$ to 2.5 $\times 10^{-28}\,\,{{\rm s}^{-1}}$.
In the present scenario, the above constraint should be applied on the modified decay
width instead of the actual decay width ($\Gamma_{\chi_{{}_{{}_3}} \rightarrow \gamma \gamma}$)
for the channel $\chi_{{}_{{}_3}} \rightarrow \gamma \gamma$, since we are working
in a framework where the entire dark sector is composed of two different scalar fields
$\chi_{{}_{{}_2}}$ and $\chi_{{}_{{}_3}}$. The modified decay width
($\Gamma^{\prime}_{\chi_{{}_{{}_3}} \rightarrow \gamma \gamma}$)
of $\chi_{{}_{{}_3}}$ for the channel $\chi_{{}_{{}_3}} \rightarrow \gamma \gamma$
is defined as,               
\begin{eqnarray}
{\Gamma^\prime}_{\chi_{{}_{{}_3}}\rightarrow \gamma \gamma} =
\xi_{\chi_{{}_{{}_3}}} \Gamma_{\chi_{{}_{{}_3}}\rightarrow \gamma \gamma}  
\end{eqnarray}
with
\begin{eqnarray}
\xi_{\chi_{{}_{{}_3}}} = \frac{\Omega_{\chi_{{}_{{}_3}}}}{\Omega_{\rm T}} \,\,
\end{eqnarray}
as the fractional contribution of the DM component $\chi_{{}_{{}_3}}$ to the
total relic density ($\Omega_{\rm T} h^2$). The Feynman diagrams for the decay channel
${\chi_{{}_{{}_3}} \rightarrow \gamma \gamma}$ are shown in Fig. \ref{x3gg}. 
The expression for the decay width for the channel
$\chi_{{}_{{}_3}} \rightarrow \gamma\gamma$,
which takes place at one loop, is given by
\begin{eqnarray}
\Gamma_{\chi_{{}_{{}_3}} \rightarrow \gamma \gamma} &=&
\frac{ G_F~m^3_{\chi_{{}_{{}_3}}}\alpha^2_{\rm em}}{128\sqrt{2}~\pi^3}\,\,
(\sin\theta_{12} \sin\theta_{23} - \cos\theta_{12}
\cos\theta_{23} \sin\theta_{13})^2 |F|^2,
 \nonumber \\ 
\end{eqnarray}
where
\begin{eqnarray}
F &=& F_W(\tau_{{}_W}) + \sum_f {n_c}_f\,Q_f^2\,F_f(\tau_{{}_f})\,\,
\end{eqnarray}
with
\begin{eqnarray}
\tau_{{}_W} &=& \frac{4 M_W^2}{M^2_{\chi_{{}_{{}_3}}}},~~
\tau_{{}_f} = \frac{4 M_f^2}{M^2_{\chi_{{}_{{}_3}}}},\nonumber \\
F_W(\tau_{{}_W}) &=& 2+3\tau_{{}_W}+3 \tau_{{}_W}(2-\tau_{{}_W})
\mathfrak{g}(\tau_{{}_W}),\nonumber\\
F_f(\tau_{{}_f}) &=& -2\tau_{{}_f}[1+(1-\tau_{{}_f})
\mathfrak{g}(\tau_{{}_f})],\nonumber\\
\mathfrak{g}(\tau) &=& \arcsin^2[\tau^{-1/2}]~.
\end{eqnarray}
Here $\alpha_{\rm em} \simeq \frac{1}{137}$ is the fine structure constant
while $Q_f$ and ${n_c}_f$ are the electric charge and the colour charge of
the fermion ($f$) \footnote{$f$ is any SM fermion} involved in the fermionic
loop of Fig. \ref{x3gg}.
\begin{figure}[h!]
\centering
\includegraphics[height=8.5cm,width=7cm,angle=-90]{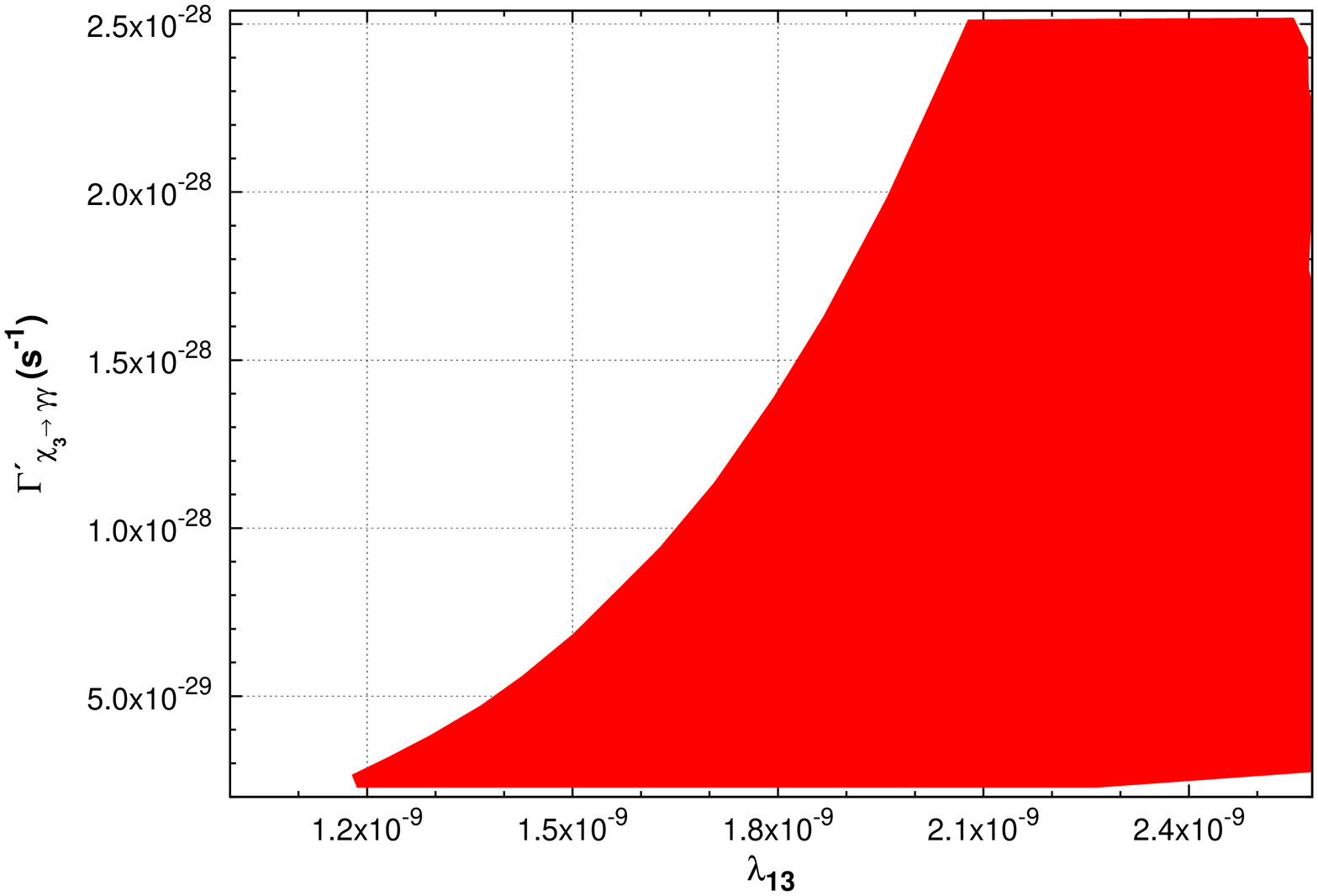}
\includegraphics[height=8.5cm,width=7cm,angle=-90]{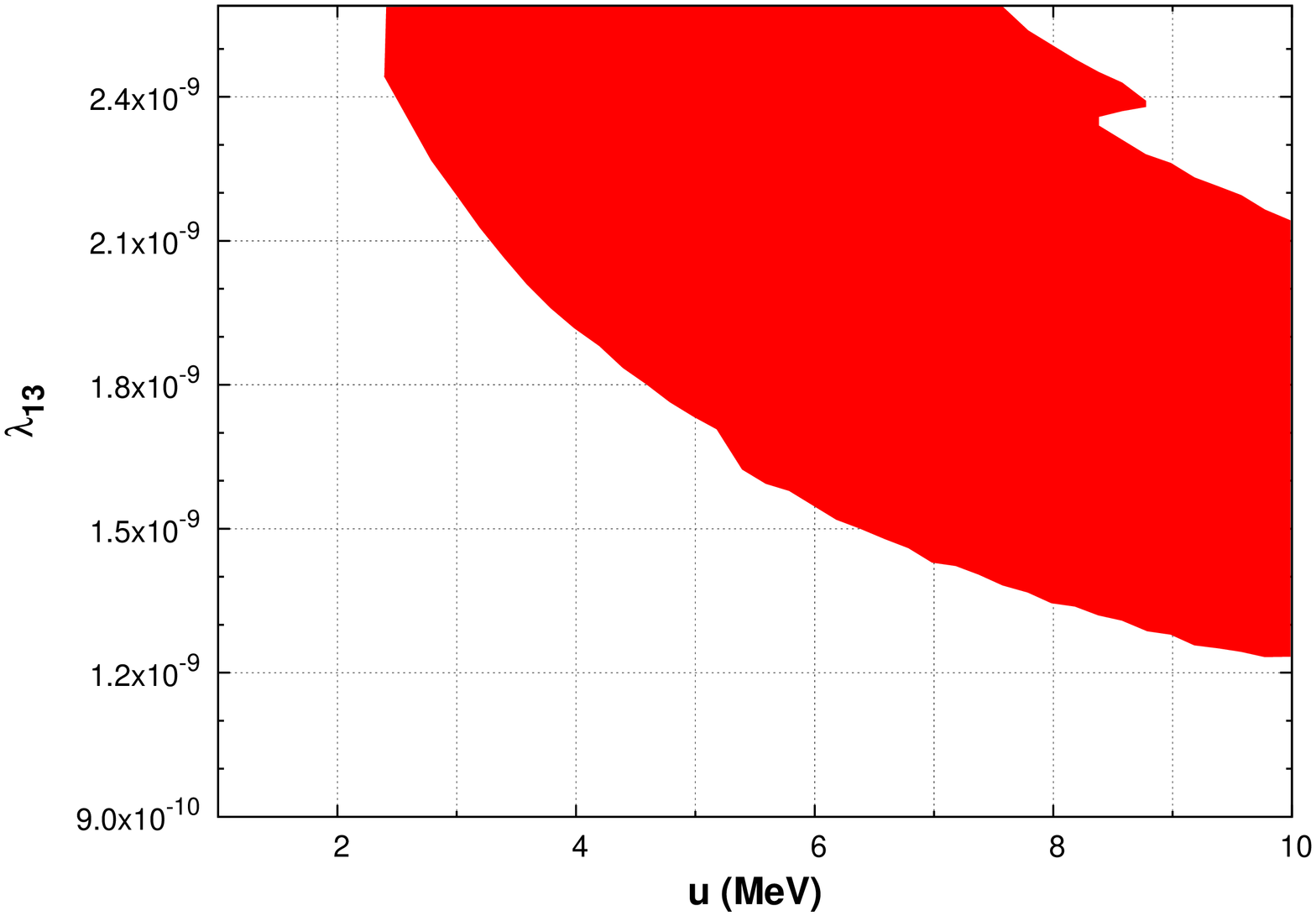}
\caption{Left panel: Variation of the modified decay width
${\Gamma^\prime}_{\chi_{{}_{{}_3}}\rightarrow \gamma \gamma}$ of the decay channel
$\chi_{{}_{{}_3}}\rightarrow \gamma \gamma$ with $\lambda_{13}$. Right panel: 
Allowed region in the $u-\lambda_{13}$ plane which produces
the observed X-ray flux as well as satisfies the PLANCK limit
on the total relic density.}
\label{xray-parameter-plot}
\end{figure}

In the left panel of Fig. \ref{xray-parameter-plot} we show the
allowed range $1.2 \times 10^{-9} \la \lambda_{13} \la 2.6 \times 10^{-9}$
of the parameter $\lambda_{13}$ for which the modified decay width (${\Gamma^\prime}_{\chi_{{}_{{}_3}}\rightarrow \gamma \gamma}$)
for the decay channel $\chi_{{}_{{}_3}} \rightarrow \gamma \gamma$
lies in the range 2.5$\times 10^{-29} \leq
{\Gamma^\prime}_{\chi_{{}_{{}_3}}\rightarrow \gamma \gamma} ({\rm s}^{-1})
\leq$ 2.5 $\times 10^{-28}$ which is necessary to produce the observed X-ray flux from the
extragalactic sources such as Perseus, Andromeda etc. The right panel of
Fig. \ref{xray-parameter-plot}  shows the allowed region in the $u-\lambda_{13}$
plane. From this plot (right panel of Fig. \ref{xray-parameter-plot}) one notices that,
in order to produce the observed X-ray flux, the VEV $u$ of the scalar field $S_3$
must be $> 2.4 $ MeV. The upper bound ($u \leq 10$ MeV) on the allowed values of $u$
comes from the domain wall constraint (see Section 3.4 of Ref. \cite{Babu:2014pxa})
which arises from the spontaneous breaking of the discrete symmetry $\mathbb{Z}^{\prime\prime}_2$.
Needless to mention here that, for all the points in both panels
of Fig. \ref{xray-parameter-plot}, the PLANCK limit on total relic density
$\Omega_{\rm T} h^2$ of the dark matter candidates is always satisfied.      
%%%%%%%%%%%%%%%%%%%%%%%%%%%%%%%%%%%%%%%%%%%%%%%%%%%%%%%%%%%%%%%%%%%%%%%%%%%%%%%%%%%%%%%%%%%%%%
\section{Conclusion}
\label{conclusion}
%%%%%%%%%%%%%%%%%%%%%%%%%%%%%%%%%%%%%%%%%%%%%%%%%%%%%%%%%%%%%%%%%%%%%%%%%%%%%%%%%%%%%%%%%%%%%%
In this paper we have presented a two-component model of nonthermal dark matter
by postulating two additional SM-singlet scalar fields $S_2$ and $S_3$ which interact
among themselves and with the SM Higgs doublet $H$. These interaction terms have been
restricted by assuming a suitable $\mathbb{Z}_2 \times \mathbb{Z}^\prime_2$ discrete
symmetry which is softly and explicitly broken to a residual $\mathbb{Z}^{\prime\prime}_2$
symmetry. The latter gets spontaneously broken when $S_3$ develops a VEV of order
MeV. Our physical scalar spectrum comprises three particles: (1) the SM-like Higgs
boson $\chi_{{}_{{}_1}}$ with $M_{\chi_{{}_{{}_1}}} \sim 125$ GeV, (2) a moderately
heavy scalar dark matter particle $\chi_{{}_{{}_2}}$ with 50 GeV
$\leq M_{\chi_{{}_{{}_2}}} \leq$ 80 GeV and (3) a very light scalar DM
particle $\chi_{{}_{{}_3}}$ with $M_{\chi_{{}_{{}_3}}} \sim 7$ keV.
We have discussed the issue of domain wall formation which arises in our model
due to spontaneous breaking of the discrete symmetry $\mathbb{Z}_2^{\prime\prime}$
and have explained why it is unimportant for us.
We have computed the $\gamma$-ray flux from the decay channel
$\chi_{{}_{{}_2}} \rightarrow {\rm b} \bar{\rm b}$ of the heavier
dark matter candidate $\chi_{{}_{{}_2}}$ while the X-ray line is generated
from the decay mode of the lighter DM candidate $\chi_{{}_{{}_3}}$ into a pair of two keV
energy photons. These two decay channels ($\chi_{{}_{{}_2}} \rightarrow {\rm b} \bar{\rm b}$,
$\chi_{{}_{{}_3}} \rightarrow \gamma \gamma$) exist due to the fact that both the DM candidates
in the present scenario possess tiny amounts of mixing with the SM-like Higgs boson $\chi_{{}_{{}_1}}$.
We have found that the $\gamma$-ray flux originating from the decay of a 70 GeV dark matter
particle ($\chi_{{}_{{}_2}}$) into a ${\rm b} \bar{\rm b}$ final state 
at the Galactic Centre with a modified decay width
${\Gamma^\prime}_{\chi_{{}_{{}_2}}\rightarrow {\rm b} \bar{\rm b}}
= 4.15\times10^{-27}\,{\rm s}^{-1}$ fits the Fermi-LAT data well.
Finally, we have shown that the modified decay width of the
lighter dark matter component $\chi_{{}_{{}_3}}$ for the channel
$\chi_{{}_{{}_3}} \rightarrow \gamma \gamma$ lies in the appropriate range of
$2.5\times 10^{-29} \,\,{\rm s}^{-1}$ to $2.5\times 10^{-28} \,\,{\rm s}^{-1}$
as long as the VEV $u$ is bounded from below by 2.4 MeV.   
%%%%%%%%%%%%%%%%%%%%%%%%%%%%%%%%%%%%%%%%%%%%%%%%%%%%%%%%%%%%%%%%%%%%%%%%%%%%%%%%%%%%%%%%%%%%%%     
\section{Acknowledgement}
%%%%%%%%%%%%%%%%%%%%%%%%%%%%%%%%%%%%%%%%%%%%%%%%%%%%%%%%%%%%%%%%%%%%%%%%%%%%%%%%%%%%%%%%%%%%%%
One of the authors A.B. would like to thank D. Adak, M. Chakraborty, A. Dutta Banik and
N. Haque for many useful suggestions and discussions. The research of A.B. has been funded
by the Department of Atomic Energy (DAE) of Govt. of India. P.R. has been supported as a
Senior Scientist by the Indian National Science Academy.  
%%%%%%%%%%%%%%%%%%%%%%%%%%%%%%%%%%%%%%%%%%%%%%%%%%%%%%%%%%%%%%%%%%%%%%%%%%%%%%%%%%%%%%%%%%%%%%%
\appendix
\section{Appendix}
%%%%%%%%%%%%%%%%%%%%%%%%%%%%%%%%%%%%%%%%%%%%%%%%%%%%%%%%%%%%%%%%%%%%%%%%%%%%%%%%%%%%%%%%%%%%%%%%
\subsection{Couplings of the physical scalars
$\chi_{{}_{{}_1}}$, $\chi_{{}_{{}_2}}$ and $\chi_{{}_{{}_3}}$}
\label{a1}
The couplings of the physical scalars $\chi_{{}_{{}_1}}$, $\chi_{{}_{{}_2}}$,
$\chi_{{}_{{}_3}}$ $-$ among themselves and with other SM particles are $-$
given below in the limit when all three mixing angles $\theta_{12}$,
$\theta_{13}$, $\theta_{23}$ are extremely small. 
\begin{eqnarray}
g_{{\!}_{\chi_{{}_{{}_1}}\chi_{{}_{{}_1}}\chi_{{}_{{}_2}}\chi_{{}_{{}_2}}}} &\simeq &
-\theta_{23}(3 \theta_{12} \theta_{13} \kappa_{{}_1} - 3 \theta_{12} \theta_{13} \lambda_{12}
- 2 \theta_{12} \theta_{13} \lambda_{13} + 4 \theta_{12} \theta_{13} \lambda_{23})
- \frac{\lambda_{12}}{2}\,\, ,  \nonumber \\
g_{{\!}_{\chi_{{}_{{}_1}}\chi_{{}_{{}_1}}\chi_{{}_{{}_3}}\chi_{{}_{{}_3}}}} &\simeq &
-\theta_{23}(-3 \theta_{12} \theta_{13} \kappa_{{}_1} + 3 \theta_{12} \theta_{13} \lambda_{12}
+ 2 \theta_{12} \theta_{13} \lambda_{13} - 4 \theta_{12} \theta_{13} \lambda_{23})
- \frac{\lambda_{13}}{2}\,\, ,  \nonumber \\
g_{{\!}_{\chi_{{}_{{}_2}}\chi_{{}_{{}_2}}\chi_{{}_{{}_3}}\chi_{{}_{{}_3}}}} &\simeq &
-\lambda_{23} - \theta_{23}(3 \theta_{12} \theta_{13} \kappa_{{}_2} - 
3 \theta_{12} \theta_{13} \lambda_{12} + 3 \theta_{12} \theta_{13} \lambda_{13} - 
6 \theta_{12} \theta_{13} \lambda_{23})\,\, , \nonumber \\
g_{{\!}_{\chi_{{}_{{}_1}}\chi_{{}_{{}_1}}\chi_{{}_{{}_1}}}} &\simeq & 
-v \kappa_{{}_1} - \theta_{13} u \lambda_{13} \,\, , \nonumber \\
g_{{\!}_{\chi_{{}_{{}_2}}\chi_{{}_{{}_2}}\chi_{{}_{{}_2}}}} &\simeq & 
\theta_{12} v \lambda_{12} - \theta_{23} (2 u \lambda_{23}-\theta_{13} v \lambda_{12})
\,\, , \nonumber \\
g_{{\!}_{\chi_{{}_{{}_3}}\chi_{{}_{{}_3}}\chi_{{}_{{}_3}}}} &\simeq &
-u \kappa_{{}_3} + \theta_{13} v \lambda_{13} - 
\theta_{23} (-2 \theta_{12} \theta_{13} u \lambda_{13} + \theta_{12} v \lambda_{13}
+ 4 \theta_{12} \theta_{13} u \lambda_{23}) \,\, , \nonumber \\
g_{{\!}_{\chi_{{}_{{}_1}}\chi_{{}_{{}_2}}\chi_{{}_{{}_2}}}} &\simeq &
-\theta_{23} \left(6 \theta_{12} \theta_{13} v \kappa_{{}_1} -
4 \theta_{12} \theta_{13} v \lambda_{12} - 2 \theta_{12} u \lambda_{13}
- 2 \theta_{12} \theta_{13} v \lambda_{13} + 4 \theta_{12} u \lambda_{23}\right)\nonumber \\
&& -~v \lambda_{12} - 2 \theta_{13} u \lambda_{23} \nonumber \\
g_{{\!}_{\chi_{{}_{{}_1}}\chi_{{}_{{}_3}}\chi_{{}_{{}_3}}}} &\simeq &
-\theta_{23} \left(-6 \theta_{12} \theta_{13} v \kappa_{{}_1} +
4 \theta_{12} \theta_{13} v \lambda_{12} + 2 \theta_{12} u \lambda_{13}
+ 2 \theta_{12} \theta_{13} v \lambda_{13} - 4 \theta_{12} u \lambda_{23}\right)\nonumber \\
&& -~v \lambda_{13} + 2 \theta_{13} u \lambda_{13} - 3 \theta_{13} u \kappa_{{}_3} \nonumber \\
g_{{\!}_{\chi_{{}_{{}_2}}\chi_{{}_{{}_3}}\chi_{{}_{{}_3}}}} &\simeq &
-\theta_{23} \left(3 u \kappa_{{}_3} + 2 \theta_{13} v \lambda_{12} 
- 3 \theta_{13} v \lambda_{13} - 4 u \lambda_{23}\right)
- 2 \theta_{12} \theta_{13} u \lambda_{13} + \theta_{12} v \lambda_{13}
\nonumber \\
&& +~4 \theta_{12} \theta_{13} u \lambda_{23}  \,\, ,\nonumber \\
g_{{\!}_{\chi_{{}_{{}_1}}\chi_{{}_{{}_2}}\chi_{{}_{{}_3}}}} &\simeq &
-\theta_{23} \left(6 \theta_{13} u \kappa_{{}_3} - 2 v \lambda_{12} - 4 \theta_{13} u \lambda_{13}
+ 2 v \lambda_{13} - 4 \theta_{13} u \lambda_{23}\right) -
6 \theta_{12} \theta_{13} v \kappa_{{}_1} \nonumber \\ 
&& +~4 \theta_{12} \theta_{13} v \lambda_{12}
+ 2 \theta_{12} u \lambda_{13} 
+ 2 \theta_{12} \theta_{13} v \lambda_{13}
- 4 \theta_{12} u \lambda_{23} \,\, , \nonumber\\
g_{{\!}_{WW\chi_{{}_{{}_1}}}} &\simeq &
\frac{2 M^2_W}{v}\,\, , \nonumber \\
g_{{\!}_{WW\chi_{{}_{{}_2}}}} &\simeq &
-\frac{2 M^2_W}{v}(\theta_{12}+\theta_{13}\theta_{23})\,\, , \nonumber \\
g_{{\!}_{WW\chi_{{}_{{}_3}}}} &\simeq &
-\frac{2 M^2_W}{v} (\theta_{13} - \theta_{12}\theta_{23})\,\, , \nonumber \\
g_{{\!}_{ZZ\chi_{{}_{{}_1}}}} &\simeq &
\frac{M^2_W}{v}\,\, , \nonumber \\
g_{{\!}_{ZZ\chi_{{}_{{}_2}}}} &\simeq &
 -\frac{M^2_W}{v}(\theta_{12}+\theta_{13}\theta_{23})\,\, , \nonumber \\
g_{{\!}_{ZZ\chi_{{}_{{}_3}}}} &\simeq &
-\frac{M^2_W}{v} (\theta_{13} - \theta_{12}\theta_{23}) \,\, , \nonumber \\
g_{{\!}_{ff\chi_{{}_{{}_1}}}} &\simeq &
-\frac{M_f}{v}\,\, , \nonumber \\
g_{{\!}_{ff\chi_{{}_{{}_2}}}} &\simeq &
\frac{M_f}{v}(\theta_{12}+\theta_{13}\theta_{23})\,\, , \nonumber \\
g_{{\!}_{ff\chi_{{}_{{}_3}}}} &\simeq &
\frac{M_f}{v} (\theta_{13} - \theta_{12}\theta_{23})\,\, . \nonumber 
\end{eqnarray}
\subsection{Masses of physical scalars $\chi_{{}_{{}_1}}$, $\chi_{{}_{{}_2}}$ and $\chi_{{}_{{}_3}}$}
\label{a2}
The masses of the physical scalar $\chi_{{}_{{}_1}}$, $\chi_{{}_{{}_2}}$ and $\chi_{{}_{{}_3}}$ in the
limit of extremely small $\theta_{12}$, $\theta_{13}$, $\theta_{23}$  are given as fellows:
\begin{eqnarray}
M_{\chi_{{}_{{}_1}}} &=& \left(2 \kappa_{{}_1} v^2 + 4  u v \theta_{13}\lambda_{13} 
+ 2\alpha \theta_{12} \theta_{13}\right)^{1/2}
+{\cal O} (\theta^2_{12}) + {\cal O} (\theta^2_{23}) + {\cal O} (\theta^2_{13})
\,\, , \nonumber \\
\,\,\, \nonumber \\
%\end{eqnarray}
%\begin{eqnarray}
M_{\chi_{{}_{{}_2}}} &=& \left\{\rho_{{}_2}^2 + \lambda_{12} v^2 + 2\lambda_{23} u^2 
+2 \theta_{23} \left(\alpha - \rho_{{}_2}^2 \theta_{12} \theta_{13} +
2 v^2 \kappa_{{}_1} \theta_{12} \theta_{13} 
- v^2 \lambda_{12} \theta_{12} \theta_{13} 
\right.\right.\nonumber\\&&\left.\left.
- 2 u v \lambda_{13} \theta_{12}  - 
2 u^2 \lambda_{23} \theta_{12} \theta_{13} \right)\right\}^{1/2} 
%\nonumber \\ &&
+ {\cal O} (\theta^2_{12}) + {\cal O} (\theta^2_{23}) + {\cal O} (\theta^2_{13})
\,\, , \nonumber \\
\,\,\, \nonumber \\
%\end{eqnarray}
%\begin{eqnarray}
M_{\chi_{{}_{{}_3}}} &=& \left\{2 \kappa_{{}_3} u^2 -2 \alpha \theta_{12} \theta_{13} 
- 4 u v \theta_{13} \lambda_{13}
- 2 \theta_{23} \left(\alpha - \rho_{{}_2}^2 \theta_{12} \theta_{13} +
2 v^2 \kappa_{{}_1} \theta_{12} \theta_{13} 
\right.\right.\nonumber\\&&\left.\left.
- v^2 \lambda_{12} \theta_{12} \theta_{13} 
- 2 u v \lambda_{13} \theta_{12}  - 
2 u^2 \lambda_{23} \theta_{12} \theta_{13} \right)\right\}^{1/2} 
%\nonumber \\ &&
+ {\cal O} (\theta^2_{12}) + {\cal O} (\theta^2_{23}) + {\cal O} (\theta^2_{13})
\,\, . \nonumber
\end{eqnarray}
%%%%%%%%%%%%%%%%%%%%%%%%%%%%%%%%%%%%%%%%%%%%%%%%%%%%%%%%%%%%%%%%%%%%%%%%%%%%%%%%%%%%%%%%%%%

\end{document}